\documentclass[preprint,12pt]{elsarticle}
\usepackage{float}
\usepackage{geometry}

\usepackage{amsmath}
\usepackage{cases}
\usepackage{hyperref}
\usepackage{amssymb}
\usepackage{graphicx}
\usepackage{subfigure}
\usepackage{pict2e}
  \usepackage{graphicx} 
  \usepackage{epstopdf}
    \usepackage{color, soul}
    \usepackage{tcolorbox}
    \usepackage{framed}
\colorlet{shadecolor}{blue!20}
  \setulcolor{red}
  \setstcolor{blue}
\sethlcolor{green}
\soulregister\cite7
\soulregister\ref7
\usepackage[bordercolor=white,backgroundcolor=gray!30,linecolor=black,colorinlistoftodos]{todonotes}
\makeatletter
\renewcommand{\@thesubfigure}{\hskip\subfiglabelskip}
\makeatother

\biboptions{numbers,sort&compress}

\begin{document}
\begin{frontmatter}



\title{On the risks of using double precision \\ in numerical simulations of spatio-temporal chaos}


\author{Tianli Hu $^{a}$}
\author{Shijun Liao $^{a, b, c, d}$  \, \corref{cor1}}
\ead{sjliao@sjtu.edu.cn} \cortext[cor1]{Corresponding author}

\address[1]{Center of Advanced Scientific Computing \\
School of Naval Architecture, Ocean and Civil Engineering, Shanghai Jiao Tong University, China}

\address[2]{State Key Laboratory of Ocean Engineering, Shanghai 200240, China}

\address[3]{State Key Laboratory of Plateau Ecology and Agriculture,  Xining 810018, China}

\address[4]{School of Hydraulic and Electric Engineering,  Qinghai University,  Xining 810018,  China}

\begin{abstract}
      Due to the butterfly-effect,   computer-generated chaotic simulations  often  deviate  exponentially from  the true solution,   so that it is very hard to obtain a reliable simulation of  chaos in a long-duration time.  In this paper,  a new strategy of the so-called Clean Numerical Simulation (CNS) in physical space is proposed for spatio-temporal chaos,  which is computationally much more efficient than its predecessor (in spectral space).   The strategy of the CNS is to reduce both of the truncation and round-off errors to a specified level by implementing high-order algorithms in multiple-precision arithmetic (with sufficient significant digits for all variables and parameters) so as to guarantee that numerical noise is below such a critical level in a temporal interval $t\in[0,T_c]$ that corresponding numerical simulation  remains reliable over the whole interval.   Without loss of generality,  the complex Ginzburg-Landau equation (CGLE)  is used to illustrate   its validity.   As a result,   a reliable long-duration numerical simulation of the CGLE is achieved in the whole spatial domain over a long interval of time $t\in[0,3000]$, which is used as a reliable benchmark solution to investigate the influence of  numerical noise by comparing it with the corresponding ones given by the 4th-order Runge-Kutta method in double precision (RKwD).   Our results demonstrate that the use of double precision in simulations of chaos might lead to huge errors in the prediction of spatio-temporal trajectories and in statistics, not only quantitatively but also qualitatively,  particularly in a long interval of time.

\end{abstract}


\end{frontmatter}


\section{Introduction}

In 1890 Poincar\'{e} \cite{Poincare1890a} discovered that the trajectories of an $N$-body system ($N \geq 3$), governed by Newtonian gravitational attraction, generally have sensitive dependance on the initial condition (SDIC), i.e. a tiny difference in initial condition might lead to a completely different trajectory.   The so-called SDIC was rediscovered in 1963 by Lorenz  \cite{Lorenz1962Deterministic} who numerically solved an idealized model of weather prediction, nowadays called the Lorenz equation, by means of a digital computer. The SDIC became popularized following the title of a talk in 1972 by Lorenz to the American Association for the Advancement of Science as ``Does the flap of a butterfly's wings in Brazil set off a tornado in Texas?". In his seminal paper, Lorenz stated that ``long-term prediction of chaos is impossible'' \cite{Lorenz1962Deterministic}.   The discovery of Poincar\'{e} \cite{Poincare1890a} and Lorenz \cite{Lorenz1962Deterministic} has helped create a completely new field in science, called chaos theory.

Furthermore,  Lorenz \cite{Lorenz1989Computational, Lorenz2006Computational}  found  that computer-generated simulations of a chaotic dynamic system using the Runge-Kutta method in the {\em double}  precision  are sensitive {\em not only}  to the initial condition {\em but also} to the numerical algorithms (including temporal/spatial discretization).  For the specific problem investigated, Lorenz \cite{Lorenz2006Computational}  reported  that the Lyapunov exponent, one of the most important characteristics of a dynamic system,  frequently  varied  its sign  even when the time-step became rather small. It is well-known that only  a positive Lyapunov exponent corresponds to chaos; a negative one does not!   This is easy to understand from the viewpoint of the so-called butterfly-effect,  because  numerical noise from truncation and round-off error is {\em unavoidable}  in {\em any}  numerical simulation, and leads to some errors  (or approximations) being introduced to the solution at each time-step.   For example, let us consider the high-order Taylor expansion
\begin{equation}
f(t+\Delta t) \approx  f(t) + \sum_{m=1}^{M} \frac{f^{(m)}(t)}{m!} \; (\Delta t)^m,   \label{TaylorExpandMethod:f[t]}
\end{equation}
where $f(t)$ is a function of time $t$, $\Delta t$ is the time-step, $M$ is the order of Taylor expansion and  $f^{(m)}(t)$ is the $m$th-order derivative of $f(t)$.  Given that $M$ is \emph {finite} in practice, we obtain a truncation error as the difference between the results of the infinite and truncated Taylor series. Moreover,  because all variables and physical/numerical parameters, such as $f(t)$, $f^{(m)}(t)$, $\Delta t$ and so on, are expressed in a precision (usually double precision) limited by a \emph{finite} number of significant digits, a round-off error invariably arises. Logically, a chaotic system with the butterfly-effect should also be rather sensitive to these {\em man-made} numerical errors.  This kind of  sensitive dependence on numerical algorithm (SDNA) for a chaotic system has been confirmed by many researchers \cite{jianping2000computational, jianping2001computational, Teixeira2005Time}, and has led to some intense arguments and serious doubts concerning the reliability of numerical simulations of chaos.  It has even been hypothesized that ``all chaotic responses are simply numerical noise and have nothing to do with the solutions of differential equations'' \cite{Yao2008Comment}.  Note that the foregoing researchers  \cite{jianping2000computational, jianping2001computational, Teixeira2005Time, Yao2008Comment}  undoubtedly used  data  in {\em double}  precision  for  chaotic  systems,  although different types of numerical algorithm were tested.  As a consequence, it was not clear how numerical noise  influenced  the reliability of numerical simulations of chaos.

  Without doubt, the reliability of computer-generated  simulation of chaotic systems is a very important fundamental problem.
Anosov \cite{ANOSOV1967} in 1967 and Bowen \cite{BOWEN1975} in 1975 showed that {\em uniformly hyperbolic} systems  have the so-called shadowing property: a computer-generated (or noisy) orbit will stay close to (shadow) true orbits with {\em adjusted} initial conditions for arbitrarily long times.   Let $\{{\bf p}_k\}_{k=a}^{b}$ denote a $\delta_f$-pseudo-orbit for ${\bf f}$ if $|{\bf p}_{k+1} - {\bf f} ( {\bf p} _k ) | < \delta_f$ for $a < k < b$, where $\bf f$ is a $D$-dimensional map and $\bf{p}$ is a $D$-dimensional vector representing the dynamical variables.  Here the term pseudo-orbit is used to describe a computer-generated noisy orbit.   The true orbit $\{{\bf y}_k\}_{k=a}^{b}$ $\delta$-shadows $\{{\bf p}_k\}_{k=a}^{b}$ on $[a, b]$ if $|{\bf y}_k - {\bf p}_k | < \delta$, where the true orbit $\{{\bf y}_k\}_{k=a}^{b}$ satisfies ${\bf y}_{k+1} = {\bf f} ({\bf y}_k)$.  For details of the shadowing theorems, please refer to \cite{ANOSOV1967, BOWEN1975, hammel1987numerical, hammel1988numerical, grebogi1990shadowing, sauer1991rigorous}.

However,  hardly any chaotic systems are uniformly hyperbolic.   Dawson et al. \cite{dawson1994obstructions} pointed out that the non-existence of such shadowing trajectories may be caused by finite-time Lyapunov exponents of a system fluctuating about zero.     For non-hyperbolic chaotic systems with unstable dimension variability, no true trajectory of reasonable length can be found to exist near any computer-generated trajectories, as reported by Do and Lai \cite{do2004statistics}.   Using examples having only two degrees of freedom,  Sauer \cite{sauer2002shadowing} showed that extremely small levels of numerical noise might result in macroscopic errors even in simulation average {\em statistics}  that are several orders of  magnitude  larger than the noise level.   Thus,  the good promise from the Shadow Lemma \cite{ANOSOV1967, BOWEN1975} for a uniformly hyperbolic system does not work for many non-hyperbolic systems \cite{Anishchenko2000}.  It seems that  we  had  to be satisfied with  finite-length shadows for non-hyperbolic systems \cite{hayes2007fast}.
Note that most of investigations on shadows are  based  on  low dimensional dynamic systems, although a shadowing algorithm for high dimensional systems (such as  the motion of one hundred stars) was proposed by  Hayes and Jackson  \cite{hayes2007fast}.
To the best of our knowledge, neither rigorous shadowing theorems nor practical shadowing algorithms have been proposed for spatio-temporal chaotic systems (such as turbulent flows), which are governed by nonlinear PDEs and thus have an infinite number of dimensions.    It seems that, although the shadowing concept is rigorous in mathematics, it is hardly used in practice for complicated systems such as spatio-temporal chaos and turbulence that are non-hyperbolic in general and besides have an infinite number of degrees of freedom.

Liao \cite{Liao2009On,  Liao2013On,  liao2017clean} suggested a numerical strategy called ``Clean Numerical Simulation'' (CNS) that could provide reliable simulations of chaotic systems over a given finite interval of time. The CNS is based on an {\em arbitrary}-order Taylor expansion method \cite{Gibbon1960, Barton1971, Corliss1977, Corliss1982Solving, Barrio2005VSVO},  multiple-precision (MP) data  \cite{Oyanarte1990} with  {\em arbitrary} numbers of significant digits, and a solution verification check \cite{oberkampf2010verification, roy2010complete} (using another simulation for the same physical parameters but with even smaller numerical noise).  Here, the word ``arbitrary'' means a value which can be ``as large as required''.   So, as long as the order $M$ of the Taylor expansion (\ref{TaylorExpandMethod:f[t]})  is \emph{sufficiently} large  and   all  data values are expressed in  multiple-precision with a \emph {sufficiently}  large number of significant digits (denoted by $N_s$),  both the truncation error and the round-off error can be held below specified thresholds throughout a prescribed simulation time $[0,T_c]$. Here the so-called  critical predictable time $T_c$ is determined through the solution verification check by comparing it to an additional simulation with the same physical parameters but smaller numerical noise, such that the difference between them remains negligible in the interval $[0,T_c]$.   Thus, a basic task of the CNS is to determine the relationship between $T_c$ and numerical noise.  This relationship needs clarification particularly as it has been largely ignored by researchers working on chaotic dynamics.

 The CNS is based on such a  hypothesis that the level of numerical noise increases exponentially (in an average meaning) within an interval of time $t\in[0,T_c]$, say,
\begin{equation}
 {\cal E}(t) =  {\cal E}_0 \; \exp(\mu \; t), \hspace{1.0cm}  t\in[0,T_c], \label{def:exponent:mu}
\end{equation}
where the constant $\mu>0$ is the so-called ``noise-growing exponent'' that is dependent upon the physical parameters of the system, ${\cal E}_0$ denotes the level of initial noise (i.e. truncation and round-off error),  ${\cal E}(t)$  is  the level of evolving noise of the computer-generated simulation at the time $t$, respectively.   The critical predictable time $T_c$ is determined by a critical level of noise ${\cal E}_c$, say,
\begin{equation}
{\cal E}_c = {\cal E}_0  \exp (\mu \; T_c),    \label{def:delta-c}
\end{equation}
which gives
\begin{equation}
T_c = \frac{\left( \ln {\cal E}_c -\ln {\cal E}_0\right)}{\mu}.     \label{Tc-delta0}
\end{equation}
So, for a given critical level of noise ${\cal E}_c$,  the smaller the level of the initial noise ${\cal E}_0$, the larger the critical  predictable time of $T_c$.

However, it is practically impossible to exactly calculate the level of evolving noise ${\cal E}(t)$, because the true solution is unknown.  So, we should propose a practical method with sufficiently high accuracy to determine $T_c$.   Let $\bf x \in \Omega$ denote a vector of  spatial  coordinates  and $\psi({\bf x},t)$ a computer-generated simulation reliable in $t\in [0,T_c]$ with the level of initial noise ${\cal E}_0$,   $\psi'({\bf x},t)$ is another computer-generated simulation reliable in $t\in [0,T'_c]$ with a level of initial noise ${\cal E}'_0$  that is several orders of magnitude smaller than ${\cal E}_0$.  According to the hypothesis of exponential growth in numerical noise of chaotic systems,  it holds that $T'_c > T_c$ and $\psi'({\bf x},t)$ in $t\in[0,T_c]$ should be much closer to the true solution than $\psi({\bf x},t)$ and thus can be regarded as a {\em reference} to approximately calculate the level of numerical noise of   $\psi({\bf x},t)$ in ${\bf x} \in \Omega$.
Note that a similar idea was used by Turchetti et al. \cite{turchetti2010asymptotic} who compared a numerical map computed for a given accuracy (single floating-point precision)  with the same map evaluated numerically with a higher accuracy (double or higher floating-point precision)  that is regarded as the reference map.
 In practice,  the value of $T_c$  can be determined by comparing $\psi({\bf x},t)$ with another better simulation $\psi'({\bf x},t)$ having smaller initial noise ${\cal E}'_0$.   In this way, the level of numerical noise of $\psi({\bf x},t)$ is not beyond the critical level ${\cal E}_c$ of noise  in the temporal interval $t\in[0,T_c]$  within the whole spatial domain $\bf x \in \Omega$.   In other words,  $\psi({\bf x},t)$ is a ``clean'' numerical simulation (CNS) in $t\in[0,T_c]$ and  $\bf x \in \Omega$.   The above-mentioned can be regarded as a heuristic explanation of the strategy of the CNS.

 Liao \cite{Liao2009On}  successfully used  this strategy of the CNS to gain reliable chaotic simulations of the Lorenz equation and indeed  found that the smaller the numerical noise, the larger the value of $T_c$. Liao also found that when the data are of high multiple-precision, such that $N_s$ is sufficiently large for round-off errors to be ignored and the truncation error dominates, $T_c$ is linearly proportional to $M$, the order of Taylor expansion. Moreover, when $M$ reaches such a sufficiently high value that round-off error dominates, then $T_c$ becomes linearly proportional to $N_s$, the number of significant digits. Using these linear relationships, it is possible to determine values for $M$ and $N_s$ that apply to \emph{any} prescribed $T_c$.  Here,  $T_c$ is finite due to computer limitations, but can be quite large, depending on the computer resources available.     For example, Liao \cite{Liao2009On} applied  CNS to obtain a reliable chaotic simulation of  Lorenz equation  until $T_c$=1000 Lorenz time unit  (LTU) by means of $\Delta t= 0.01$, $M=400$ and $N_s$=800.  Furthermore, Liao and Wang \cite{LIAO2014On}  used the CNS to obtain, for the first time,  a convergent/reliable chaotic simulation of Lorenz equation up to $T_c$=10000 LTU by means of $M = 3500$ and $N_s = 4180$, using 1200 CPUs to the National Supercomputer TH-1A  in  Tianjin, China.  Note that the reliability of this  long-term chaotic simulation was verified by means of  another better simulation using $M = 3600$ and $N_s = 4515$.   These showed the validity of the CNS.   Note that, exactly for the same physical parameters and the same initial conditions of the Lorenz equation, by means of the 4th-order Runge-Kutta's method and many other algorithms in {\em double} precision,  a reliable chaotic simulation is often obtained over a rather small interval, approximately $[0,30]$ LTU, which is only 0.3 \% of the interval [0,10000]. This may be  the underlying reason why there is such controversy about the reliability of numerical simulations of chaos  \cite{Yao2008Comment}, noting that  most researchers neglect the influence of round-off error  and  use \emph {double} precision in their computer-generated simulations.

 In CNS,  numerical noise arising from truncation and round-off errors is much smaller than the values of physical variables under consideration,  provided $t < T_c$,  where $T_c$ is a specified parameter that can be as large as deemed necessary by the user. Consequently, many complicated chaotic systems can be studied by means of the CNS, as illustrated in \cite{liao2014can, liao2014physical, liao2015inherent, Li2017More, li20171223, Lin2017On, Li2019-NewAstronomy}.   For example,  according to Poincar\'{e} \cite{Poincare1890a}, a three-body system can often be chaotic.   In physics, the initial positions of the three-bodies  have inherent micro-level physical uncertainty, at scales below the Planck length scale. Such micro-level uncertainty in the initial position is much smaller than the round-off error  caused by the use of  double precision, and so its influence on macroscopic  trajectories  cannot be investigated by  traditional algorithms using double precision arithmetic.  However, using CNS, Liao and Li \cite{liao2015inherent}  investigated the propagation of  micro-level physical uncertainty in the initial condition for a chaotic three-body system, and found that the uncertainty grows to become  macroscopic, leading to random escape and symmetry-breaking behaviour of the three-body system.  This implies that micro-level physical uncertainty might be the origin of  macroscopic randomness  in the three-body system.   Besides,  given that  micro-level  uncertainty  is  physically  inherent,  the escape and  symmetry-breaking behaviour of the chaotic three-body system  can  happen  even {\em without}  any  external  forces  present.   In other words,   such behaviour is self-excited. This suggests that macroscopic randomness, self-excited random escape, and self-excited symmetry-breaking  of a chaotic three-body system are unavoidable. (Liao  and Li \cite{liao2015inherent} provide further details.)   Notably,  although the three-body problem can be traced back to Newton in 1680s,  only three families of periodic orbits were found in the 300 years since then.  In 1890  Poincar\'{e} \cite{Poincare1890a}  pointed out  that a three-body system  is  chaotic in general and its closed-form solution does not exist.   This is  why so few periodic orbits have been discovered.  However,  by undertaking CNS on China's national supercomputer, Li, Jing and Liao \cite {Li2017More, li20171223, Li2019-NewAstronomy}  successfully  found  more than 2000 new periodic orbits of three-body system.  The new periodic orbits were profiled in New Scientists \cite{NewScientist2017, NewScientist2018}.   All of these illustrated the usefulness of CNS as a powerful tool for reliable and accurate investigation of chaotic systems in physics.

The Lorenz equation is a greatly simplified form of the Navier-Stokes equations that are widely used to describe turbulent flows. In 2017  Lin \emph{et al} \cite{Lin2017On} applied the CNS to study the relationship between  inherent  micro-level thermal fluctuation and the macroscopic  randomness of a two-dimensional  Rayleigh-B\'{e}rnard convection turbulent flow in a long enough interval of time, taking micro-level thermal fluctuation (expressed by Gaussian white noise) to be the initial condition,  with no external disturbances applied.  Using CNS with numerical noise set to be even smaller than the micro-level thermal fluctuations, Lin {\em et al} \cite{Lin2017On}   proved theoretically that inherent  micro-level thermal fluctuations are the root source of macroscopic  randomness of Rayleigh-B\'{e}rnard turbulent convection flows. Unlike the Lorenz system and three-body system,  which are described mathematically by ordinary differential equations (ODEs),  the Rayleigh-B\'{e}rnard convection system is governed by partial differential equations (PDE) as a spatio-temporal chaotic system.  This case illustrated the validity of the CNS for reliable simulation of spatio-temporal chaotic systems governed by nonlinear PDEs.   However, Lin \emph {et al} \cite{Lin2017On}  used a Galerkin-Fourier spectral method with CNS to solve the system in spectral space. This involved mapping the original nonlinear PDE in physical space onto a huge system of nonlinear ODEs in spectral space, proving to be rather time-consuming and impractical on present-day computers for simulation for spatio-temporal chaotic systems.

In this paper, we propose a new strategy that greatly increases the computational efficiency of CNS for spatio-temporal chaotic systems. Without loss of generality,  we consider the one-dimensional complex Ginzburg-Landau equation(CGLE) as an example of a spatio-temporal chaotic system, to outline the basic features of the strategy and then illustrate its validity and efficiency.  Section 2 describes two CNS algorithms: one in spectral space using the Galerkin-Fourier spectral method; the other in physical space. The performance of these algorithms is compared in terms of computational efficiency, validity, etc.   Section 3 discusses  the influence of numerical noise on spatio-temporal  trajectories and statistics of the chaotic system.  Section 4 summarizes the discussions and conclusions.

\section{CNS algorithms for spatio-temporal chaos\label{Numerical_algorithms}}

  Spatio-temporal chaos, characterized by irregular behaviour in both space and time, arises when a spatially extended system is driven away from its equilibrium state \cite{Cross1993Pattern}.  The one-dimension complex Ginzburg-Landau equation (CGLE),  which describes oscillatory media near the Hopf bifurcation, is commonly used in studies of spatio-temporal chaos \cite{sakaguchi1990breakdown,  Sch1991Small,  shraiman1992spatiotemporal, Bohr1998Dynamical, brusch2000modulated,   van2001ordered, aranson2002world,   degond2008time, garcia2012complex, uchiyama2014birth}, and is given by
\begin{equation}
   A_{t}=A+(1+ i \, c_{1})\;A_{xx}-(1- i \, c_{3})\;\left| A\right|^2A,   \label{cgle}
\end{equation}
subject to the initial condition
 \[  A(x,0)=f(x) \]
and the periodic boundary condition
 \[  A(x,t) = A(x+L, t), \]
 where $i = \sqrt{-1}$,   $A$ is an unknown complex function, the subscript denotes the partial derivative,  $t$ and $x$ denote the temporal and spatial coordinates,  $c_1$ and $c_3$ are physical parameters, respectively.
 The CGLE has spatio-temporal chaotic solutions in cases when  $c_{1}c_{3}\geq1$,  corresponding to Benjamin-Feir unstability \cite{Benjamin1967The}.   Depending upon the values of  $c_{1}$ and $c_{3}$, the CGLE exhibits two distinct chaotic phases, namely  ``phase chaos'' when $A$ is bounded away from zero,  and ``defect chaos'' for $A=0$ when the phase exhibits singularities \cite{sakaguchi1990breakdown, shraiman1992spatiotemporal, brusch2000modulated,brusch2001modulated}.
  As pointed out by Shraiman {\em et al} \cite{shraiman1992spatiotemporal},  the crossover between phase and defect chaos is invertible only when $c_{1} > 1.9$.   Furthermore, the CGLE solution occupies a bichaos region when $c_{1}<1.9$, where phase chaos and defect chaos coexist. Chat\'{e} \cite{Chate1994c} examined the relation with spatio-temporal intermittency, and defined an intermittency regime where defect chaos and stable plane waves coexist.   By considering modulated amplitude waves (MAWs), Brusch {\em et al} \cite{brusch2000modulated} found that the crossover between phase and defect chaos take place when the periods of MAWs are driven beyond their saddle-node bifurcation.

Despite intensive numerical investigations into solutions of the CGLE \cite{Torcini1996Order, sakaguchi1990breakdown, shraiman1992spatiotemporal, brusch2000modulated, Chate1994c, Montagne1997Wound,Montagne1996Winding} for values of  $c_{1}$ and $c_{3}$ ranging from $L=500$ to $L=4000$,  the sensitive  dependence on initial conditions (SDIC) of the CGLE has made it impossible to obtain  reliable, long duration numerical simulations of the spatio-temporal chaotic solution by means of the traditional algorithms using {\em double} precision arithmetic.   Here, we use  Clean Numerical Simulation (CNS) to obtain reliable computer-generated simulations in a given specified finite interval of time.  Taking the one-dimension complex Ginzburg-Landau equation as an example, we outline two  different  CNS algorithms for spatio-temporal chaotic systems: one in spectral space, the other in physical space. We will demonstrate that the latter is computationally much more efficient than the former.

First of all, we  \emph {temporally} expand the unknown complex function $A$ in the CGLE by means of the following high-order Taylor expansion
\begin{equation}\label{taylor1}
A(x,t+\Delta t) = \sum^{M}_{m=0} A^{[m]}(x,t) \;  (\Delta t)^{m},
\end{equation}
where  $\Delta t$ is the time step, the order $M$  is a (sufficiently large) positive integer, and
\begin{equation}
A^{[m]}(x,t) =\frac{1}{m!}\;\frac{\partial^{m} A(x,t)}{\partial t^{m}}.  \label{def:A[m]}
\end{equation}
Note that $\left| A\right|^2 = \overline{A}\;A$, where $ \overline{A}$ is the complex conjugate of $A(x,t)$.
From Eqs. (\ref{cgle}) and (\ref{def:A[m]}), we have
\begin{eqnarray}
 A^{[m]}(x,t)&= & \frac{1}{m}\; A^{[m-1]}(x,t)+\frac{(1+i \, c_{1})}{m}\;A_{xx}^{[m-1]}(x,t)  \nonumber \\
&-&\frac{(1-i \, c_{3})}{m}\sum^{m-1}_{j=0}\sum^{m-1-j}_{n=0}\overline{A}^{[j]}(x,t) \; A^{[n]}(x,t) \; A^{[m-1-j-n]}(x,t).  \hspace{0.5cm} \label{geq:A:mth}
\end{eqnarray}
Note that  $A^{[m]}(x,t)$  is  solely dependent upon  $A^{[n]}(x,t)$  and  its  spatial  derivative  $A_{xx}^{[n]}(x,t)$, where $0\leq n\leq m-1$,  and thus can be calculated consecutively,  up to a high-enough order $M$ ensuring that the  truncation error does not exceed a prescribed level.  Hence, $A(x+\Delta t)$ is calculated to a required precision,  provided the  spatial  derivative  term $A_{xx}^{[n]}(x,t)$ is calculated to sufficient accuracy.  This is a key to using the CNS in problems involving spatio-temporal chaos.

\subsection{The CNS algorithm in spectral space}

In CNS combined with a Galerkin Fourier spectral method, the  unknown  complex  function $A$  of Eq. (\ref{cgle})  is  expressed by Fourier series (see e.g. Finlayson {\em et al} and Isaacson {\em et al.} \cite{finlayson2013method,isaacson2012analysis}), as follows
\begin{equation}
A(x,t)\approx \sum_{k=-\frac{N}{2}}^{\frac{N}{2}-1}a_{k}(t)\;\varphi_{k}(x), \label{def:A(x,t):spectral}
\end{equation}
where $N$ is the mode number of the spatial Fourier expansion with the base function
\begin{equation}
 \varphi_{k}(x)= \frac{1}{\sqrt{L}}\;e^{i \,  k \alpha  x}.  \label{def:varphi}
 \end{equation}
 Here $\alpha = 2 \pi/L$, and $L$ is the spatial period of the solution.    Then, according to   (\ref{def:A[m]}),
\begin{equation}
A^{[m]}(x,t)\approx \sum_{k=-N/2}^{N/2-1}a^{[m]}_{k}(t)\;\varphi_{k}(x).  \label{def:A[m]:1st}
\end{equation}
After transformation, Eq. (\ref{geq:A:mth}) becomes
 \begin{eqnarray}\label{CGLEG}
&&\int^{L}_{0} A^{[m]}(x,t)\;\overline{\varphi}_{k}(x) dx = \frac{1}{m} \int^{L}_{0} A^{[m-1]}(x,t)\;\overline{\varphi}_{k}(x) dx \nonumber\\
&+& \frac{(1+i \, c_{1})}{m}\;\int^{L}_{0} A^{[m-1]}_{xx}(x,t)\;\overline{\varphi}_{k}(x) dx \nonumber\\
&-&\frac{(1-i \, c_{3})}{m}\;\sum_{j=0}^{m-1}\sum_{n=0}^{m-1-j}\int^{L}_{0} \overline{A}^{[j]}(x,t)\; A^{[n]}(x,t) \; A^{[m-1-j-n]}(x,t)\;\overline{\varphi}_{k}(x) dx, \hspace{0.95cm}
\end{eqnarray}
where  $\overline{\varphi}_{k}(x)$ is the complex conjugate of  the basis function $\varphi_k(x)$.   Substituting (\ref{def:A[m]:1st}) into the above equation, we have for $m\geq 1$ that
\begin{eqnarray}
a^{[m]}_{k}(t) &=& \frac{1}{m}\;a^{[m-1]}_{k}(t)- \frac{(1+i \, c_{1})}{m} \left(\frac{2 k \pi}{L}\right)^{2} a^{[m-1]}_{k}(t) \nonumber\\
&-&\frac{(1-i \, c_{3})}{m L}\sum_{-k_{1}+k_{2}+k_{3}=k}
\sum^{m-1}_{j=0} \sum_{n=0}^{m-1-j} \overline{a}^{[j]}_{k_1}(t)\;a^{[n]}_{k_2}(t)\;a^{[m-1-j-n]}_{k_{3}}(t),\nonumber\\
&& \hspace{1.0cm}\hspace{1.0cm}-\frac{N}{2} \leq  k,k_{1},k_{2},k_{3} \leq \frac{N}{2}-1,
\label{geq:a[m]}
\end{eqnarray}
where $\overline{a}^{[n]}_k(t)$ is the complex conjugate of $a^{[n]}_k(t)$, with $a_k^{[0]}(t)=a_k(t)$ and $\overline{a}_k^{[0]}(t)=\overline{a}_k(t)$.   The initial condition of $a_{k}(t)$ is given by
\begin{equation}
a_{k}(0)=\int^{L}_{0}A(x,0)\;\overline{\varphi}_{k}(x) dx, \hspace{1.0cm}  -\frac{N}{2} \leq k \leq \frac{N}{2}-1.  \label{ic:a[k]}
\end{equation}
In the frame of CNS, we apply the temporal high-order Taylor expansion
\begin{equation}
a_k(t + \Delta t) \approx \sum_{m=0}^{M} a_k^{[m]}(t) (\Delta t)^m,  \hspace{1.0cm}  -\frac{N}{2} \leq k \leq \frac{N}{2}-1, \label{geq:evolution:a(t)}
\end{equation}
subject to the initial condition (\ref{ic:a[k]}), to decrease the truncation error to a required level. We also ensure that calculations are performed in multiple-precision using a sufficient number  $N_s$ of significant digits to limit the round-off error to below a specified level, and carry out a verification check on the solution by considering results from an additional simulation with even smaller numerical noise to determine the critical predictable time $T_c$. Here  $a_k^{[m]}(t)$ is given by (\ref{geq:a[m]}),  $\Delta t$ is the time step, $M$ is the order of the Taylor expansion, and  $N$ is the mode number  of the spatial Fourier expansion, respectively.    So  long  as  all values of $a_k(t)$ in spectral space are known, the solution $A(x,t)$ can be determined in physical space by means of (\ref{def:A(x,t):spectral}), as and when required.  As mentioned in the Introduction, this CNS algorithm  in spectral space was  successfully  used by Lin {\em et al.} \cite{Lin2017On}  to study the influence of inherent micro-level thermal fluctuations on the macroscopic randomness of turbulent  Rayleigh-B\'{e}rnard turbulent convection. Lin {\em et al.} found that the accuracy of this Galerkin Fourier spectral method is very high, provided $N$, $M$, and $N_s$  are  all sufficiently large.  However, as reported by Lin {\em  et al.} \cite{Lin2017On},  calculation of the nonlinear term in (\ref{geq:a[m]}) is very time consuming when $N$ is large.

\subsection{The CNS  algorithm  in physical space \label{2nd-algorithm}}

To overcome the shortcoming of the above-mentioned CNS algorithm in spectral space, we propose the following CNS algorithm in physical space.   Instead of mapping the original governing equation and initial condition onto spectral space, via a collocation method \cite{Peyret2002Spectral, Peyret2002SpectralBook}, we  directly  solve  the original  equation  in physical space by first  dividing  the spatial domain $[0,L]$ into a uniform grid, such that
\[
x_k = \frac{k  L}{N} = k \Delta x, \hspace{1.0cm} k = 0,1,2,3, \cdots, N,
\]
and then approximating  $A(x,t)$ by a set of discrete unknown variables
\begin{equation}
\left\{ A(x_0,t), A(x_1,t), A(x_2,t), \cdots, A(x_N,t)    \right\},  \label{discretization}
\end{equation}
where $A(x_N,t) = A(x_0,t)$ in order to satisfy the periodic condition.   Thus, we have only $N$ unknowns $A(x_k,t)$ ($0\leq k \leq N-1$), whose  temporal evolution is given by the high-order Taylor  expansion
\begin{equation}
A(x_{k}, t+\Delta t) \approx  \sum_{m=0}^{M} A^{[m]}(x_k,t) (\Delta t)^m,  \hspace{0.75cm} k = 0,1,2,\cdots, N-1, \label{geq:evolution:A}
\end{equation}
 where $\Delta t$ is the time step, and $A^{[m]}(x_k,t)$ defined by (\ref{def:A[m]}) is given by
\begin{eqnarray}
  A^{[m]}(x_{k},t) &=& \frac{1}{m}\; A^{[m-1]}(x_{k},t)+\frac{(1+i \, c_{1})}{m}\;A_{xx}^{[m-1]}(x_{k},t)  \nonumber \\
&-&\frac{(1-i \, c_{3})}{m}\sum^{m-1}_{j=0}\sum^{m-1-j}_{n=0}\overline{A}^{[j]}(x_{k},t) \; A^{[n]}(x_{k},t) \; A^{[m-1-j-n]}(x_{k},t), \hspace{0.5cm} \label{geq:A:space}
\end{eqnarray}
with the specified initial condition  $A(x_k,0) = f(x_k)$.

Note that there exists the spatial partial derivative $A^{[n]}_{xx}(x_k,t)$ in (\ref{geq:A:space}), where $0\leq n\leq m-1$ and $0\leq k \leq N-1$ are positive integers.   To evaluate accurately this spatial partial derivative term from  the set of the  {\em known}  discrete  variables  $A^{[n]}(x_j,t)$,  we first invoke the Fourier expansion in space,
\begin{equation}
A^{[n]}(x,t) \approx  \sum^{\frac{N}{2}-1}_{k=-\frac{N}{2}} b_{k}^{[n]}(t) \;e^{i \, k \, \alpha\, x },  \label{expression:A:2nd}
\end{equation}
where $\alpha = 2\pi/L$ and
\begin{equation}
b_{k}^{[n]}(t) \approx \frac{1}{N}\sum^{N-1}_{j=0}A^{[n]}(x_{j},t)\;e^{-i \, k \, \alpha \, x_{j}},  \hspace{0.75cm}   -\frac{N}{2} \leq k \leq \frac{N}{2}-1.  \label{def:b[m]}
\end{equation}
This is given by the set of the known variables $A^{[n]}(x_{j},t)$ at discrete  points $x_j$ ($0\leq j \leq N-1$).   Then, we have the spatial partial derivative term
\begin{equation}\label{partial}
A^{[n]}_{xx}(x_{j},t) \approx -\alpha^{2}\;\sum^{\frac{N}{2}-1}_{k=-\frac{N}{2}} k^2\; b^{[n]}_{k}(t) \;e^{i \,  k\, \alpha\, x_j}, \hspace{0.75cm} 0\leq j\leq N-1,
\end{equation}
where $b^{[n]}_{k}(t)$ is given by (\ref{def:b[m]}).  The Fast Fourier Transform (FFT) \cite{Cooley1965An} is used to increase computational efficiency, given that the  discrete  points  $x_j$ are equidistant.

Obviously, the larger the order $M$ of the Taylor expansion (\ref{geq:evolution:A})  in time  and the mode number $N$ of the Fourier expansion (\ref{expression:A:2nd}) in space, the smaller the corresponding truncation errors.  To decrease the round-off error to a required level, \emph {all}  variables  and  physical parameters  are calculated in multiple-precision with $N_s$ significant digits, where $N_s$ is a sufficiently large positive integer.  In this way, both the truncation and round-off errors are reduced to a required limiting level.  Finally, an additional numerical simulation with smaller numerical noise is carried out to determine the critical predictable time $T_c$ (i.e. the maximum time of reliable simulation)  by comparing the results of both simulations; this ensures that the CNS result is reliable within the temporal interval $t\in[0,T_c]$ over the whole spatial domain.

Comparing (\ref{geq:A:space}) with (\ref{geq:a[m]}), it is obvious that  the CNS algorithm in physical space is numerically much more efficient than the CNS algorithm in spectral space, especially for a large mode number $N$ of the spatial Fourier expansion.   This is indeed true, as shown in \S 2.5.

Note that the high order Taylor expansion method  \cite{Gibbon1960, Barton1971, Corliss1977, Corliss1982Solving, Barrio2005VSVO},  the Fourier expansion method  \cite{Cooley1965An},  the multiple precision \cite{Oyanarte1990}, and the solution verification check \cite{oberkampf2010verification, roy2010complete} are widely used.   However, their combination may give reliable simulations of spatio-temporal chaotic systems, which provide us benchmarks  that can be used to study the  influence of  numerical noise on chaotic simulations given by traditional algorithms in {\em double} precision.

\subsection{The optimal time-step}

In CNS, the temporal truncation error is determined by the order $M$ of the Taylor expansion (\ref{geq:evolution:a(t)})  or  (\ref{geq:evolution:A}), and the spatial truncation error is determined by the mode number $N$ of the spatial Fourier expansion (\ref{def:A(x,t):spectral}) or (\ref{expression:A:2nd}).  As long as both $M$ and $N$ are large enough, the  temporal and spatial truncation errors are reduced to below a target level. In addition,  the round-off error is reduced to a specified level by using multiple-precision with sufficiently large number $N_s$ of significant digits.

To calculate the $M$th-order Taylor expansion  (\ref{geq:evolution:a(t)}) or (\ref{geq:evolution:A}) in time dimension,  we use  an optimal  time-step \cite{Barrio2005VSVO}
\begin{equation}
 \Delta t = \min \left( \frac{g(tol,M-1)}{\big\|A^{(M-1)}(x_{k},t)\big\|_{\infty}^{1/(M-1)}},\frac{g(tol,M)}{\big\|A^{(M)}(x_{k},t)\big\|_{\infty}^{1/M}} \right),  \label{def:optimal-timestep-1}
\end{equation}
where   $tol$  is  an  allowed  tolerance at each time step, $\big\|A^{(M)}(x_{k},t) \big\|_{\infty}$ is the infinite norm of the $M$th-order Taylor expansion of the modules of $A(x_{k},t)$,   $k=0,1,2,3, \cdots, N-1$  and
\begin{equation}
g(tol,M) \approx tol^{1/(M+1)}.    \label{def:g}
\end{equation}
To save computer resources and increase computational efficiency,  it is reasonable to enforce the allowed  tolerance  to be at the same level as  the round-off error,  say,
\begin{equation}\label{tolNs:rela}
 tol = 10^{-N_s},
\end{equation}
Thus, we have the optimal time-step as
\begin{equation}
 \Delta t= \min \left(\frac{10^{-{N_s}/{M}}}{\big\|A^{(M-1)}(x_{k},t)\big\|_{\infty}^{1/(M-1)}},\frac{10^{-{N_s}/{(M+1)}}}{\big\|A^{(M)}(x_{k},t)\big\|_{\infty}^{1/M}}  \right).  \label{def:optimal-timestep-2}
\end{equation}
 Therefore, for a given number $N_s$ of significant digits in multiple-precision, one has the freedom to choose the order $M$ of temporal Taylor expansion.   According to (\ref{def:optimal-timestep-2}), raising the order of the temporal  Taylor expansion usually leads to an increased optimal time-step $\Delta t$.  So,  in practice,  it makes sense to set $M$ to higher order because it  can also  improve  the  speed-up ratio of the parallelized algorithm.

In practice, we first choose a sufficiently large values for $N_s$ and $M$ to meet the error requirements, from which the optimal time-step $\Delta t$ is evaluated  by (\ref{def:optimal-timestep-2}) at each time-step.   In this way,  the temporal truncation error is kept to the same level of the round-off error.

For  a  spatio-temporal  chaotic  system (\ref{cgle}), it is also necessary to limit the spatial   truncation  error,  determined by the mode number  $N$ of the  spatial  Fourier expressions  (\ref{def:A(x,t):spectral}) or (\ref{expression:A:2nd}).   Obviously, the larger the value of $N$, the smaller the spatial truncation error.   In theory, to save computer resources and improve computational efficiency,  it is better to let the the spatial truncation error  be equal to the temporal truncation error.  Unfortunately,  it is presently unknown  how to do this.   In practice, we often choose reasonably large values for $N_s$  and $N$ in order to guarantee that the round off error, temporal truncation error, and spatial truncation error are \emph {all} below a specified level.
Finally, the solution verification check is implemented by comparing the numerical result to that of an additional simulation with even smaller numerical noise; this determines the finite time interval $[0,T_c]$, over which the spatio-temporal chaotic numerical results exhibit no distinct differences  at  {\em all}  spatial grid-points and thus are reliable.

\subsection{Criteria of reliability and convergence \label{Criteria-convergence}}

Due to the sensitivity dependence on initial conditions of chaotic systems, unavoidable numerical noise from round-off and truncation errors can increase exponentially.   To guarantee the reliability of chaotic numerical simulations, we compare our CNS simulation of interest against one using the same physical parameters and the same algorithm but with even smaller numerical noise; this determines the finite time interval $[0,T_c]$ over which the spatio-temporal chaotic numerical results display no distinct differences  at  {\em all}  spatial grid-points.

 Take the one-dimensional CGLE as an example.  Let $A(x,t)$ denote  a numerical simulation of the one-dimensional CGLE, given by CNS using the  $M$th-order of Taylor expansion in the time dimension, the mode number $N$ of the  Fourier expansion in the spatial  dimension,  the  number $N_s$ of significant digits in the multiple precision  scheme.  Now, let $A'(x,t)$ be another CNS result with even smaller numerical noise, given by the $M'$th-order  of the  temporal Taylor expansion, the mode number $N'$ of the spatial Fourier expansion, and number  $N'_s$ of significant digits in the multiple precision scheme, where $M'\geq M$, $N'>N$ and $N'_s \geq N_s$.   Since the level of numerical noise increases exponentially,  $A'(x,t)$ (with smaller numerical noises) should be closer to the true solution than $A(x,t)$ and thus can be used as a reference solution to check the reliability of  $A(x,t)$.

If $A(x,t)$ and $A'(x,t)$ have the same spatial discretization, i.e. the same $N$, their deviation can be measured by
 \begin{equation}
 \delta(t) = \frac{\sqrt{\sum\limits_{k=0}^{N-1}\Big|A(x_k,t) - A'(x_k,t) \Big|^2}}{\sqrt{\sum\limits_{k=0}^{N-1}\Big|A'(x_k,t)\Big|^2}}\; .  \label{def:delta(t)}
 \end{equation}
 However, $A'(x,t)$ often has different values of $N$ from $A(x,t)$ in practice.    In this case, it is more convenient to compare their spatial spectrum.  Rewrite $ A(x,t)$ and $A'(x,t) $  in the spatial Fourier expressions
 \begin{equation}
A(x,t)  \approx \sum^{\frac{N}{2}-1}_{k=-\frac{N}{2}} B_{k}(t) \;e^{i \, k\, \alpha\, x},  \hspace{0.5cm}  A'(x,t)  \approx \sum^{\frac{N'}{2}-1}_{k=-\frac{N'}{2}} B'_{k}(t) \;e^{i \, k\, \alpha\, x}.
\end{equation}
Note that $\sum\limits_{k=-\frac{N'}{2}}^{\frac{N'}{2}-1}|B'_k|^2$ has often physical meaning, such as the total energy of the system.  So, from a physical viewpoint, it is important for a numerical simulation of a spatio-temporal chaos to have an accurate  spatial  spectrum at given time $t$.   Thus, we define the so-called  ``spectrum-deviation''
\begin{equation}\label{delta}
 \delta_s(t) = \frac{\sum\limits_{k=-\frac{N}{2}}^{\frac{N}{2}-1}\Big|  |B'_k|^2-|B_k|^2 \Big|}{\sum\limits_{k=-\frac{N'}{2}}^{\frac{N'}{2}-1}|B'_k|^2}
 \end{equation}
to quantify the difference between $A(x,t)$ and $A'(x,t)$ at a given time $t$.  Obviously,  the smaller the spectrum-deviation $\delta_s$, the better the two simulations agree with each other and the greater their reliability. So,  it  is  reasonable to  define the reliability criterion as
\begin{equation}
\delta_s  < \delta_s^{c}, \label{criteria:convergence}
\end{equation}
where $\delta_s^{c}>0$ is a reasonably small number, called the ``critical spectrum-deviation''.    For the problem under consideration, it is found that $\delta_s^{c} = 0.01$ is reasonable, which is used for all the remaining cases in this paper.  Equation (\ref{criteria:convergence}) provides a criterion of reliability, which determines  the  so-called  ``critical predictable  time'' $T_c$ and the temporal interval $[0,T_c]$, in which the reliability-criteria (\ref{criteria:convergence}) is satisfied and the differences between the two simulations $A(x,t)$ and $A'(x,t)$ at all spatial grid-points are negligible.

\subsection{Computational efficiency of the two CNS algorithms}

  \begin{table}[b]
\begin{center}
\def\temptablewidth{1\textwidth}
{\rule{\temptablewidth}{1.5pt}}
\begin{tabular*}{\temptablewidth}{@{\extracolsep{\fill}}cccccc}
 $N$	&	$T_{1}$ 		& $T_{2}$  	& $T_{1} / T_{2}$ \\
 	&	(in seconds) 	& (in seconds)	&	\\
\hline
16	&  694	& 7.93		& 87  \\
32	&  5530	& 12.8		& 432\\
64	&  44366	& 25	& 1774\\
128  &  346490	&  51	& 6793\\
\end{tabular*}
{\rule{\temptablewidth}{1.5pt}}
\end{center}
 \caption{ CPU times of CNS simulations of the one-dimension CGLE for $c_1=2$, $c_{3}=1$ and $L=256$ in the temporal interval $t\in [0,10]$, in spectral and physical spaces with the same $N_s = 20$,  $M=14$, a fixed time step $\Delta t=0.01$ and the same mode number $N$ for the spatial Fourier expansion.  $T_{1}$ and  $T_{2}$ are the elapsed CPU times for the CNS algorithms in the spectral space; and the physical space, respectively.}
\label{table:ratio}
\vspace*{-20pt}
\end{table}

In theory the CNS  algorithm  in spectral space undertakes about  $O\left(M^{3} N^{3}\right)$ elementary operations, whereas the CNS algorithm in physical space undertakes $O\;(NM\;log_{2}N+N\;M^{3})$  elementary operations. This is  partly due to the fewer nonlinear terms in the right-hand side of (\ref{geq:A:space})  than  (\ref{geq:a[m]}),  and partly due to the implementation of the FFT in evaluating (\ref{partial}).  Obviously, using the CNS algorithm in physical space can greatly improve the calculation speed at $O\;(N^{2})$, where $N$ is the mode number of the spatial Fourier expansion. The resulting speed-up is huge in practice, especially for large $N$. This is confirmed by the results listed in Table~{\ref{table:ratio}} which compares the CPU times taken by the two CNS algorithms, for the same values of  $N$,  $M$, and  $N_{s}$.   For the purposes of comparison, a fixed time-step  $\Delta t=0.01$ is used. Table~\ref{table:ratio} shows that the ratio $T_1/T_2$ of required CPU times for the 1st and 2nd  CNS algorithms rises exponentially as $N$ increases. Notably, when $N=128$,  the required CPU time  of the CNS algorithm in spectral space is  6,000 times more than that of the CNS algorithm in physical space!   This illustrates that the CNS algorithm in physical space described in Section 2.2 is clearly much more efficient than that in its counterpart spectral space, and so is used for all the cases considered in the remainder of this paper (unless otherwise stated).

\subsection{Validity of the CNS algorithm in physical space \label{two_CNS} }

The one-dimensional CGLE  (\ref{cgle}) is now used as an example to verify the validity  of our CNS algorithms in physical space. First, the one-dimensional  CGLE has closed-form plane-wave solutions \cite{Bensimon1992The,sakaguchi1990breakdown,shraiman1992spatiotemporal,Chate1994c} given by
 \begin{equation}\label{planewave}
A(x,t)= \sqrt{1-q^{2}}\; e^{i \, (w_{q} \; t+q \; x)},
\end{equation}
where $w_{q}=c_{3}-(c_{3}+c_{1})\;q^{2}$. Provided $c_{1}\;c_{3}<1$ , these solutions are linearly stable. Along the line $c_{1}\;c_{3}=1$, the band of wave-vectors shrinks to zero corresponding to Benjamin-Feir or modulation instability of the uniform oscillatory solution \cite{Cross1993Pattern, Benjamin1967The}.  Beyond Benjamin-Feir  instability, when $c_{1}\;c_{3}>1$,  the system enters the phase chaos and defect chaos regimes.

\begin{figure}[tbhp]
\centering
  \includegraphics[width=9cm]{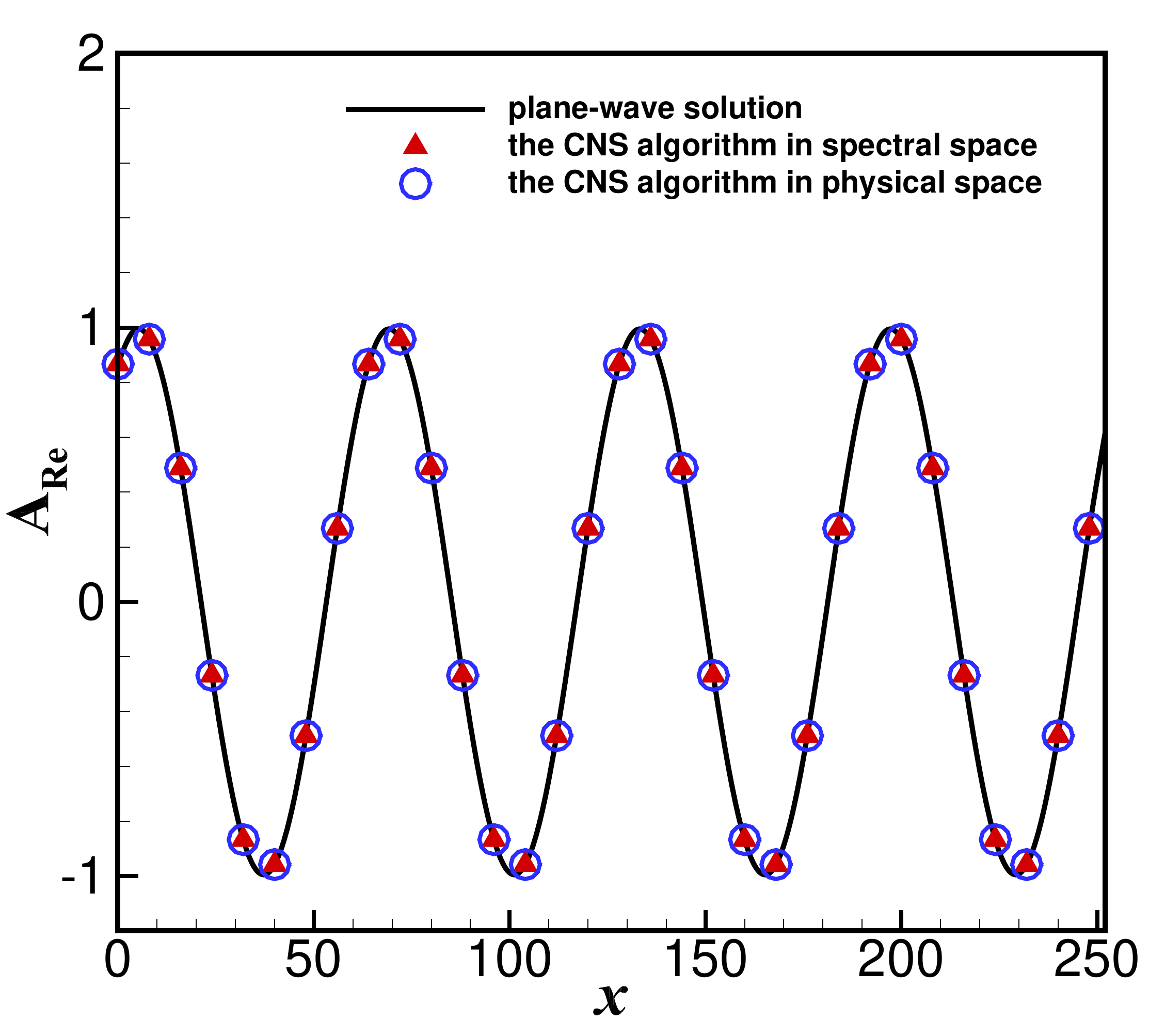}\
\renewcommand{\figurename}{Fig.}
  \caption{Numerical simulations of  the real part of $A(x,t)$ at  $t=500$ given by the two CNS algorithms using the mode-number  $N=64$ for the spatial Fourier expansion, the order $M = 8$ of temporal Taylor expansion and the number $N_{s} = 24$ of significant digits in multiple precision, compared against the exact plane-wave solution: plane-wave solution (solid line); result from CNS algorithm in spectral space (red symbols); and result from CNS algorithm in physical space (blue symbols). }
\label{m12T500}
\end{figure}

To verify the CNS algorithm in physical space, as mentioned in Section 2.2, let us first consider the plane-wave solution (\ref{planewave}) when $c_{1}=1.1$ and $c_{3}=0.2 $, with the initial condition
  \begin{equation}\label{initial1}
A(x,0)= \sqrt{1-q^2} \;e^{i \,  \; q\; x},
\end{equation}
in which $q = 8\pi/L$ and $L=256$. Given that $c_1 c_3 < 1$,  the solution should be in the form of  (\ref{planewave}) and linearly  stable. We solve this problem by means of the two CNS algorithms  using  the  same values of $N$, $M$, and $N_s$.  As shown in Fig.~\ref{m12T500}, the numerical simulations produced by the two CNS algorithms  agree well with the exact plane-wave solution  (\ref{planewave}) at $t=500$.    This verifies  the validity of the two CNS algorithms in the spectral and physical spaces, described in Section 2.1 and Section 2.2.

\begin{figure}[tbhp]
\centering
  \includegraphics[width=0.5\linewidth]{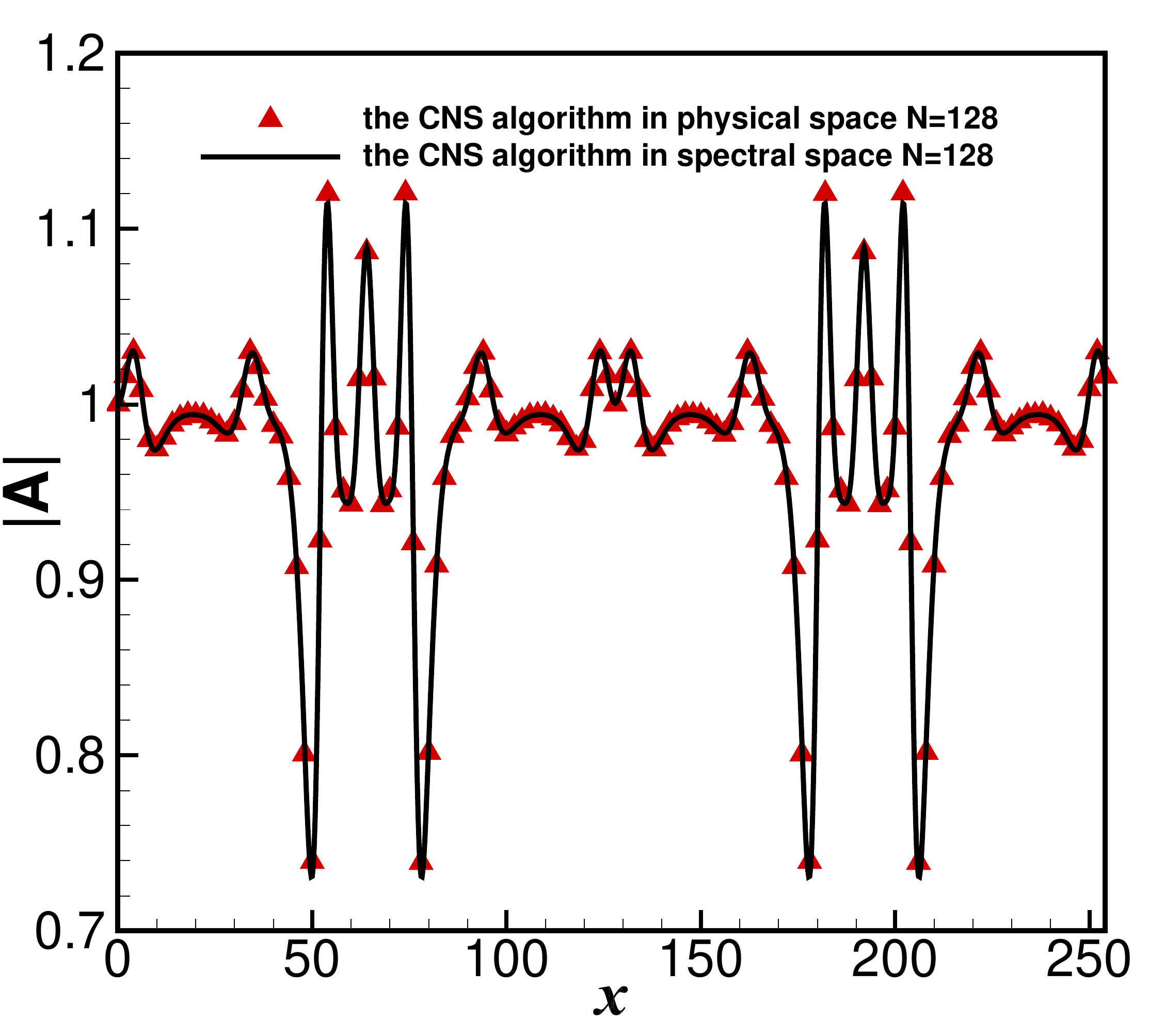}\
\renewcommand{\figurename}{Fig.}
  \caption{ Comparison of  $|A(x,t)|$ profiles (at $t = 25$) obtained by the two CNS algorithms using  the mode number $N = 128$ for the spatial Fourier expansion and the number $N_{s} = 20$ of significant digits in multiple-precision: converged results from the CNS algorithm in spectral space (solid line); and corresponding results from the CNS algorithm in physical space (red symbols).  }
\label{m12T3005}
\end{figure}

Next, let us again consider the one-dimensional CGLE, but for the case  $c_{1}=2$ and $c_{3}=1$, with the initial condition
\begin{equation}\label{initial2gg}
A(x,0) = \cos\left(\frac{4\pi x}{L}\right) + \cos\left(\frac{8\pi x}{L}\right)+i \,  \cos\left( \frac{4\pi x}{L}\right).
\end{equation}
 Here $c_1 c_3 >1$, and so the  solution is unstable and becomes chaotic.  The problem  was solved using both CNS algorithms for $N=128$ and $N_s=24$. The results shown in Fig.~\ref{m12T3005} obtained using the two CNS algorithms are in close agreement with each other.   This validates  the CNS algorithm in physical space for problems involving the chaos regime.

\begin{figure}[tbhp]
\centering
	\subfigure[]{
		\label{level.sub.3}
		\includegraphics[width=0.5\linewidth]{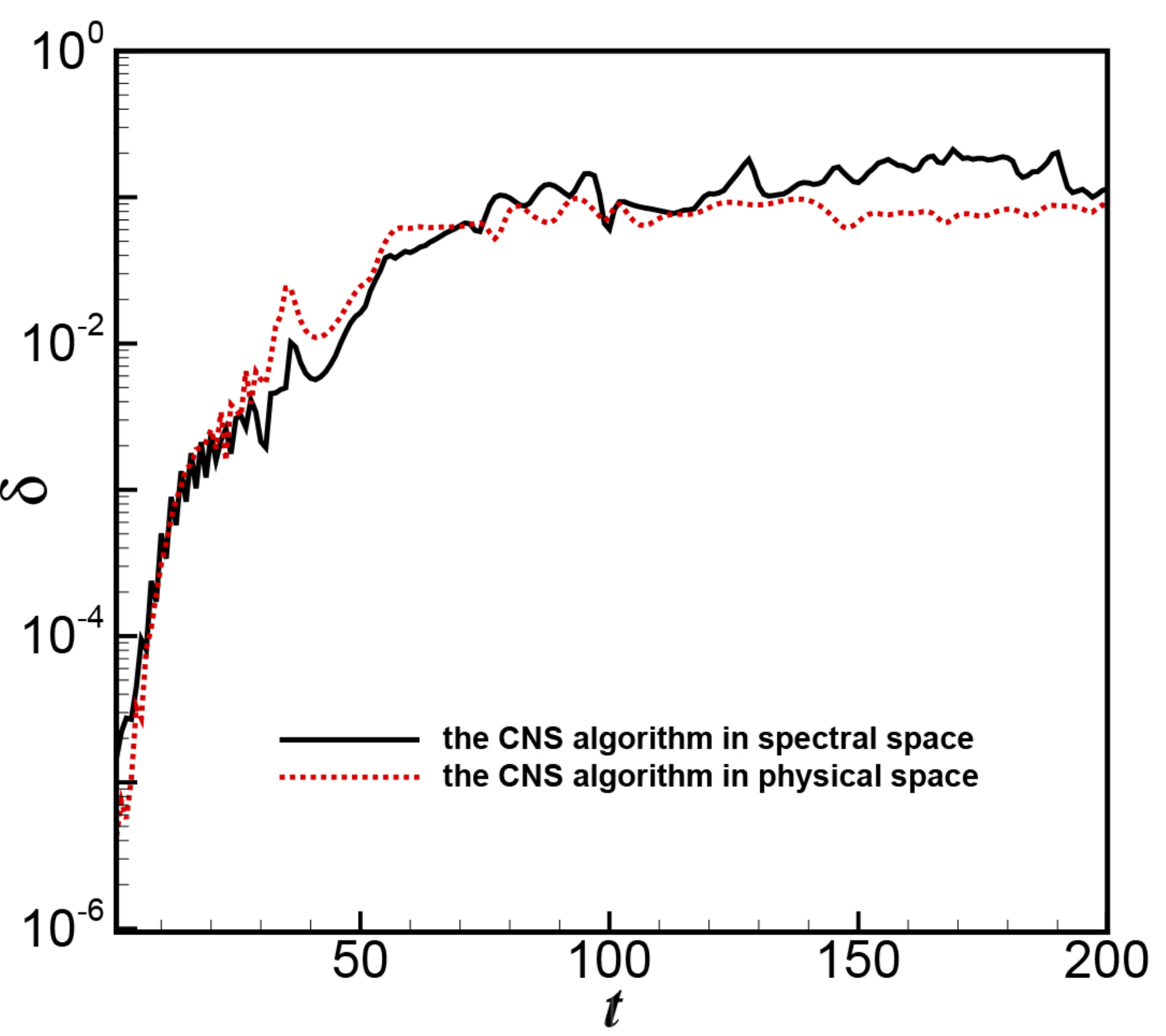}
\label{level.sub.4}
		\includegraphics[width=0.5\linewidth]{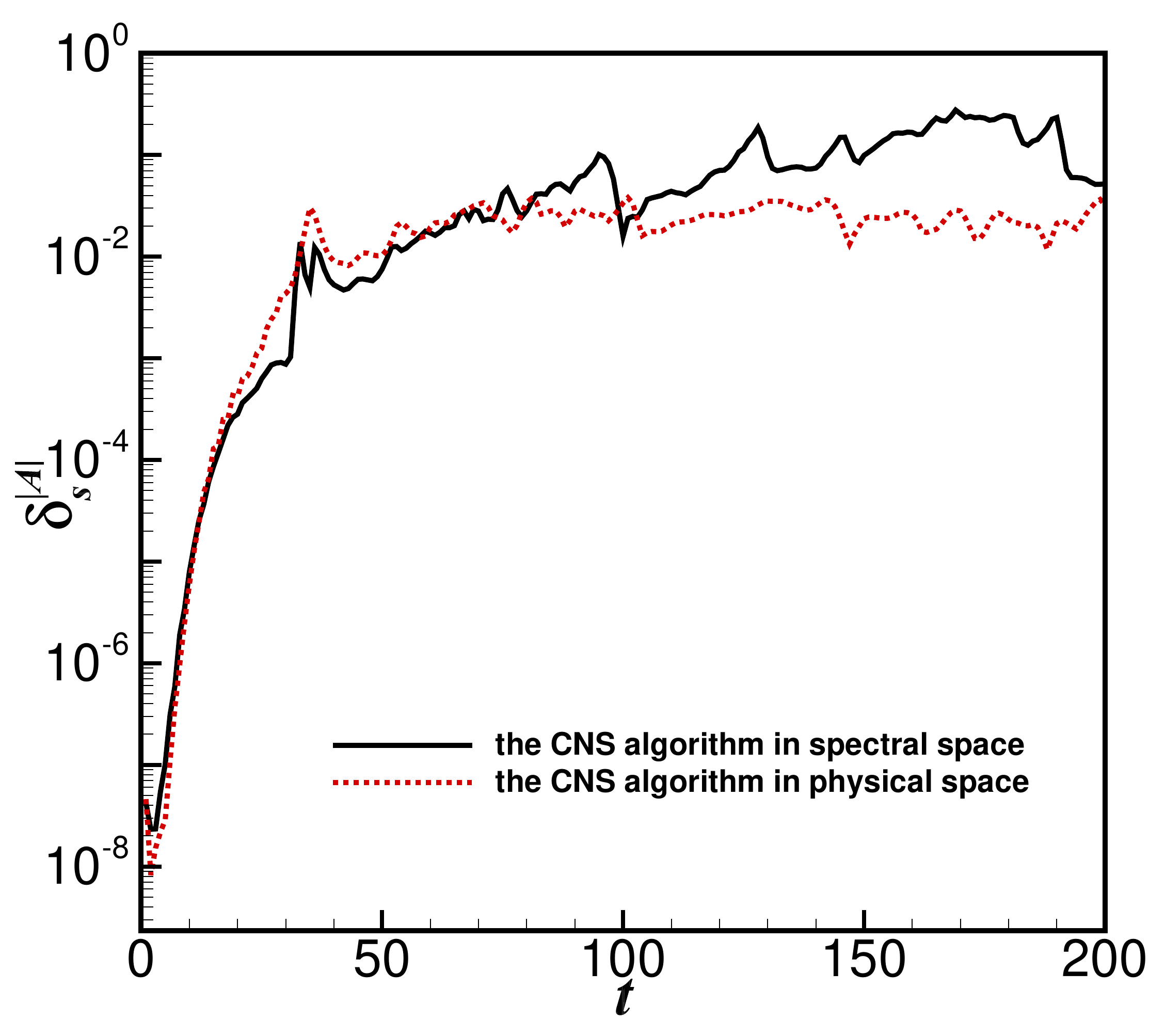}
                }	
\renewcommand{\figurename}{Fig.}
  \caption{Left: Comparison of the deviation $\delta(t)$ of $|A(x,t)|$ obtained by the two CNS algorithms using the mode number $N=128$ for the spatial Fourier expansion and the number $N_s =24$ of significant digits in multiple-precision, with reliability check using  $N=2048$ and the same $N_s$. Right: Comparison of the corresponding spectrum-deviation $\delta_s^{|A|}(t)$ of $|A(x,t)|$.  Results  given by the CNS algorithm in spectral space (black solid line); and by the CNS algorithm in physical space (red dashed line).}
\label{N64}
\end{figure}

Let $\tilde{A}(x,t)$ denote the numerical simulation for  $c_{1}=2,\; c_{3}=1$ with the initial condition (\ref{initial2gg}),  given by the CNS  algorithm in physical space with much smaller numerical noise by setting $N=2048$.  Hence, $\tilde{A}(x,t)$   can be regarded as much more accurate than the numerical simulations given by the two CNS algorithms in the spectral/physical space with $N=128$.  So,  $\tilde{A}(x,t)$  can be used as a benchmark (or a reference solution).   According to (\ref{def:delta(t)}),
  \begin{eqnarray}\label{delta1}
\delta(t) =\frac{\sqrt{\sum\limits^{N-1}_{k=0}\Big(\left|\tilde{A}(x_{16k},t)\right|-\left|A(x_{k},t)\right|\Big)^2}}{\sqrt{\sum\limits^{N-1}_{k=0}\left|\tilde{A}(x_{16k},t)\right|^2}}
\end{eqnarray}
is the deviation of $|A(x_{k},t)|$ (given by the CNS algorithm using $N=128$) from the much more accurate simulation $|\tilde{A}(x,t)|$ (given by the same CNS algorithm using $N=2048$).   As shown in Fig.~\ref{N64}, the  deviation  $\delta(t)$ defined by (\ref{delta1}) and the spectrum-deviation $\delta_s(t)$ defined by (\ref{delta}) of the numerical simulation $|A(x,t)|$ given by the CNS algorithm in physical space  grow  in a similar fashion to those  given by the CNS algorithm in spectral space; in both cases, $N = 128$ and $N_s = 24$, with the reliability check using $N=2048$ and $N_s = 24$.

The foregoing has shown that the CNS algorithm in physical space not only  can greatly  improve  computational efficiency (see Section 2.5),  but also can attain  the same ``critical predictable time'' $T_{c}$  as that of the CNS algorithm in the spectral space.

\begin{figure}[tbhp]
\centering
  \includegraphics[width=0.5\linewidth]{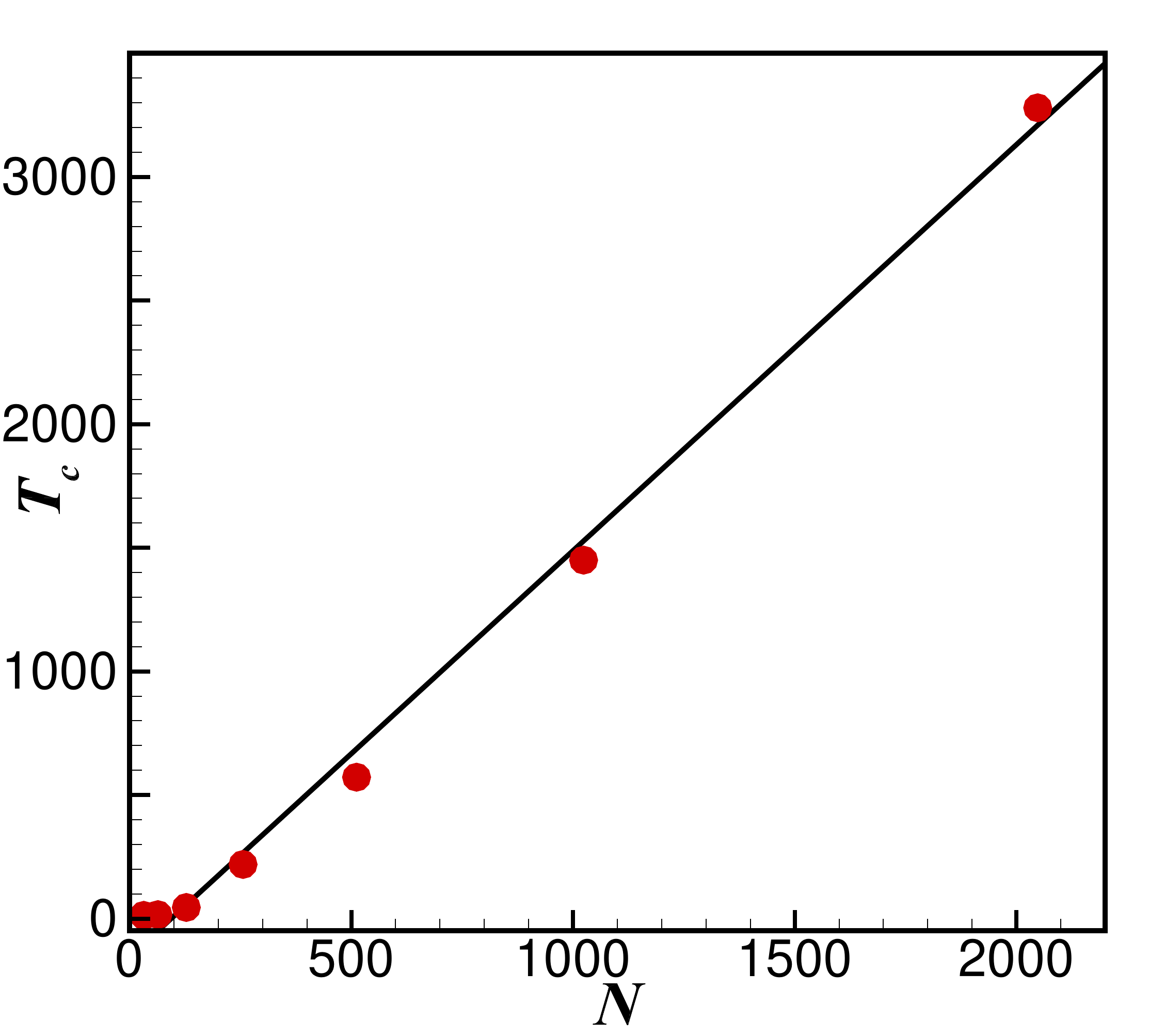}\
\renewcommand{\figurename}{Fig.}
  \caption{Critical predictable time $T_{c}$ versus the mode number $N$ of the spatial Fourier expansion in a chaotic case of $c_{1}=2$ and $c_{3}=1$ with the initial condition (\ref{initial2gg}), obtained by the CNS algorithm in  physical space using the number $N_{s}=105$ of significant digits in multiple-precision and different values of mode number $N$ (from 32 to 2048).  $T_c$ given by the CNS (symbols); and  analytic approximation formula (\ref{chaosTcNe}) (solid line).}
\label{chaosTcN}
\end{figure}

\subsection{ Relationship between  $T_c$ and the level of numerical noises  \label{Propagation}}

In summary, the basic idea of the CNS is to reduce the temporal/spatial truncation error and round-off error to such a required level that the computer-generated  simulations are reliable throughout the entire spatial domain over a specified finite time interval $[0,T_c]$,  where $T_c$ is the critical predictable time.   It is found that $T_c$ is often dependent upon the level of numerical noises, dominated by the order $M$ of temporal Taylor expansion, the  mode number $N$ of spatial Fourier expansion,  and number $N_s$ of significant digits in multiple-precision data for all variables and parameters.   Thus, for a given  critical predictable time $T_c$, we should choose sufficiently large values of $M$, $N$ and $N_s$ to guarantee such a reliable simulation in the whole spatial domain within $t\in [0,T_c]$.    Given that a universal relationship between $T_{c}$ and $M$, $N$, and $N_{s}$ is unknown, it is often necessary to have to carry out an additional CNS simulation with even smaller noise in order to determine $T_{c}$ in practice. Furthermore, because we have the freedom to choose the order $M$ of the temporal Taylor expansion according to (\ref{def:optimal-timestep-2}) for the optimal time-step, we need determine the relationship between $T_c$ and $N, N_s$ only.   Here,  let us again consider the one-dimensional CGLE for a chaotic case when $c_{1}=2, c_{3}=1$ with the initial condition (\ref{initial2gg}) as an example to illustrate how to evaluate such kind of relationships.

 To investigate the relationship between  $T_{c}$  and $N$,  we first set $N_s = 105$ to guarantee that the round-off and temporal truncation errors are below a given threshold, noting that both errors are at the same level in the present CNS algorithm, whereby $tol = 10^{-N_s}$ (see Section 2.3).  In this case, the spatial truncation error should be larger than other errors and thus make up the bulk of the numerical noise. Then,  for the same problem, we alter $N$ ( $N = 2^{n}$ with $n$ being a positive integer, in order to implement the FFT) from $32$ to $2048$ to obtain a series of CNS results at different levels of spatial truncation error, and then determine the corresponding critical predictable time $T_c$ of each CNS result  by comparing it with another CNS result using a larger $N$ (and hence smaller spatial truncation error).   Fig.~\ref{chaosTcN} shows that the critical predictable time $T_{c}$ in this case  is   almost directly  proportional to $N$, such that
 \begin{equation}\label{chaosTcNe}
 T_{c}\approx1.643\;N - 154.
\end{equation}

\begin{figure}[tbhp]
\centering
  \includegraphics[width=0.5\linewidth]{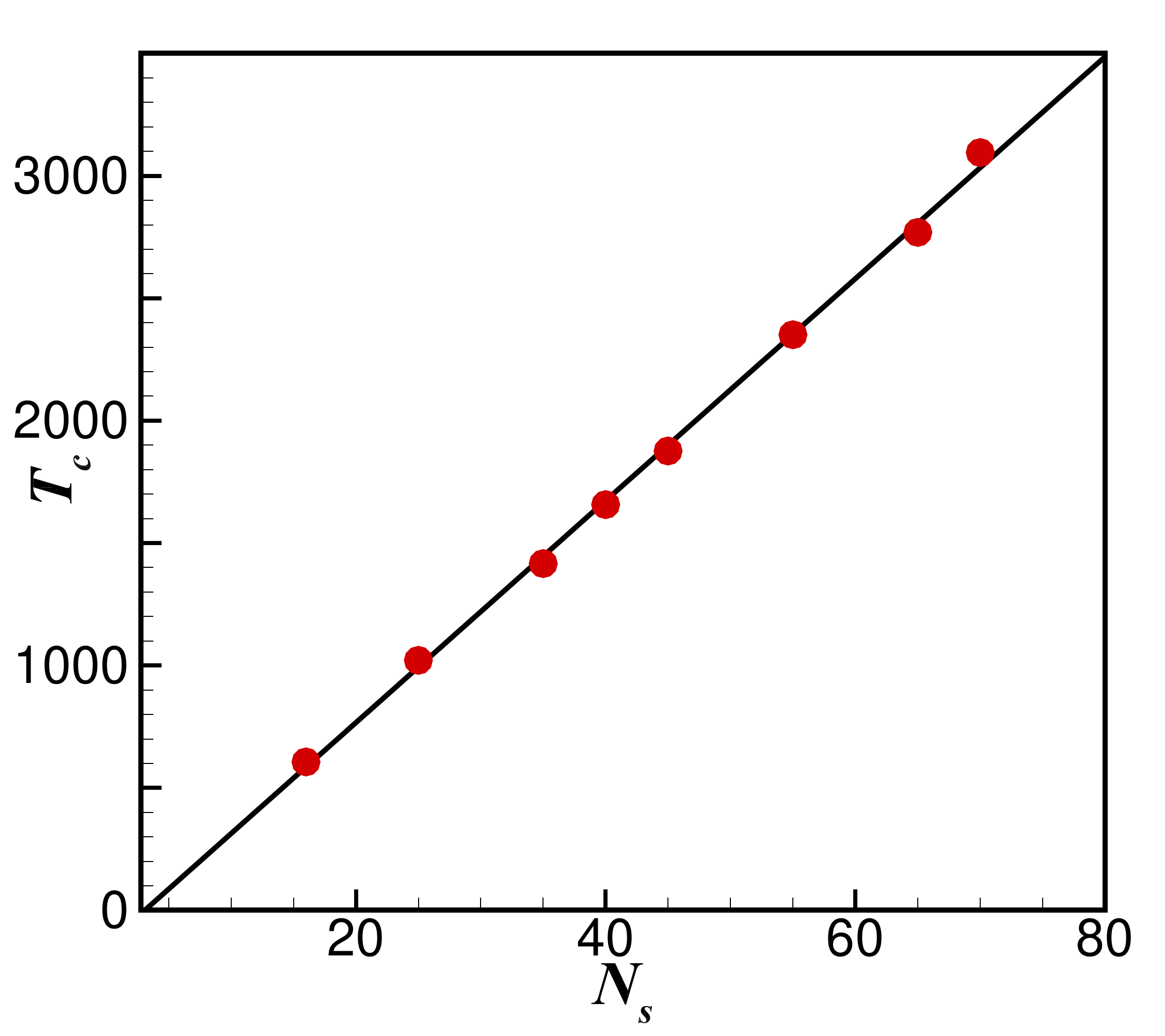}\
\renewcommand{\figurename}{Fig.}
  \caption{Critical predictable time $T_{c}$ versus the number $N_{s}$ of significant digits in multiple precision in a chaotic case of $c_{1}=2$ and $c_{3}=1$ with the initial condition (\ref{initial2gg}), given by the CNS algorithm in physical space using the mode number $N=4096$ for the spatial Fourier expansion,  the tolerance $tol=10^{-105}$ for the temporal Taylor expansion and different values of $N_s$ ranging from 16 to 70: $T_c$ given by the CNS (symbols); and analytic approximation formula (\ref{chaosTcNe}) (solid line).}
\label{chaosTcNs}
\end{figure}

In addition, to investigate the relationship between $T_c$ and $N_s$,  we set mode number $N=4096$ for the spatial Fourier expansion and the  tolerance $tol =10^{-105}$ for the temporal Taylor expansion and vary the number $N_s$ of significant digits in multiple-precision from 16 to 70 with $10^{-N_s} > tol$  in order to ensure that the round-off error is the major source of numerical noise.    In this way, for the same problem, we obtain a series of  simulations given by the CNS, from which the corresponding critical predictable time $T_c$ of each simulation is gained by comparing it against others with a larger $N_s$.   Fig.~\ref{chaosTcNs} shows that the critical predictable time $T_{c}$  is  almost directly  proportional to  $N_s$ in this case, such that
 \begin{equation}\label{Tc-Ns}
 T_{c}\approx 45\;N_{s}-140.
\end{equation}
 Notably, when $N_s = 16$, corresponding to  double precision, one has $T_c \approx  580$, say, using double precision, one can gain a reliable computer-generated simulation at most  in  $t\leq 580$,   even if truncation error is very small by using a rather high order of temporal Taylor expansion and a rather large mode number $N$ for the spatial Fourier expansion.  This is exactly the same reason why high-order algorithms in double precision are useless for modifying   convergence of chaotic computer-generated results, as previously mentioned by many researchers  \cite{Lorenz1962Deterministic, Lorenz1989Computational, jianping2000computational, Teixeira2005Time, Lorenz2006Computational}.

From (\ref{chaosTcNe}) and (\ref{Tc-Ns}), we have the linear relationship
\begin{equation}
T_c \approx \min \left\{ 1.643\;N - 154,  45\;N_{s}-140\right\}.  \label{Tc-(N,Ns)}
\end{equation}
Note that (\ref{Tc-(N,Ns)}) provides an approximate relationship between $T_c$ and $N, N_s$, which is important, although valid only for the case under consideration.   Fortunately, it has been found that  $T_c$ is {\em always} directly  proportional to  $N_s$ or $N$, approximately.     Note that a similar linearity was reported by Turchetti et al. \cite{turchetti2010asymptotic} about  a  chaotic map.  So, this kind of linearity between $T_c$ and $N, N_s$ should exist in general.   In practice,  we can first gain such kind of linear relationship by means of  relatively small values of $N$ and $N_s$ (corresponding to small values of $T_c$), and then further use this relationship to estimate the required  (large)  values of $N$ and $N_s$ for a given large value of $T_c$.   In this way,  the reliability check of the CNS simulation in a large interval of time may be avoidable,  so that  much CPU times is cut down.

According to (\ref{chaosTcNe}) and (\ref{Tc-Ns}), for the same $T_{c}$, the required value for the mode number $N$ of spatial Fourier expansion  grows  faster than the required value for the number $N_s$ of significant digits in multiple-precision.  Thus, it is much more expensive and thus more challenging to obtain a reliable numerical simulation of a spatio-temporal chaotic system (governed by a nonlinear PDE) than  a temporal chaotic one (by a set of nonlinear ODEs).

\section{Influence of numerical noise on trajectories and statistics}

It has been widely conjectured that long-term reliable computer-generated simulation of a chaotic system is impossible \cite{Lorenz1962Deterministic} even when the initial condition is exactly specified, because numerical noise from truncation and round-off errors inevitably occurs during each time-step of numerical simulation, and increases exponentially due to the butterfly-effect.  Moreover, although \emph {double} precision is widely used in computer-generated simulations, the influence of round-off error on reliable simulation of chaotic systems has been grossly underestimated.  As several researchers have previously mentioned \cite{Lorenz1989Computational, jianping2000computational, Teixeira2005Time, Lorenz2006Computational, liao2014can,Liao2009On,Liao2013On},  it is \emph {impossible} to achieve reliable, long-duration numerical simulations of chaotic systems by means of high-order algorithms in \emph {double} precision. The importance of verification and validation (V \& V) of computer-generated simulations is well known, and established techniques exist by which to determine the reliability of a numerical simulation (see e.g. \cite{oberkampf2010verification,roy2010complete}).  The key point of the CNS  is  to ensure the reliability of computer-generated simulations:  unlike traditional algorithms, CNS acts to decrease \emph {both} the truncation and round-off errors to a required level for reliable simulations of chaotic systems over a specified, large but finite interval of time, as illustrated by Liao and Wang  \cite{Wang2012Computational, Liao2009On} for the Lorenz equation.   Therefore, the CNS can provide us reliable, convergent simulations of chaotic systems  as benchmarks  that make it possible to investigate the influence of  numerical noise over a given  interval of  time  by comparing  these  benchmarks  with those produced by  traditional  algorithms in the \emph {double} precision.

\begin{figure}[tbhp]
	\centering
	\vspace{-0.35cm}
	\subfigtopskip=2pt
	\subfigbottomskip=2pt
	\subfigcapskip=-5pt
	\subfigure[(a)]{
		\label{level.sub.1}
		\includegraphics[width=0.5\linewidth]{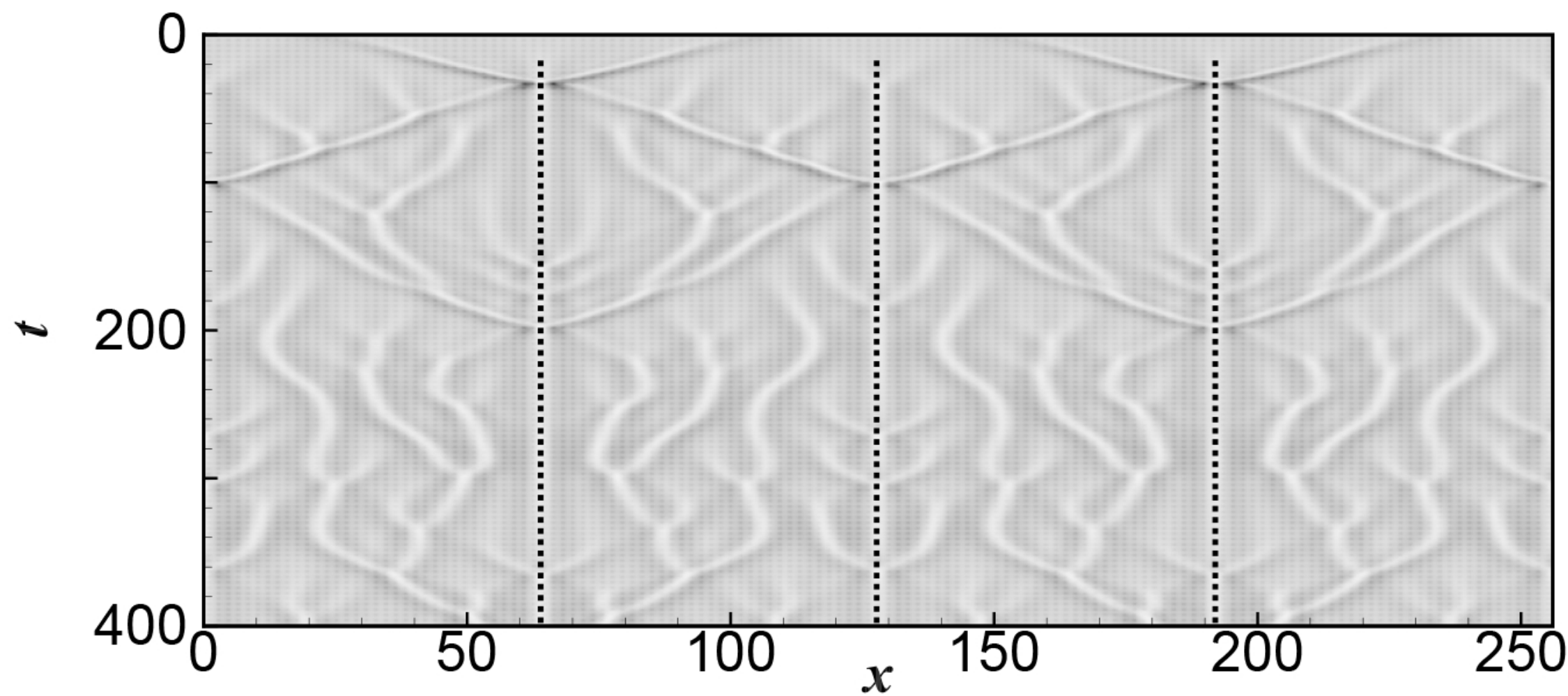}
\label{level.sub.2}
		\includegraphics[width=0.5\linewidth]{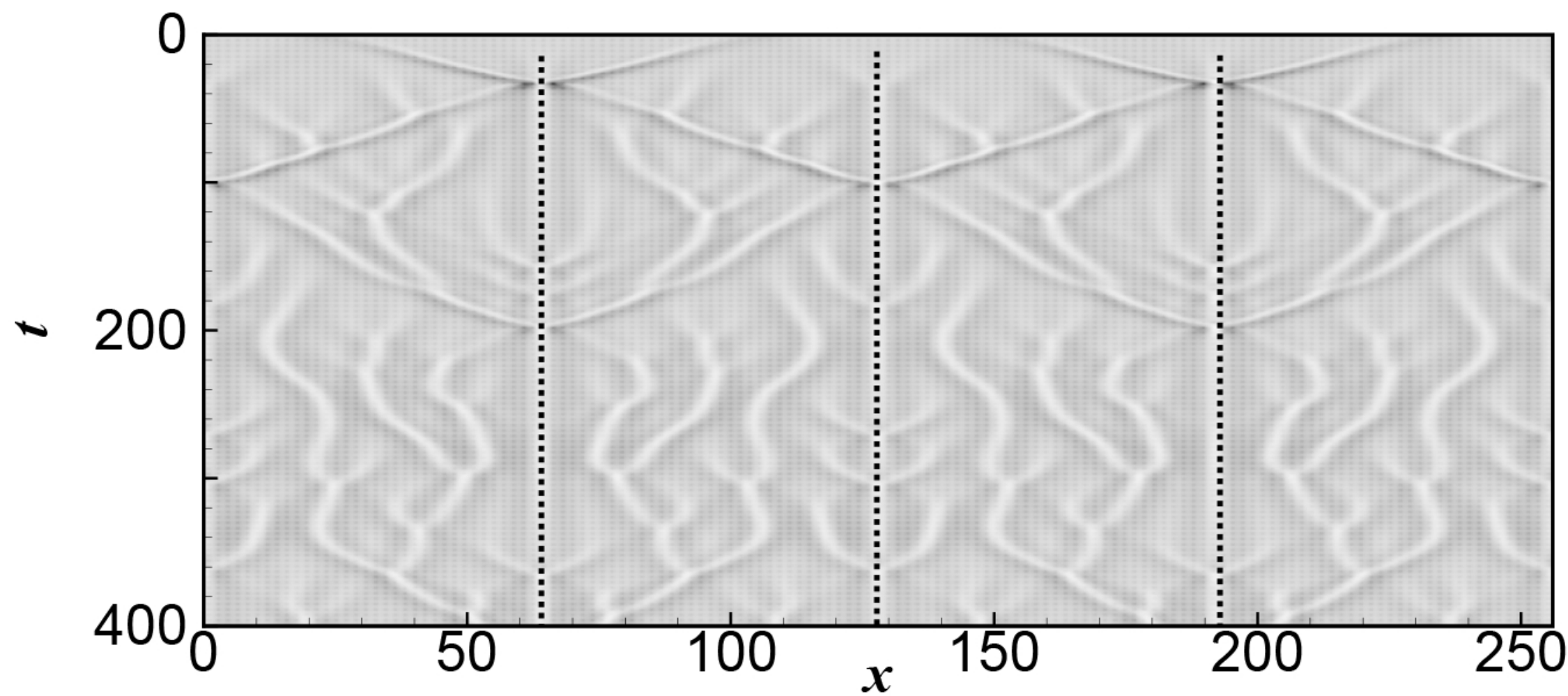}
}
	\subfigure[(b)]{
		\label{level.sub.3}
		\includegraphics[width=0.5\linewidth]{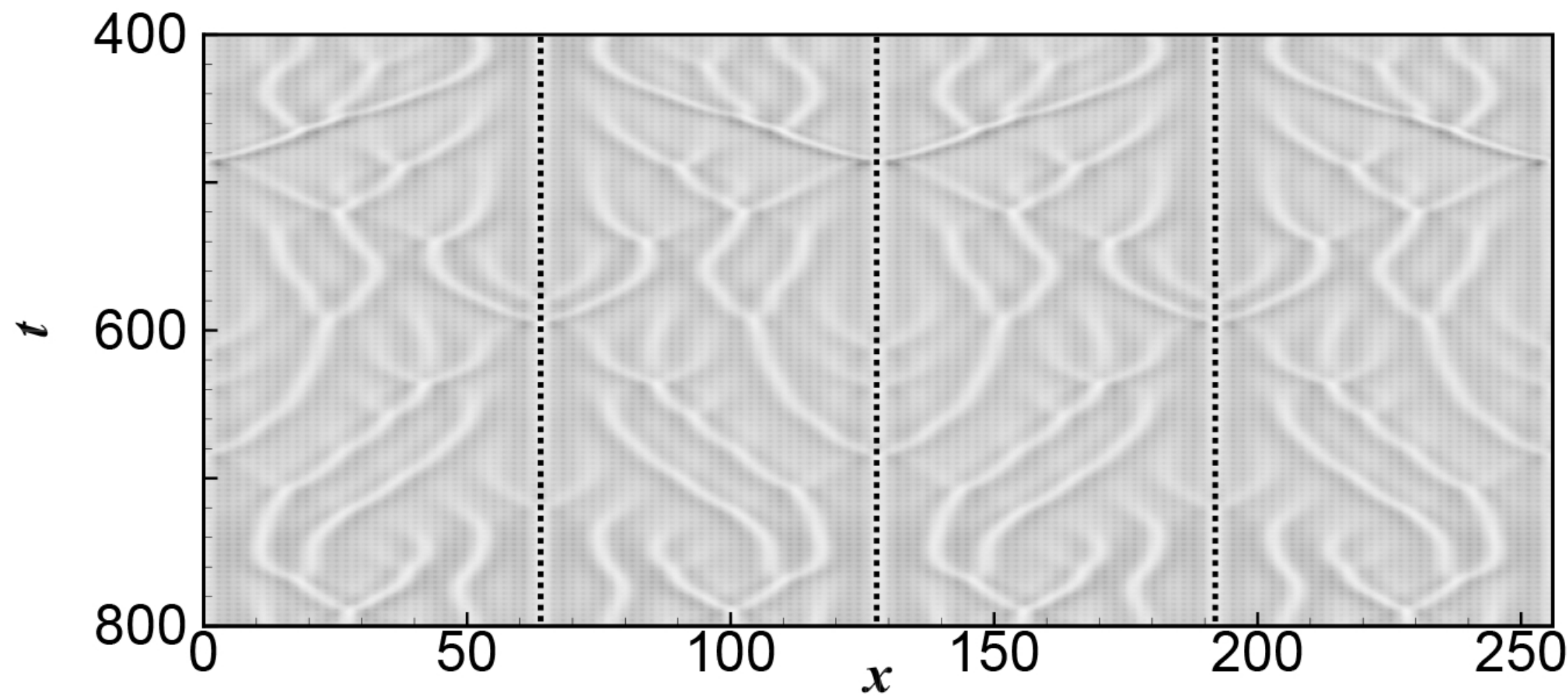}
\label{level.sub.4}
		\includegraphics[width=0.5\linewidth]{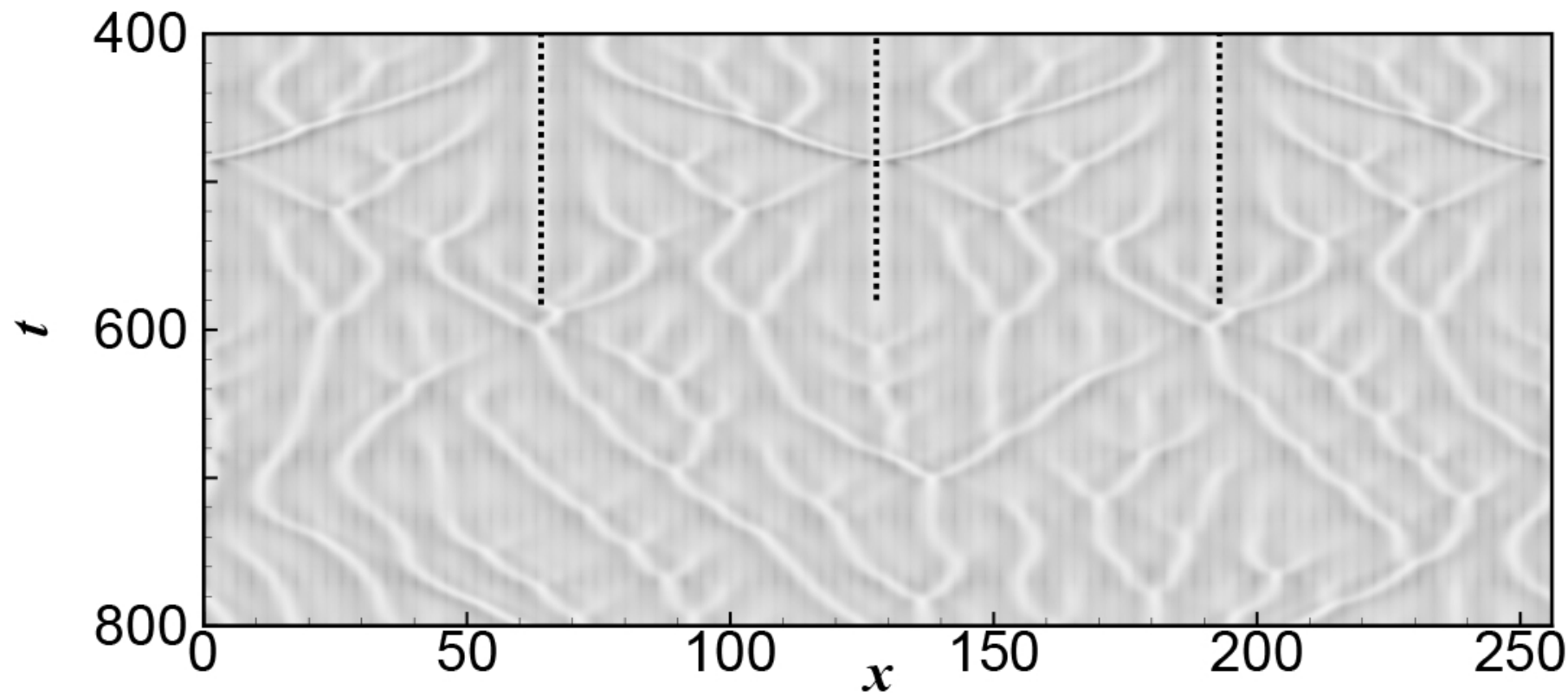}
}
	\subfigure[(c)]{
		\label{level.sub.3}
		\includegraphics[width=0.5\linewidth]{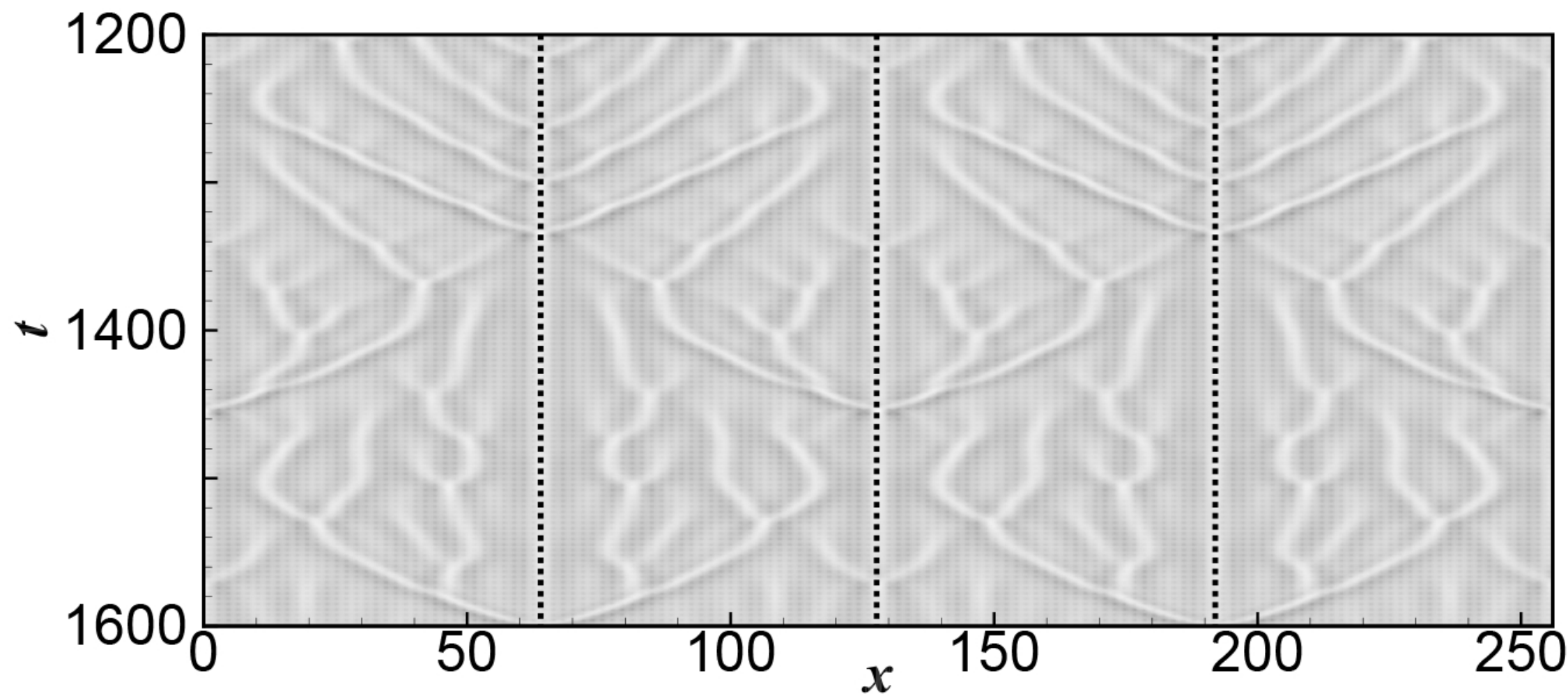}
\label{level.sub.4}
		\includegraphics[width=0.5\linewidth]{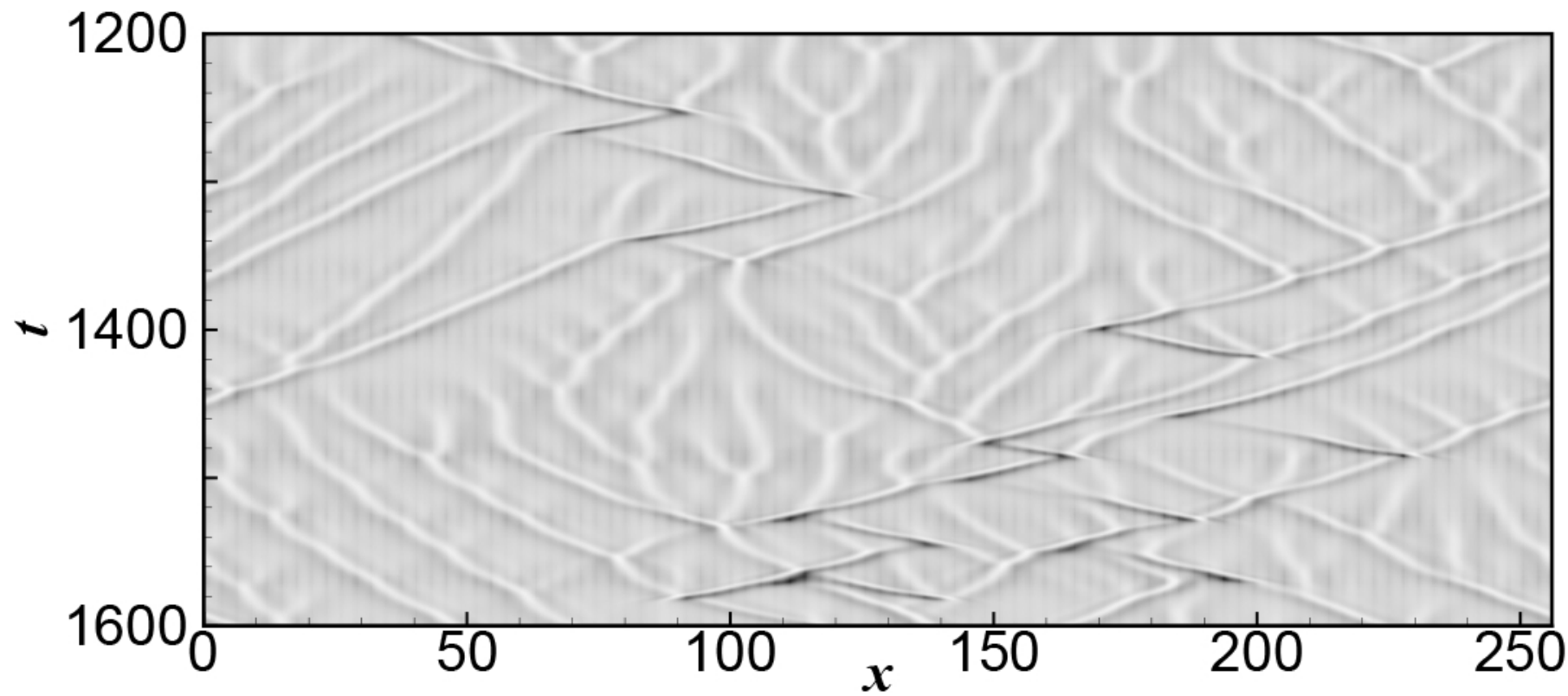}
}	
	\subfigure[(d)]{
		\label{level.sub.3}
		\includegraphics[width=0.5\linewidth]{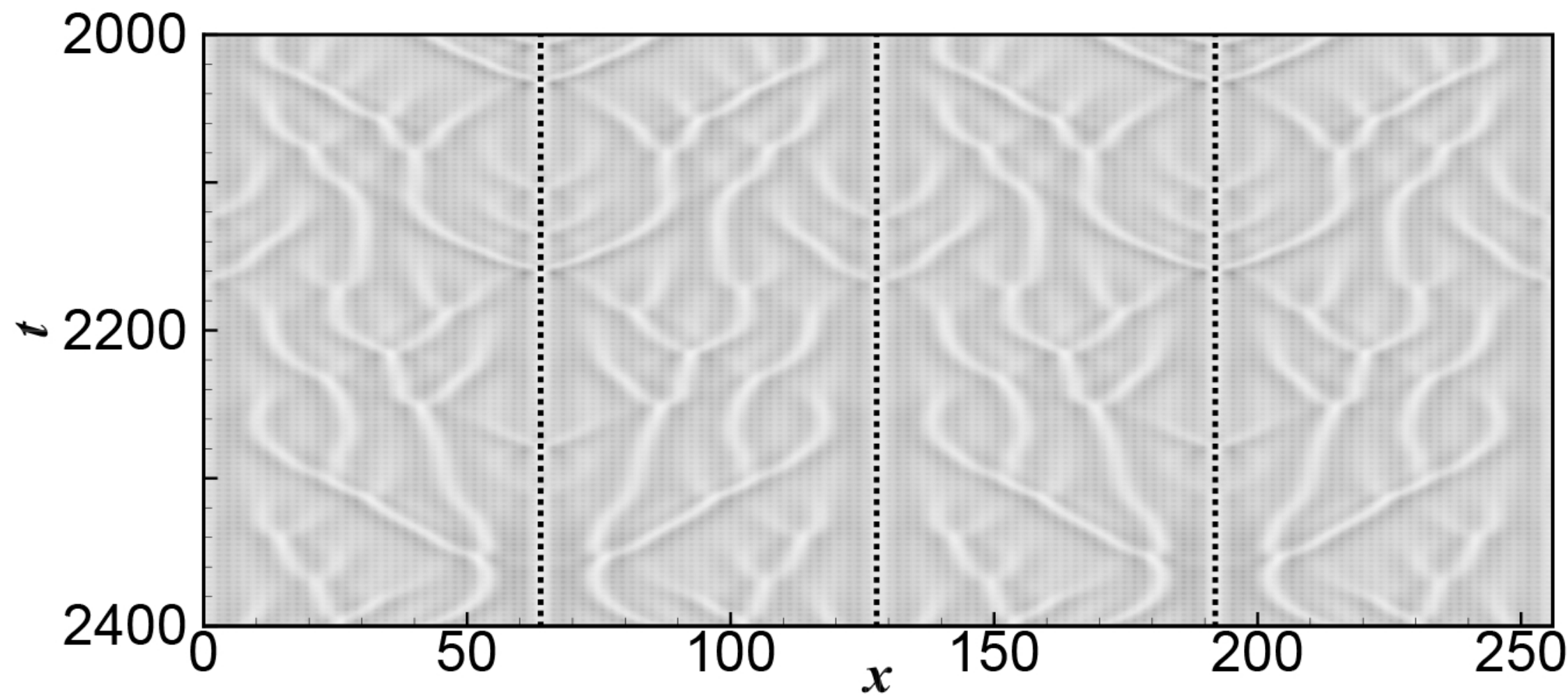}
\label{level.sub.4}
		\includegraphics[width=0.5\linewidth]{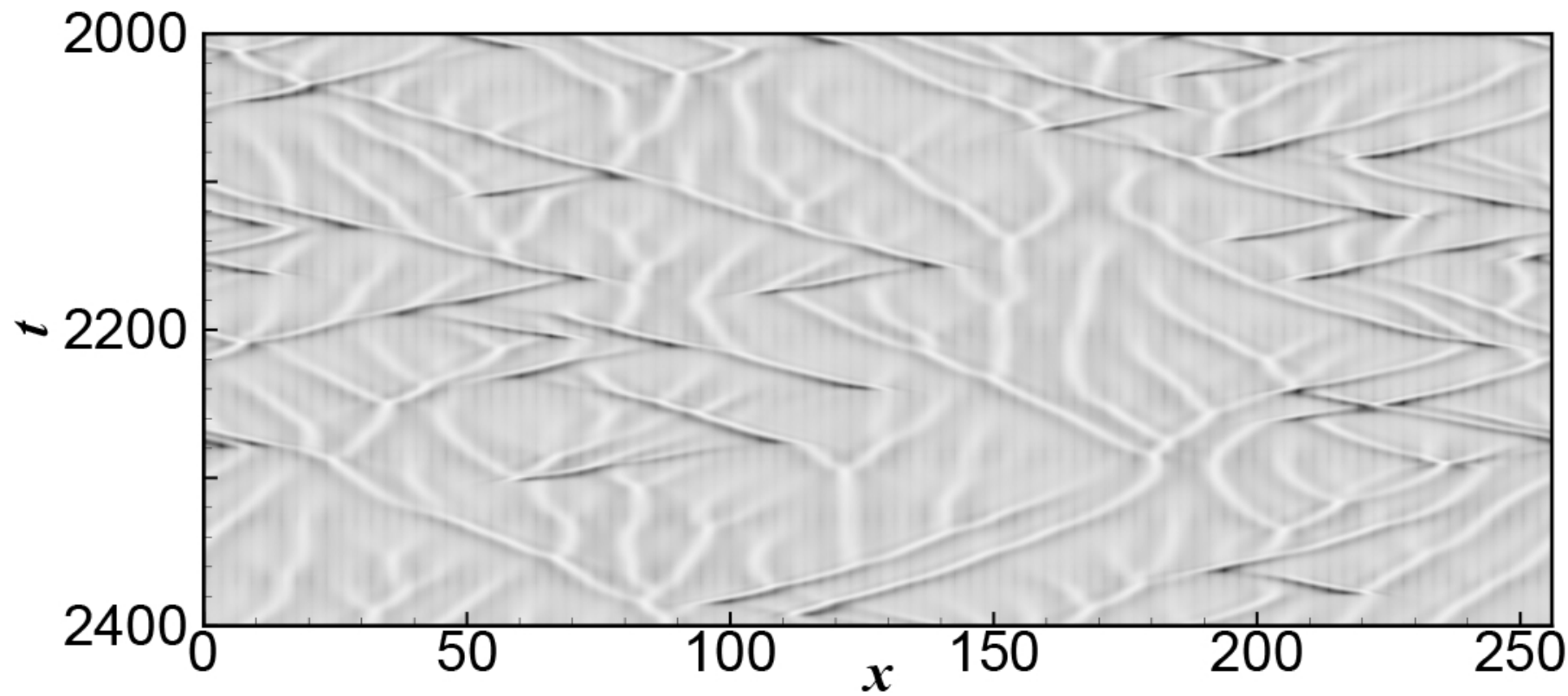}
}
	\subfigure[]{
 \qquad
 \quad

		CNS
 \qquad
  \qquad
  \qquad
  \qquad
  \qquad
  \qquad
  \qquad
  \qquad
  \qquad
RKwD
}
\caption{Spatio-temporal plot of $|A(x,t)|$ using grey level representation from $|A| = 1.25$ (white) to $|A|$ = 0 (black): (a)  $0 \leq t \leq 400$; (b) $400 \leq t \leq 800 $; (c)  $1200 \leq t \leq 1600$; and (d) $2000 \leq t \leq 2400$. CNS results (left panels); and RKwD results (right panels).}
	\label{spatio-temporal-|A|}
\end{figure}

\begin{figure}[tbhp]
\centering
  \subfigure[(a)]{
 \centering
    \includegraphics[width=16cm,height=6cm]{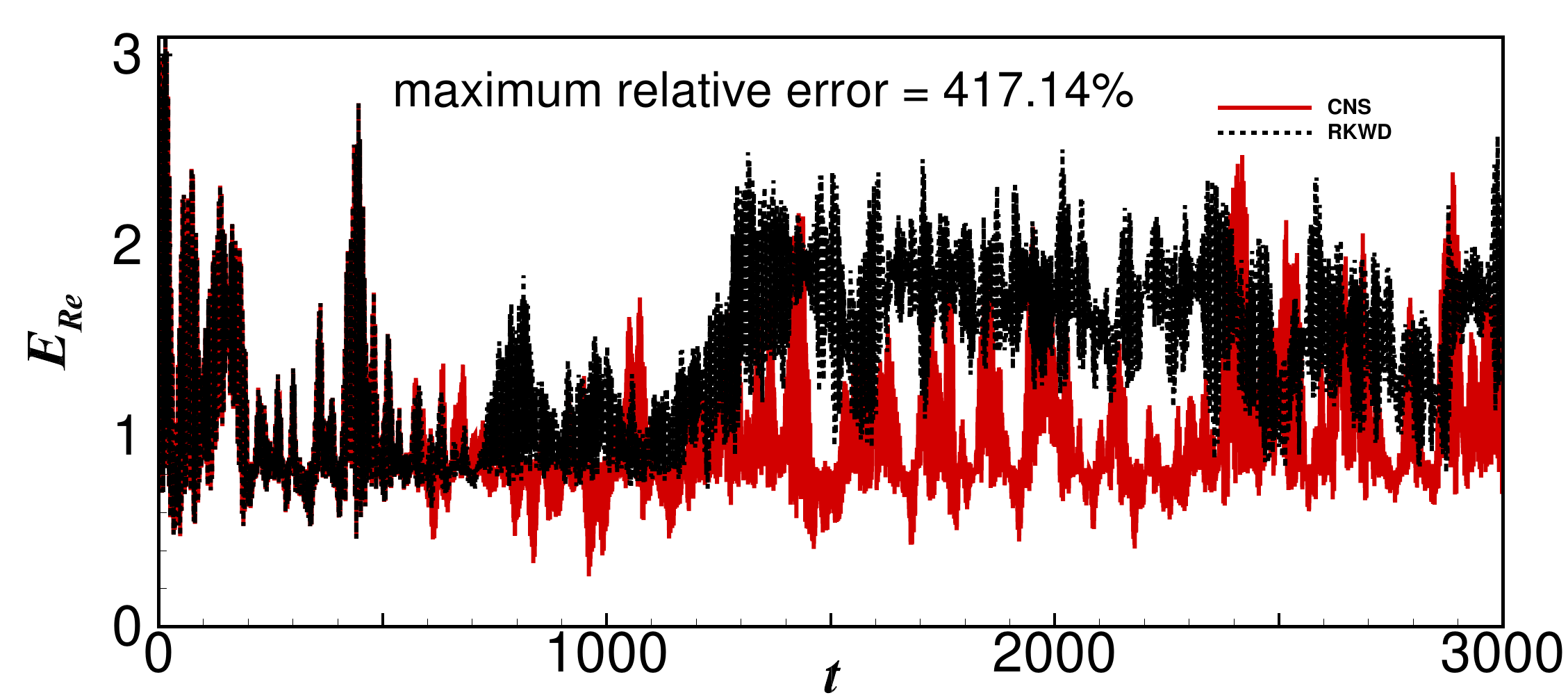}}
\centering
  \subfigure[(b)]{
 \centering
    \includegraphics[width=16cm,height=6cm]{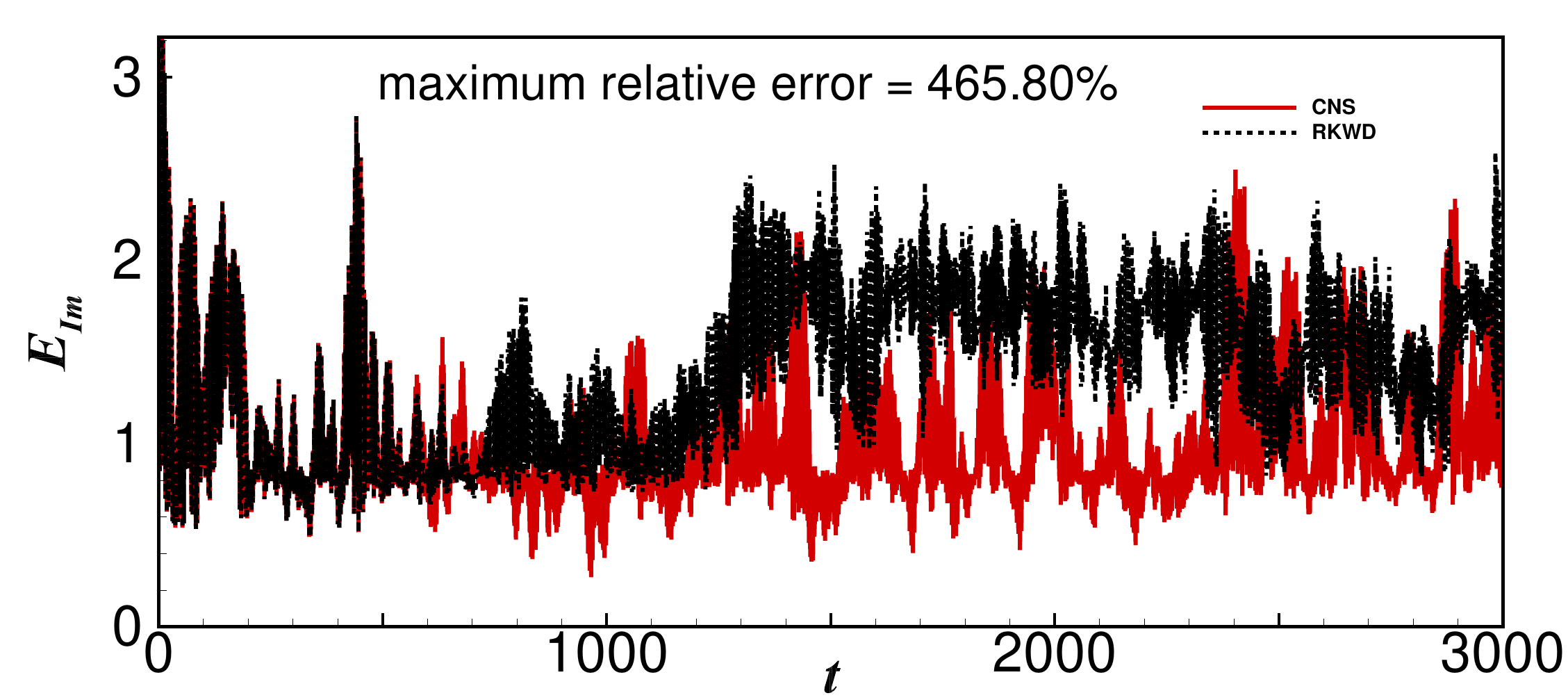}}

  \caption{Time histories of total spectrum-energy of the real and imaginary  parts of the numerical simulations obtained by CNS and RKwD in the chaotic case of  $c_{1}=2$ and  $c_{3}=1$ with the initial condition (\ref{initial2gg}): (a) real part,  with maximum relative error 417.14\%; and (b) imaginary part, with maximum relative error 465.80\%. CNS results (red line); and RKwD results (black dashed line).}
  \label{Energy-spectrum-Re-Im}
\end{figure}

\begin{figure}[tbhp]
 \centering
    \subfigure[(a)]{
    \includegraphics[width=16cm,height=6cm]{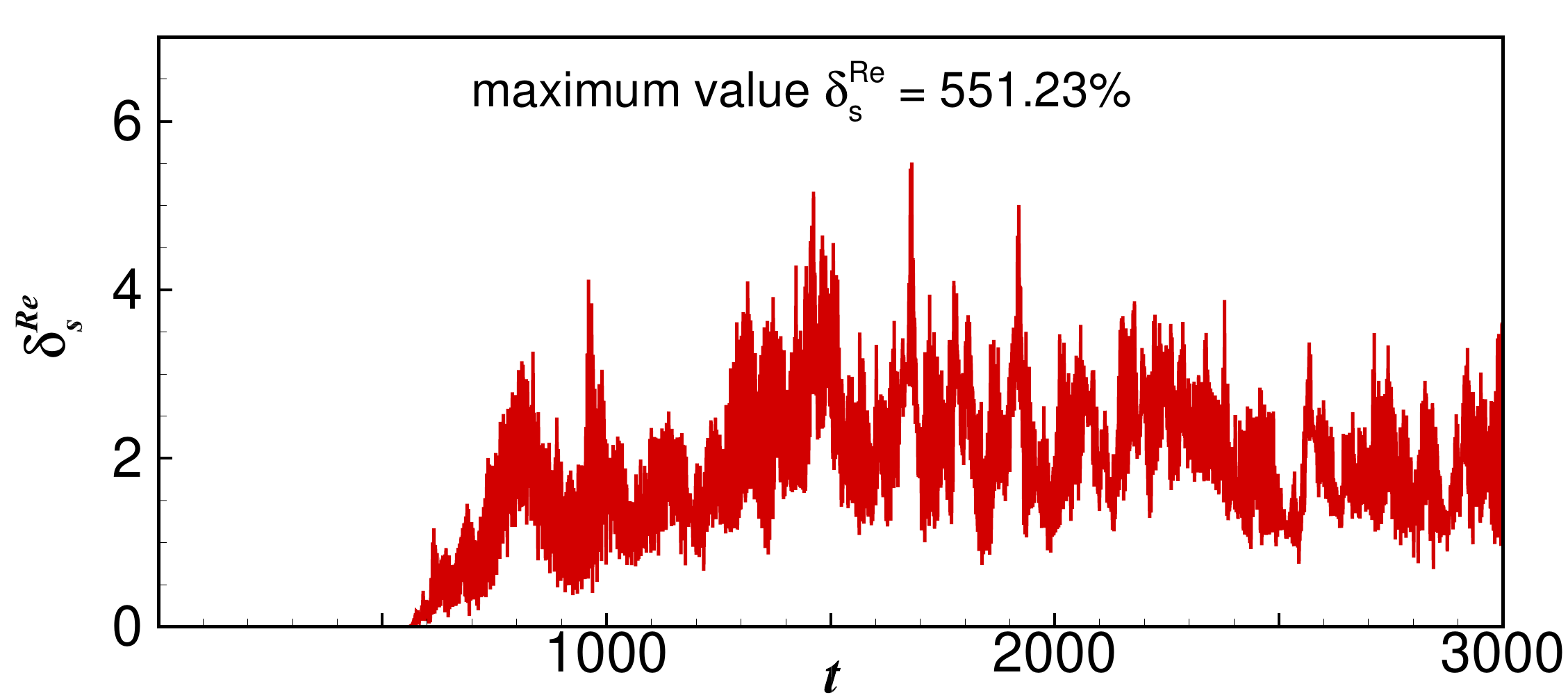}}
\subfigure[(b)]{
 \centering
    \includegraphics[width=16cm,height=6cm]{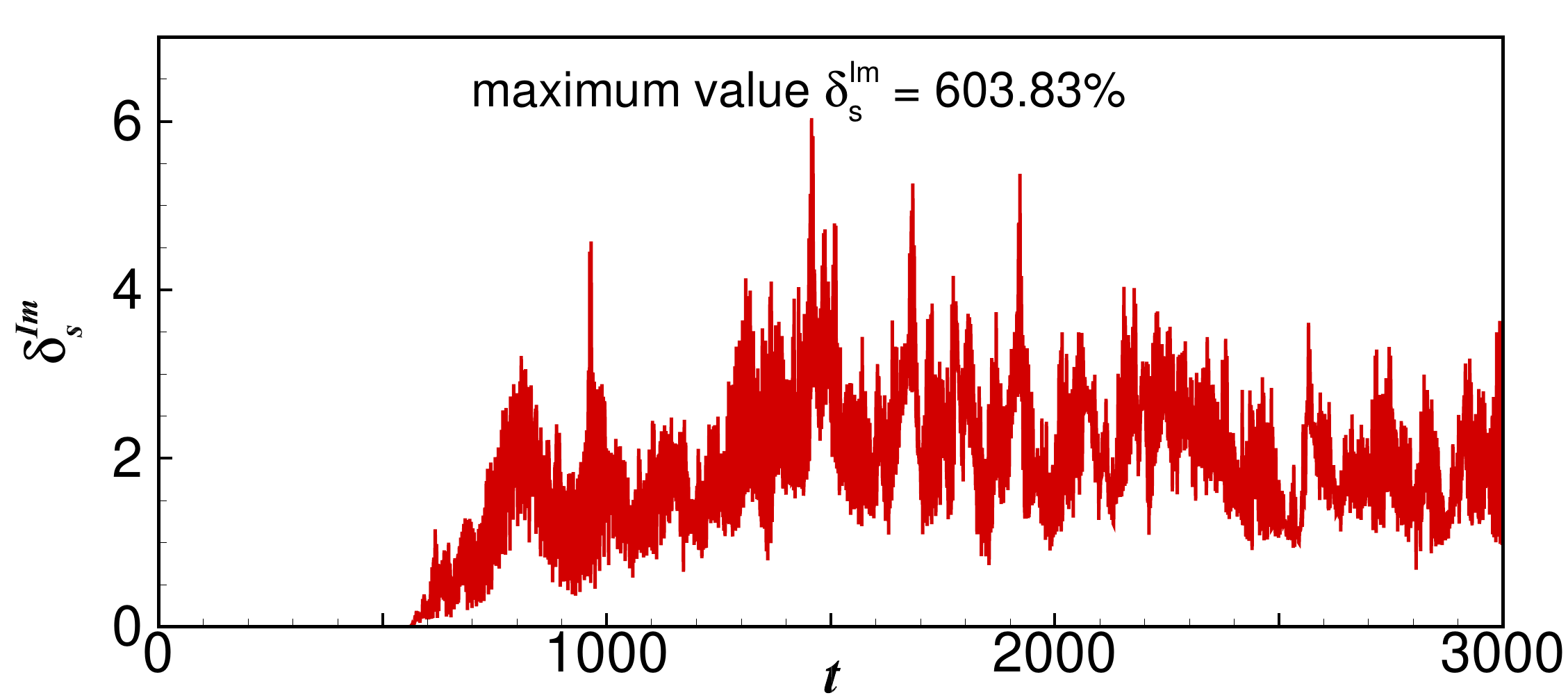}}
  \caption{Time histories of spectrum-deviation  of the real and  imaginary  parts  of the RKwD simulation in the chaotic case of  $c_{1}=2$ and  $c_{3}=1$ with the initial condition (\ref{initial2gg}), compared with the  benchmark given by CNS: (a) real part with maximum  value $\delta_s^{Re} = 551.23\% $; and (b) imaginary part with maximum value $\delta_s^{Im} = 603.83\%$. }
  \label{spectrum-deviation }
\end{figure}

\begin{figure}[tbhp]
\centering
	\subfigure[(a)]{
		\label{level.sub.3}
		\includegraphics[width=0.4\linewidth]{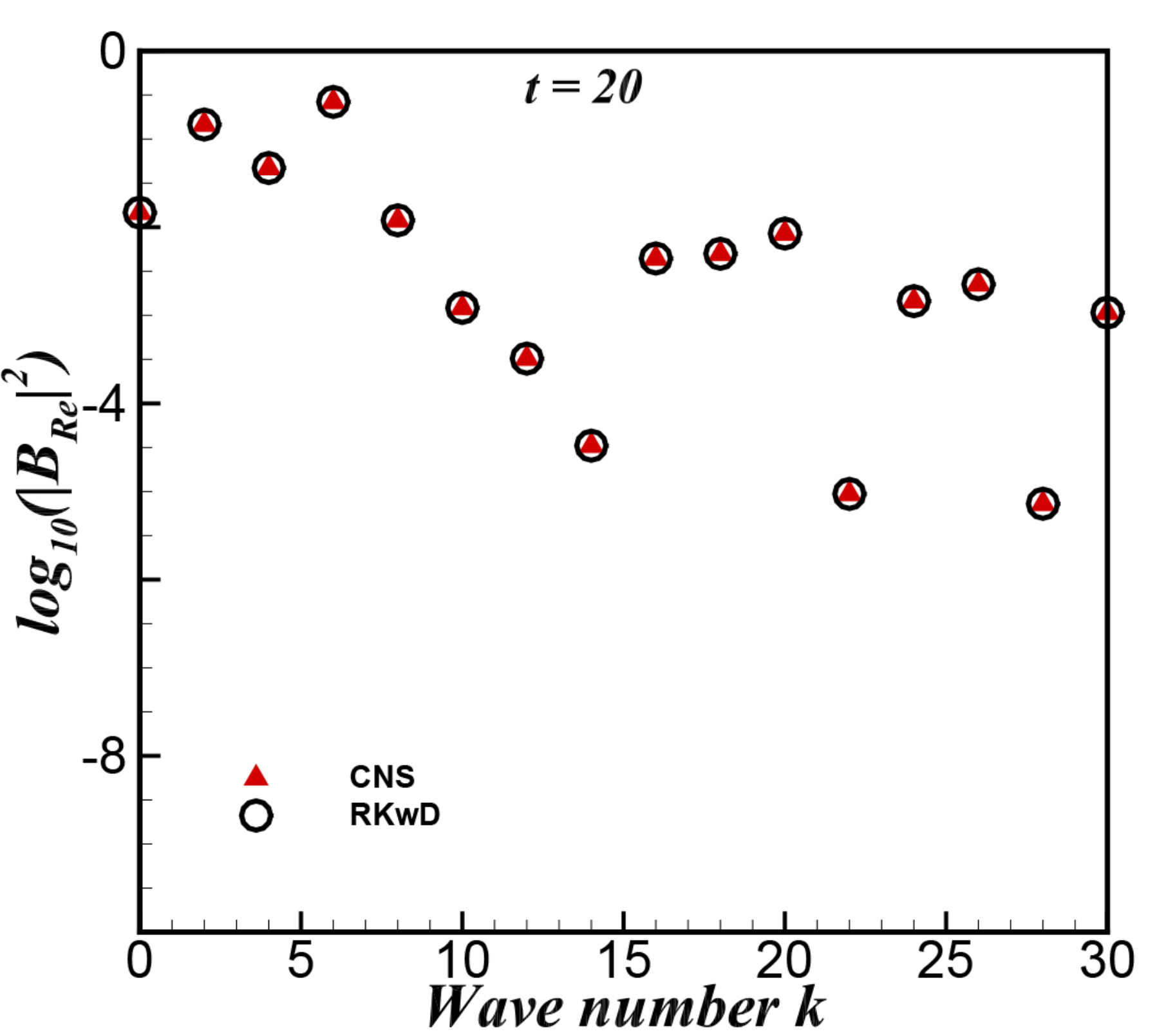}}
\subfigure[(b)]{
\label{level.sub.4}
		\includegraphics[width=0.4\linewidth]{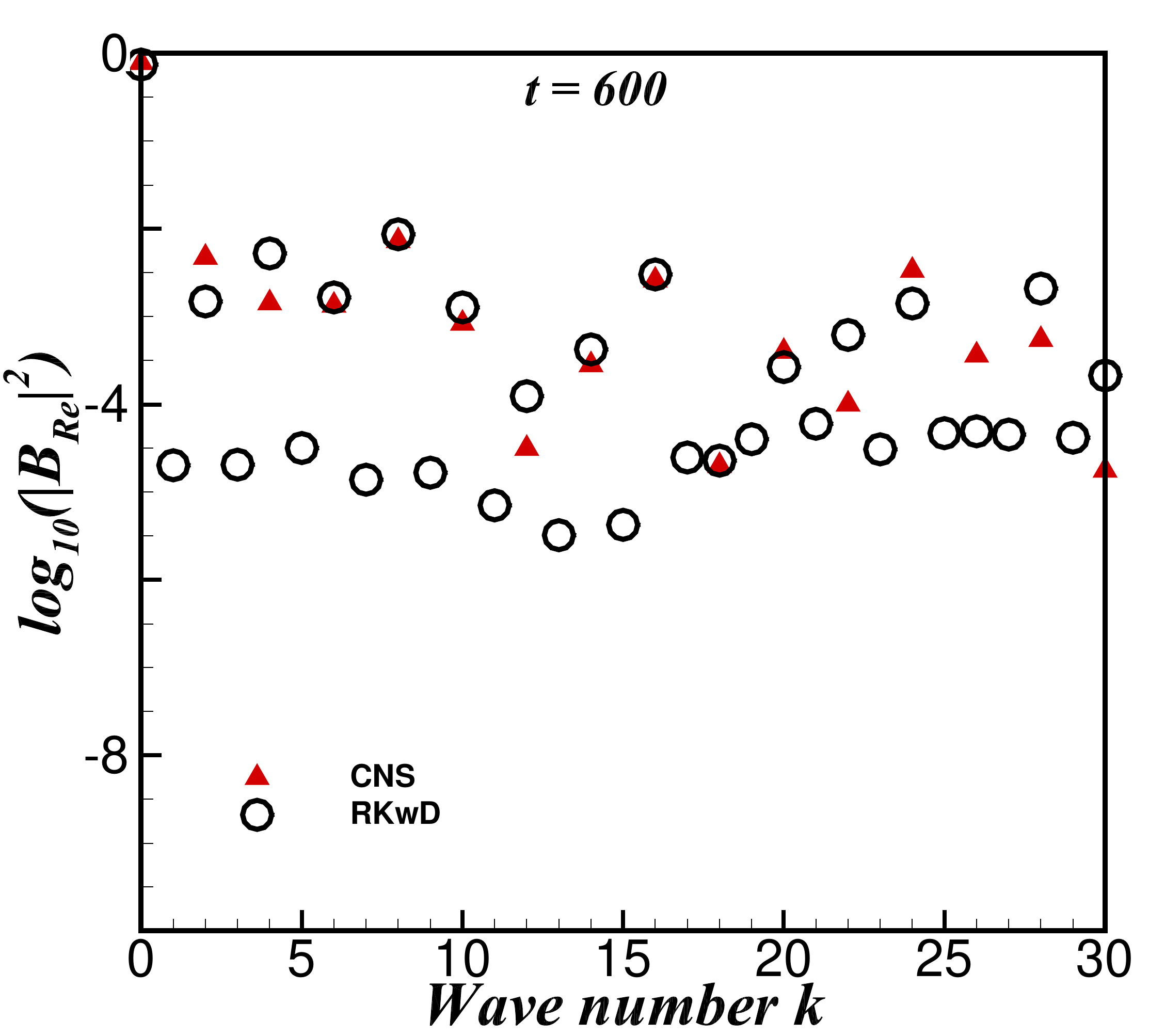}
}

	\subfigure[(c)]{
		\label{level.sub.3}
		\includegraphics[width=0.4\linewidth]{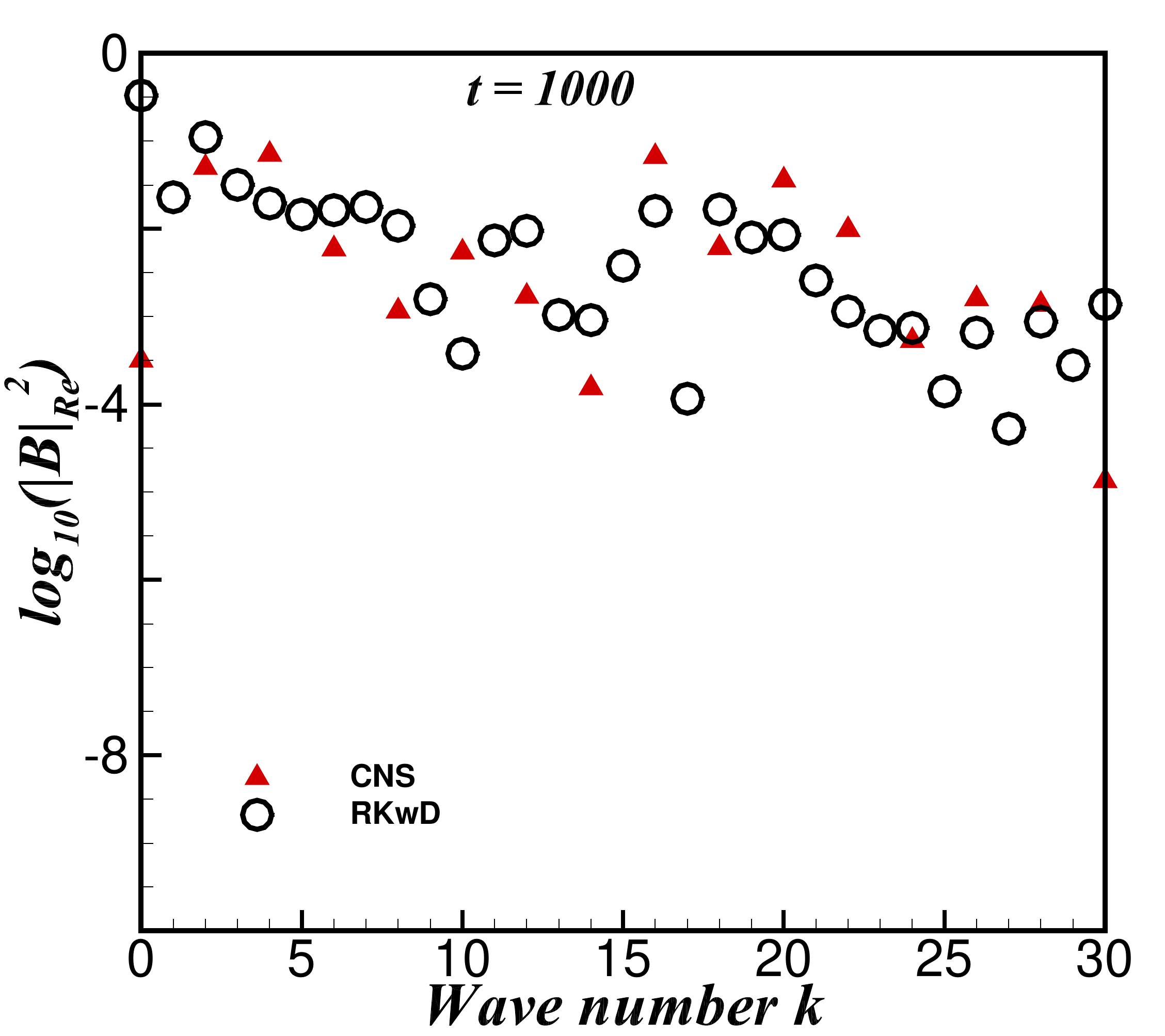}}
	\subfigure[(d)]{
		\label{level.sub.3}
		\includegraphics[width=0.4\linewidth]{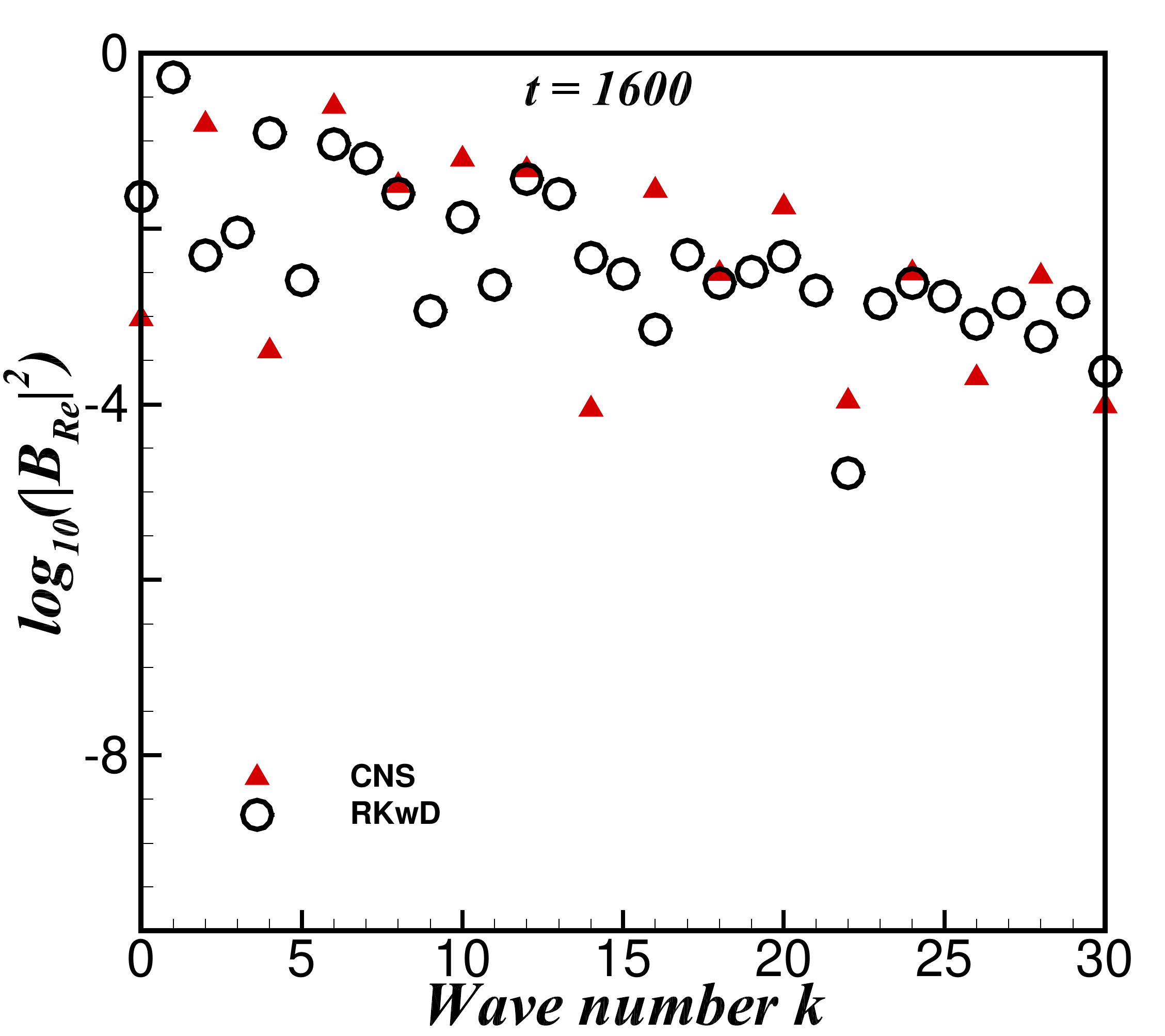}}
	
  \caption{Comparison of the spatial Fourier energy spectra of the real part of the computer-generated simulation obtained by the CNS and RKwD in the chaotic case of $c_{1}=2$ and  $c_{3}=1$ with the initial condition (\ref{initial2gg}): (a) $t = 20$; (b) $t = 600$; (c) $ t = 1000$; and (d) $ t = 1600$. CNS results (solid triangles); and RKwD results (open circles). }
  \label{Re_spectrum}
\end{figure}

\begin{figure}[tbhp]
\centering
	\subfigure[(a)]{
		\label{level.sub.3}
		\includegraphics[width=0.4\linewidth]{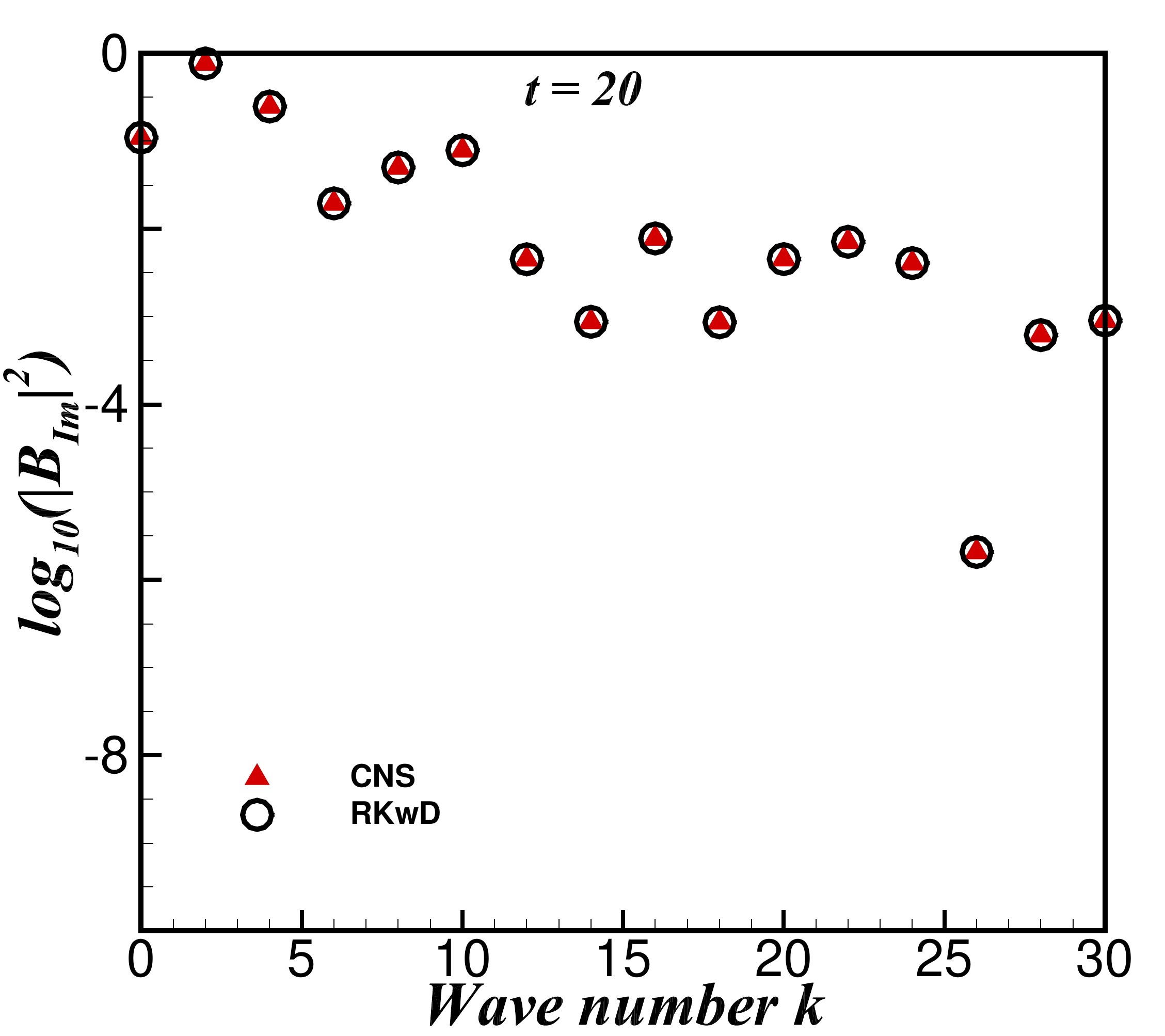}}
\subfigure[(b)]{
\label{level.sub.4}
		\includegraphics[width=0.4\linewidth]{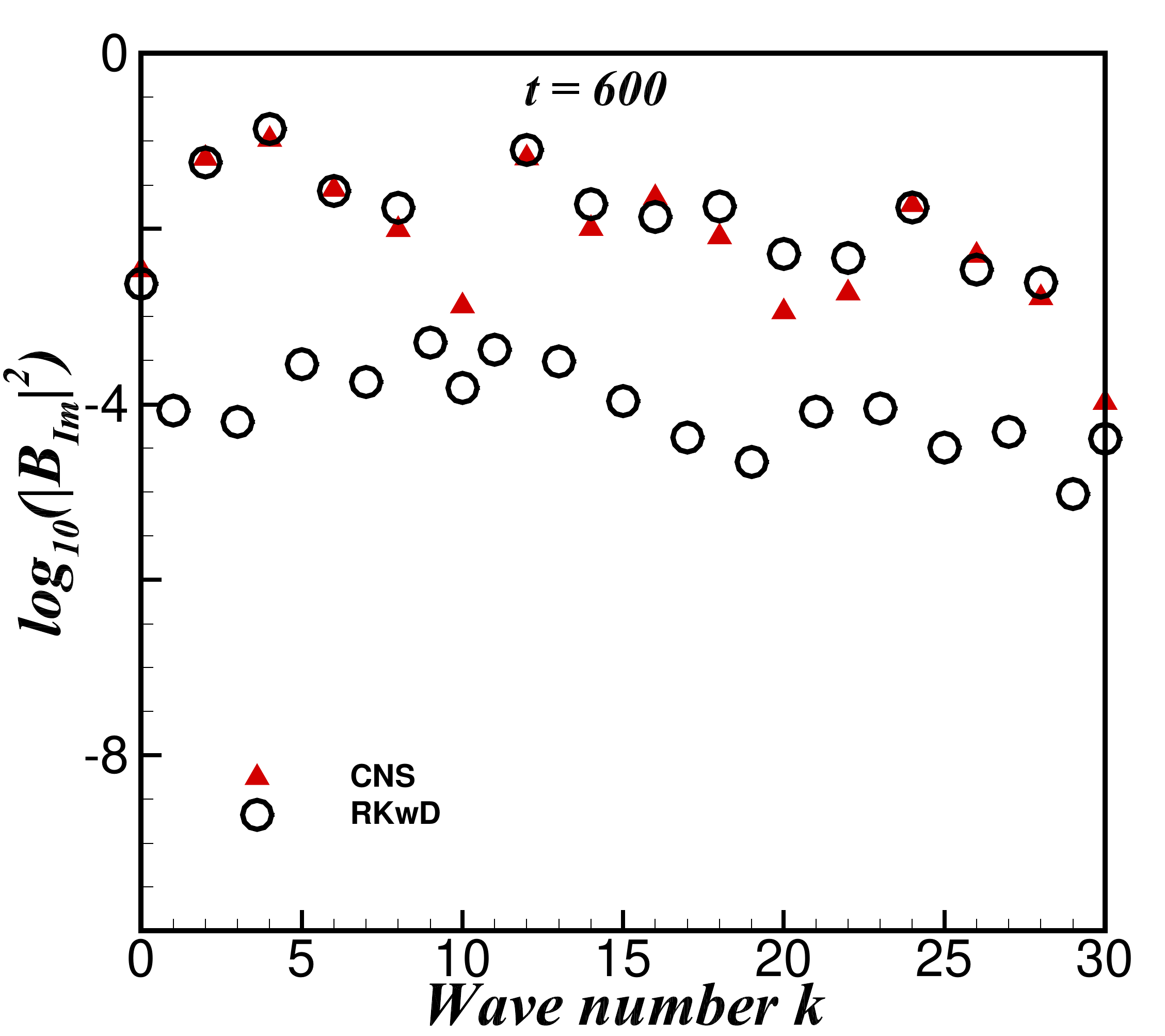}
}

	\subfigure[(c)]{
		\label{level.sub.3}
		\includegraphics[width=0.4\linewidth]{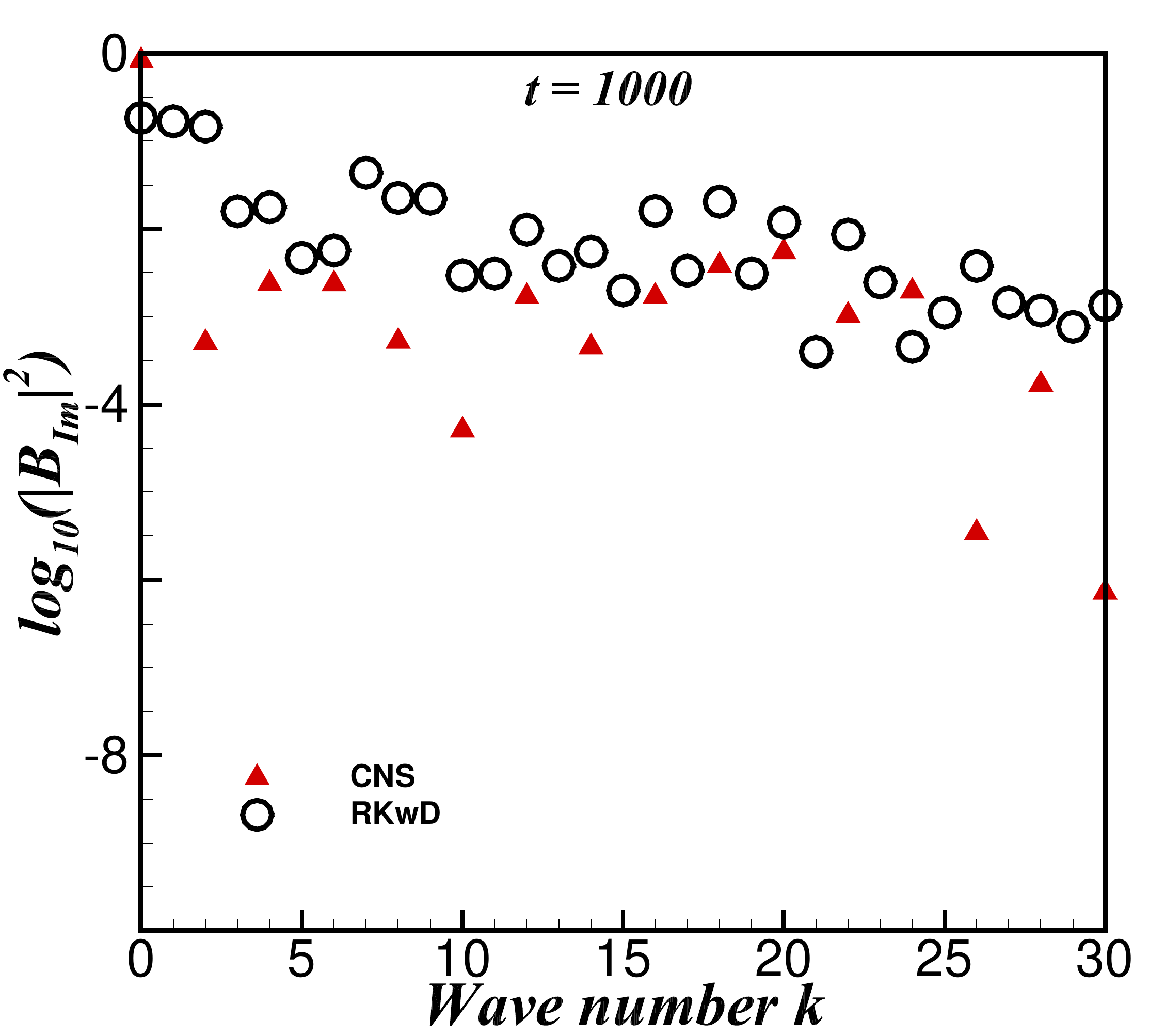}}
	\subfigure[(d)]{
		\label{level.sub.3}
		\includegraphics[width=0.4\linewidth]{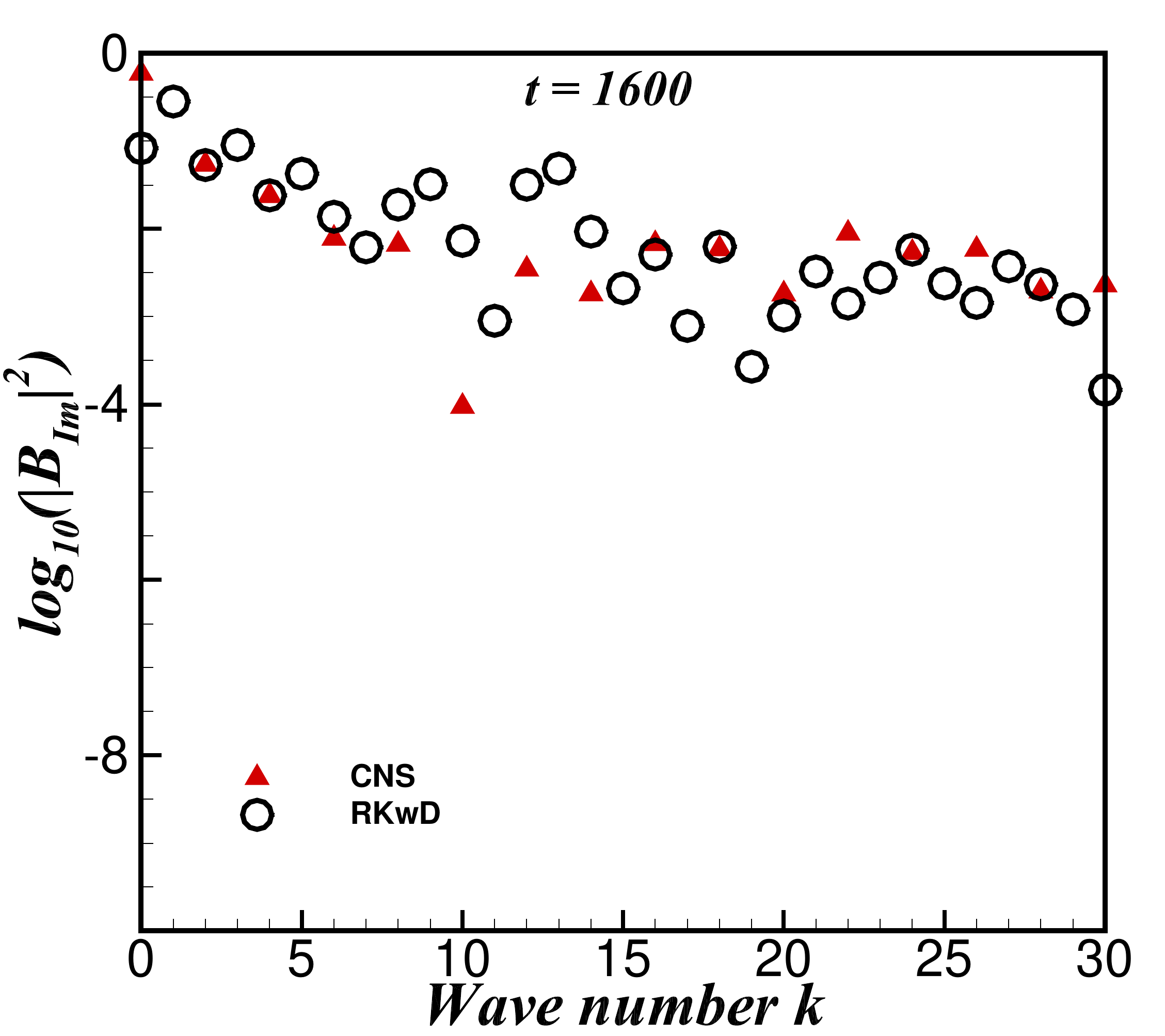}}
	
  \caption{ Comparison of the spatial Fourier energy spectra of the imaginary part of the computer-generated simulations obtained by the CNS and RKwD in the chaotic case of $c_{1}=2$ and  $c_{3}=1$ with the initial condition (\ref{initial2gg}): (a) $t = 20$; (b) $t = 600$; (c) $ t = 1000$; and (d) $ t = 1600$. CNS results (solid triangles); and RKwD results (open circles).  }
  \label{Im_spectrum}
\end{figure}

Without loss of generality,  let us consider the one-dimensional CGLE within the chaotic regime when $c_{1}=2$ and  $c_{3}=1$ for the same initial condition (\ref{initial2gg}). In this case, the solution exhibits periodicity and symmetry in space.  The reliable numerical simulations obtained by the CNS algorithm in physical space (marked as ``CNS'') using $N_s=105$ are used as benchmarks, which are compared with the corresponding results obtained by the temporal 4th-order Runge-Kutta method with double precision (marked as  ``RKwD'').  Both the CNS and RKwD results have the same mode number $N=4096$ for spatial Fourier expansion  so that they have the same  spatial  truncation error.  For time evolution, the CNS algorithm in physical space uses a high-order Taylor expansion with tolerance $tol=10^{-105}$, whereas the RKwD utilises a time-step $\Delta t=10^{-4}$ with associated temporal truncation error of order $O(10^{-16})$ (which is at the same level of the round-off error of the RKwD due to the use of double precision).  According to (\ref{Tc-(N,Ns)}), the CNS result (with $N_s = 105$ and $N=4096$) has critical predictable time $T_{c} > 4500$,  which is much larger than $T_{c} \approx 580$ for the RKwD simulation (with $N_s = 16$ and $N=4096$).  Notably, the  CNS result is reliable over an interval of time about 7.7  times longer than that of the RKwD one.   To guarantee the reliability, we use the CNS result in a smaller interval of time, say, $t\in[0,3000]$, which therefore can be  certainly used as a benchmark for comparisons described below.

  Note that the  initial condition (\ref{initial2gg}) has a kind of spatial symmetry, which, according to the governing equation, should be retained by the solution (if evaluated correctly).  As shown in Fig.~\ref{spatio-temporal-|A|},  this is indeed true for the CNS result in the {\em whole} spatial domain $x\in[0, L]$ throughout the {\em entire} interval $t\in[0,3000]$.    However, the numerical simulation given by the RKwD  loses the spatial  symmetry at  $t \approx 580$.  For $t< 580$, the result given by the RKwD agrees quite well with that by the CNS.  However, thereafter,  the deviation  between the two simulations becomes larger and larger, and the spatial symmetry in the RKwD result even breaks.   This occurs mainly because, due to the butterfly-effect, numerical noise in the RKwD result  increases  exponentially. For  $t>580$,  the RKwD result appears to become a mixture of the  true solution and  numerical noise at the {\em same} level of magnitude:  as the numerical noise  increases exponentially, the randomness in the noise  first creates a distinct difference in the spatio-temporal trajectories and then causes a spatial {\em symmetry-breaking}.  This indicates that numerical noise in the RKwD solvers might generate {\em not only} quantitative differences in the spatio-temporal trajectories {\em but also} some qualitative  deviations  in  computer-generated simulation of chaotic system.

We now investigate the influence of numerical noise on the real and imaginary parts of the RKwD result, separately. At any given time,  the result for each part can be expressed as a  spatial Fourier expansion.  Note that a spatial energy-spectrum of a spatio-temporal chaos at a given time often has important  physical meanings. With this in mind, Fig.~\ref{Energy-spectrum-Re-Im} depicts the comparison of the total spectrum-energy of the real and  imaginary  parts of the two simulations given by the CNS (in physical space) and RKwD, respectively. The two methods give almost the same total spectrum-energy when $t\leq 580$.  However, thereafter, the deviation  becomes increasingly distinct, with the maximum relative error reaching 417.14\% for the real part and  465.80\% for the imaginary  part.  The corresponding spectrum-deviation of the RKwD result (compared with the CNS one), defined by (\ref{delta}), becomes distinct after $t > 580$, with a maximum value of 551.23\% for the real part and 603.83\% for the imaginary part, as shown in Fig.~\ref{spectrum-deviation }. These results demonstrate that numerical noise can lead to great deviations not only in the spatio-temporal trajectories  but also in the total spatial spectrum-energy of a chaotic system!

Why does numerical noise have such a huge influence on the total spectrum-energy of the real and imaginary  parts of  the RKwD simulation?    To answer this question, let us compare the spatial Fourier spectra of the real part of the CNS and RKwD results at different times, as shown in Fig.~\ref{Re_spectrum}.  Note that, the odd wave numbers, i.e. $k = 1,3,5, \cdots$, of the spatial Fourier spectra of the CNS result do {\em not} contain any energy at any stage throughout the {\em whole} interval of time $0\leq t \leq 3000$.  In other words, the odd wave number components of the spatial Fourier spectrum of the CNS result {\em always} remain zero.  This is exactly the reason why the CNS results invariably retain the spatial symmetry over the entire time interval $0 \leq t \leq 3000$.  However, this  property of the true solution does not hold for the RKwD simulation. For small time, such as $t = 20$, the two spectra agree quite well.  When $t=600$, tiny differences can be discerned between the two spectra, and some odd wave number components in the spatial Fourier spectrum of the RKwD simulation have become energetic. Furthermore,  Fig.~\ref{Re_spectrum} (c) and (d) show that as time increases, the deviation between the two spectra grows, and the odd wave number components in the spatial Fourier spectrum of the RKwD simulation contain an increasing amount of energy.   At  a  sufficiently large time, such as $t = 1600$,  the odd wave number components in the spatial Fourier spectrum of the RKwD simulation reach the {\em same} level of energy as the even components, as shown in Fig.~\ref{Re_spectrum} (d).  This reveals a fundamental  mistake in the RKwD simulation methodology.  By comparing the spatial Fourier spectra  (Fig.~\ref{Im_spectrum}) of the imaginary  part of the RKwD simulation with that of the CNS one,  we reach the same conclusion.   These results demonstrate that the random numerical noise of a chaotic system can rapidly increase to the {\em same} level  (in both the real and  imaginary  parts) of the  true solution and besides could transfer  to {\em all} wave numbers of the spatial Fourier spectrum!   The foregoing comparisons explain why the RKwD simulation  might not only  lose  the spatial symmetry  but also lead  to great  deviations  in total spectrum-energy, and why  numerical noise can lead to huge deviations in the spatio-temporal trajectories, the total spectrum-energy, and certain fundamental properties (such as spatial symmetry) of a chaotic system.

\begin{figure}[htbp]
\centering
	\subfigure[(a)]{
		\label{level.sub.3}
		\includegraphics[width=0.4\linewidth]{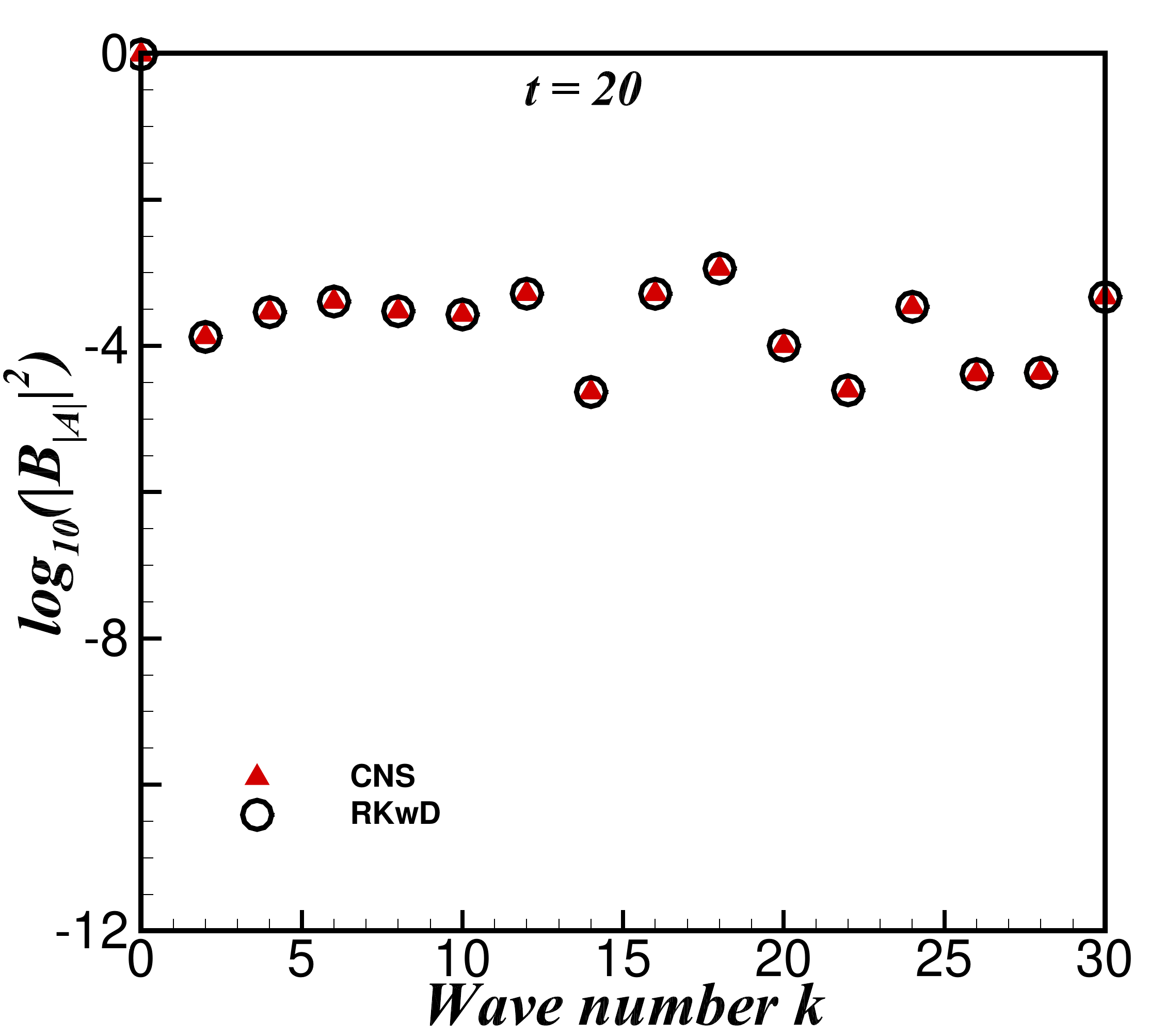}}
\subfigure[(b)]{
\label{level.sub.4}
		\includegraphics[width=0.4\linewidth]{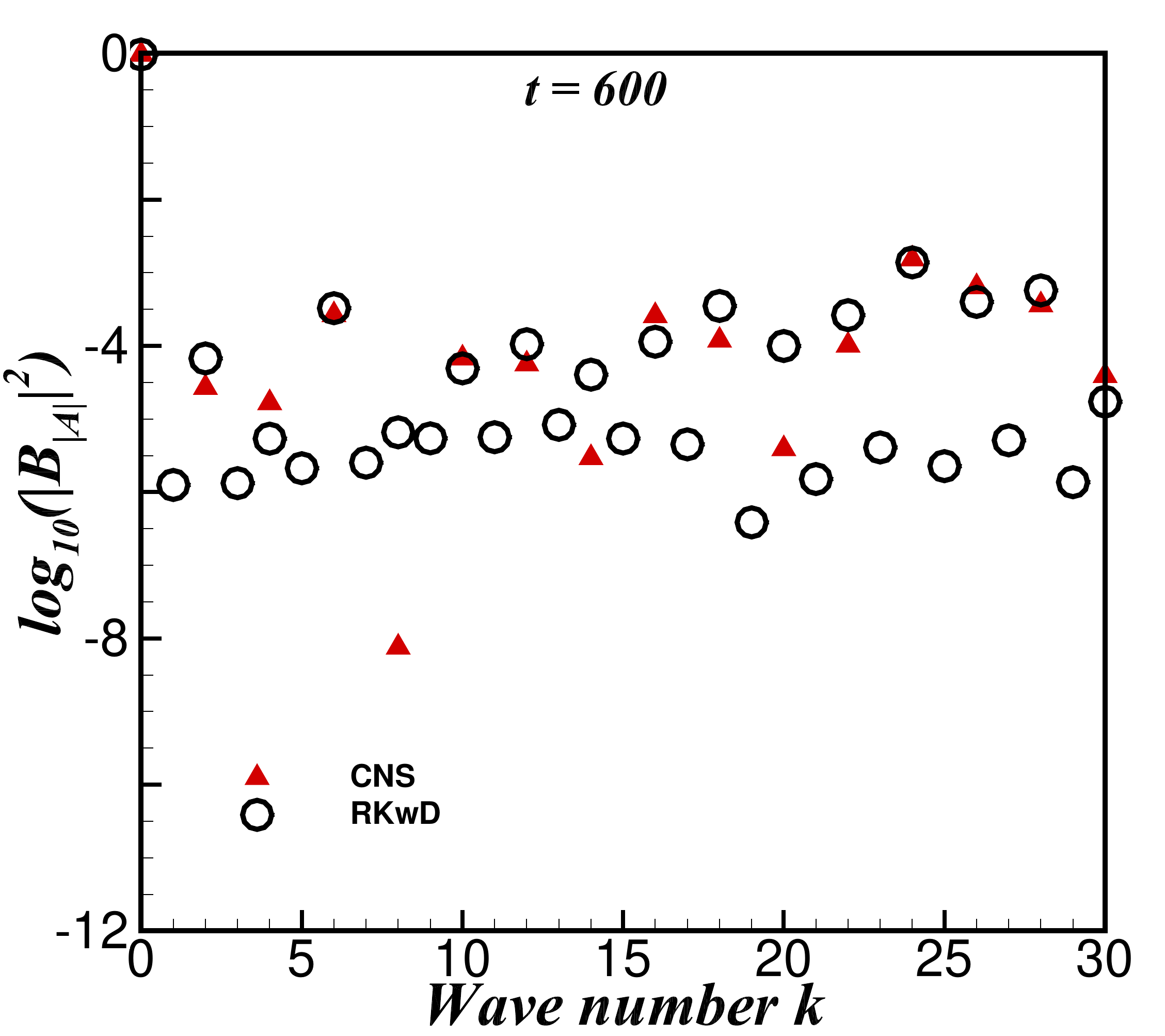}
}

	\subfigure[(c)]{
		\label{level.sub.3}
		\includegraphics[width=0.4\linewidth]{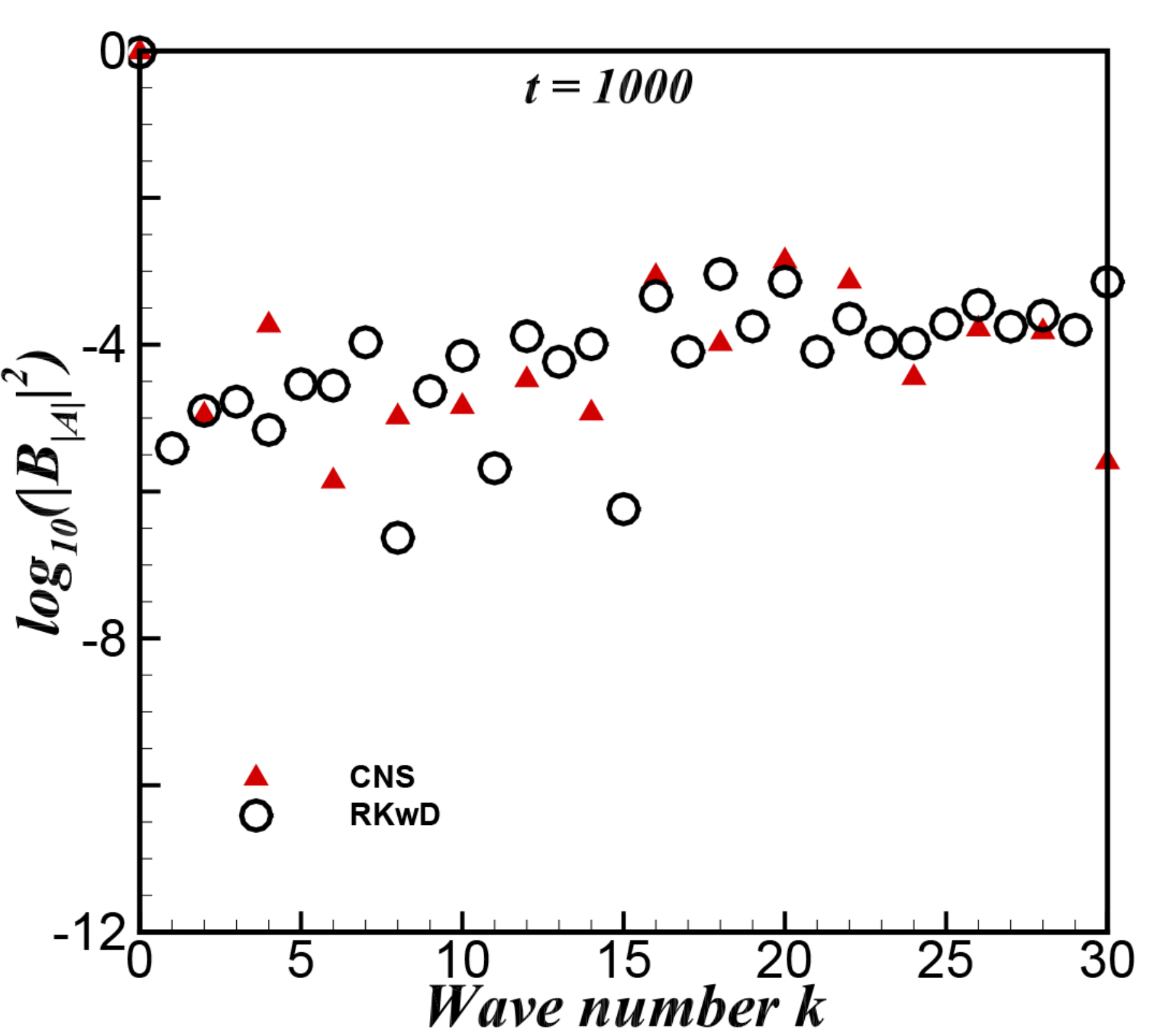}}
	\subfigure[(d)]{
		\label{level.sub.3}
		\includegraphics[width=0.4\linewidth]{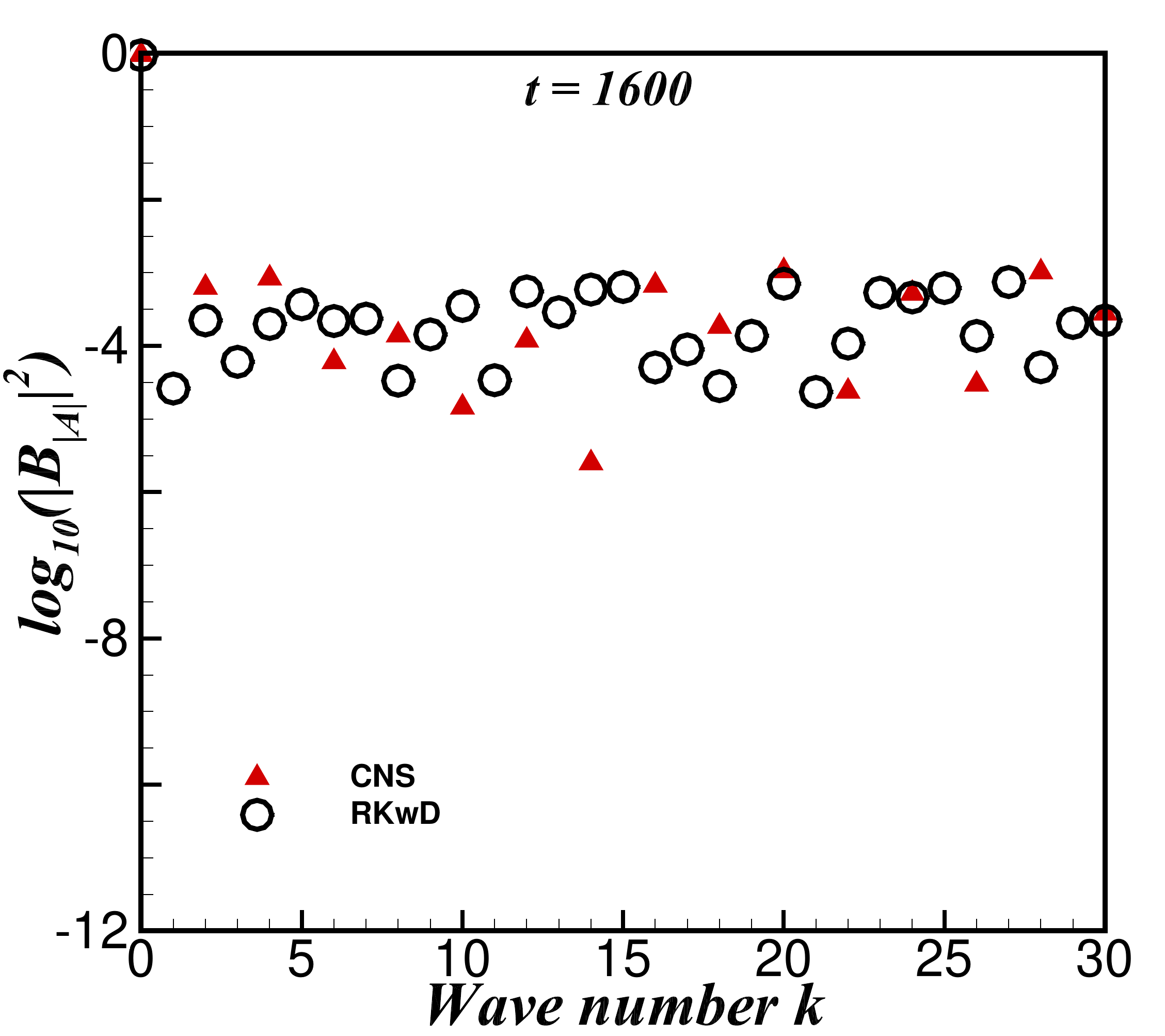}}
\caption{Comparison of the spatial Fourier energy spectra  of  the computer-generated simulation $|A(x,t)|$ obtained by CNS and RKwD in the chaotic case of $c_{1}=2$ and  $c_{3}=1$ with the initial condition (\ref{initial2gg}): (a) $ t = 20$; (b) $t = 600$; (c) $ t = 1000$; and (d) $ t = 1600$.  CNS results (solid triangles); and RKwD results (open circles). }
  \label{energy_spectrum-|A|}
\end{figure}

Fig.~\ref{energy_spectrum-|A|} presents  a comparison between the spatial Fourier spectra of $|A(x,t)|$ obtained using the CNS and RKwD at different times.  Again,  the odd wave number components in the CNS spectrum contain no energy throughout the entire time interval, corresponding to the spatial symmetry of the true chaotic solution.   At small times, such as $t=20$, the two spectra are in close agreement.  However, as time further increases, the odd wave number components in the spatial Fourier spectrum of $|A|$ obtained using RKwD amplify to the point at which they could reach the same energy level as the even components (Fig.~\ref{energy_spectrum-|A|} (b), (c) and (d)) at $t=600$, $t=1000$ and $t=1600$.  Again, this is incorrect, and is an artefact of numerical noise as it grows in the system as predicted by RKwD.

\begin{figure}[tbp]
\centering
	\subfigure{
		\label{level.sub.3}
		\includegraphics[width=0.5\linewidth]{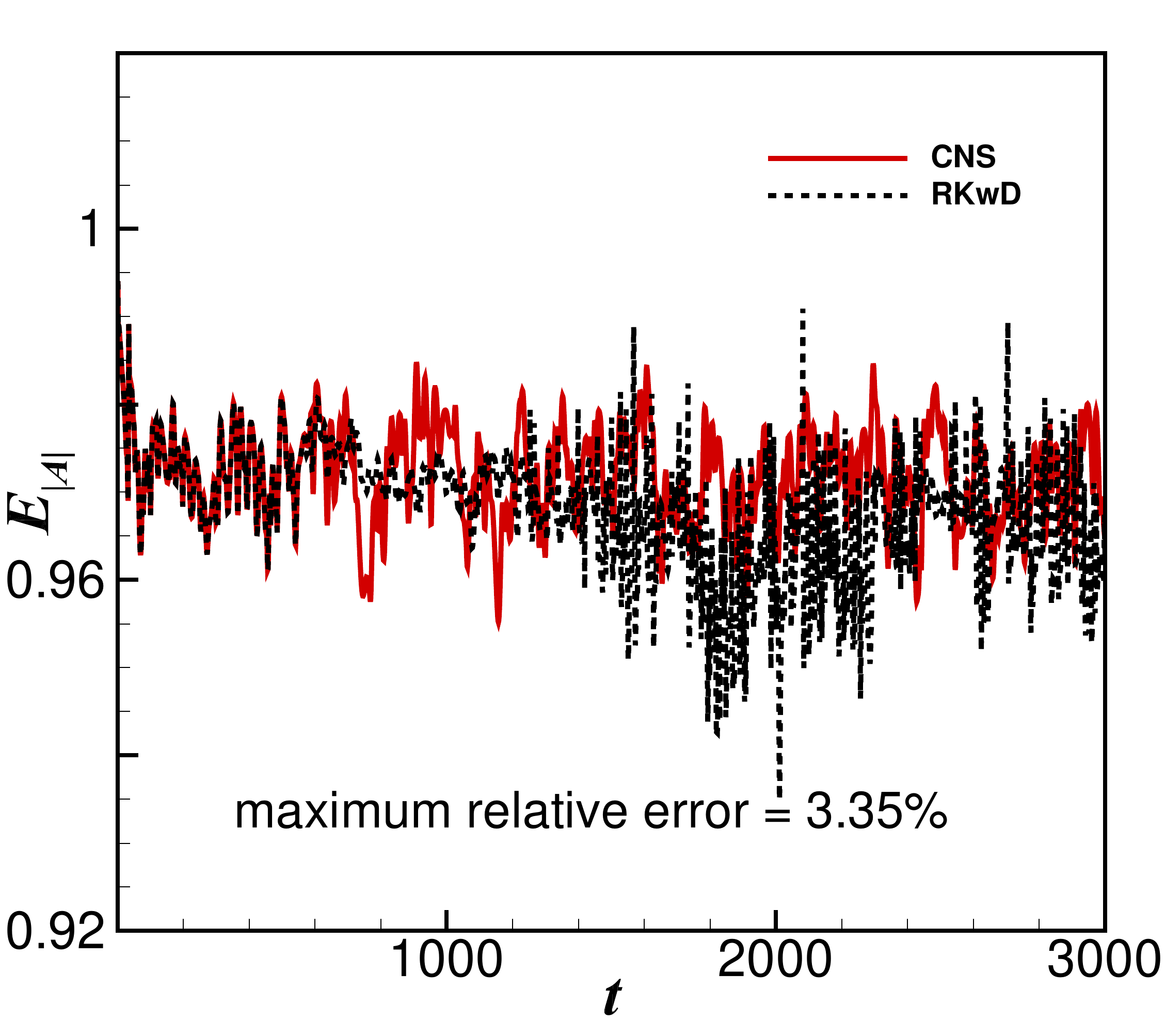}
\label{level.sub.4}
		\includegraphics[width=0.5\linewidth]{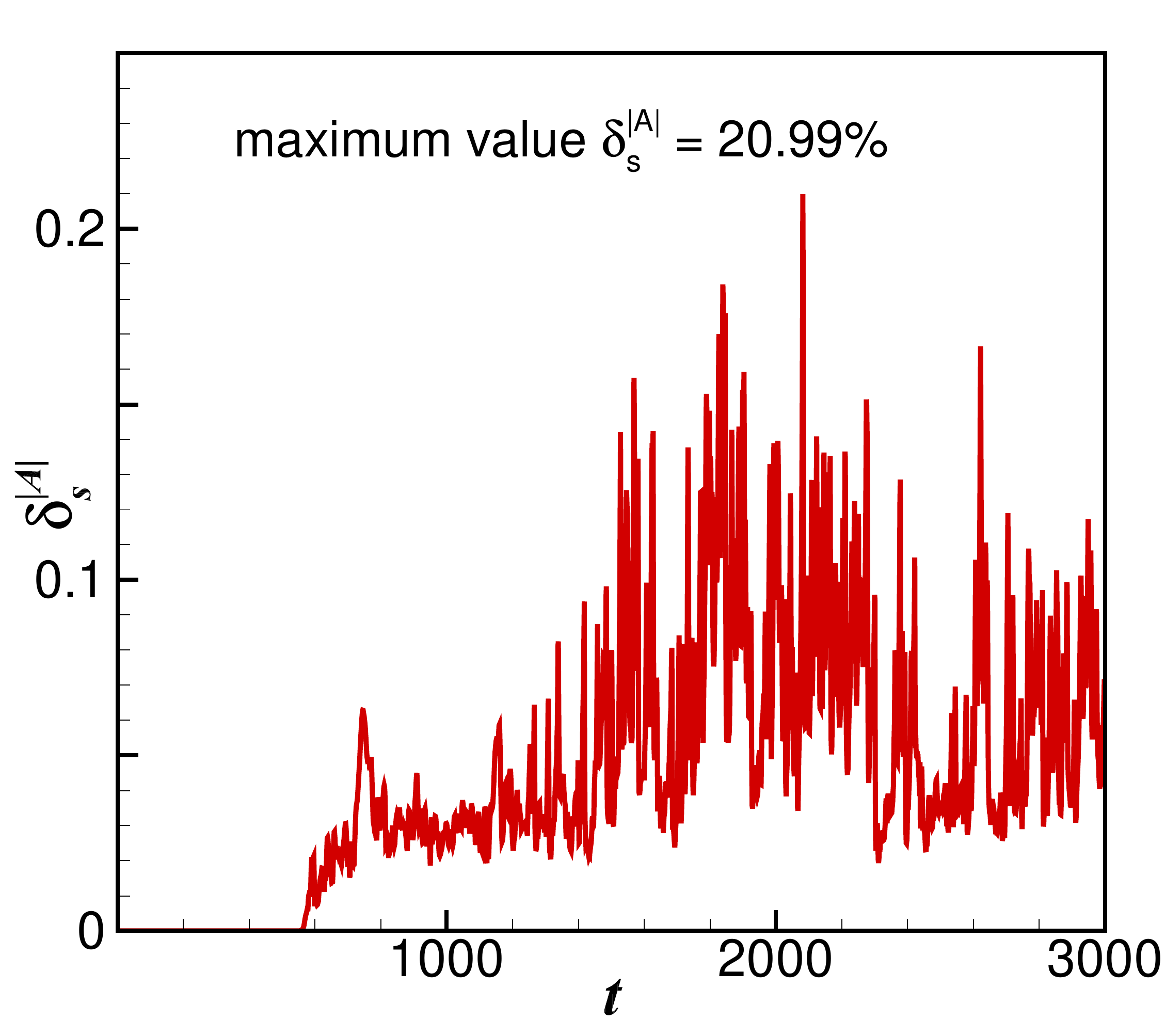}
}
 \caption{ Left:  Time histories of total spectrum-energy of the computer-generated simulations $|A(x,t)|$ obtained by CNS and RKwD in the chaotic case of $c_{1}=2$ and  $c_{3}=1$ with the initial condition (\ref{initial2gg}): CNS result (red line); and RKwD result (black dashed line).  Right: Time histories of corresponding spectrum-deviation $\delta_s^{|A|}$, with a maximum value of 20.99\%.   }
  \label{spectrum-energy-|A|}
\end{figure}

However, Fig.~\ref{spectrum-energy-|A|} indicates that $|A(x,t)|$ of the RKwD simulation has a  maximum relative error 3.35\%  of the total spatial spectrum-energy with a maximum spectrum-deviation 20.99\%, which are much smaller than those obtained above for the real and imaginary parts of $A(x,t)$.   Note that $|A(x,t)|$ is a function of the real and  imaginary  parts of $A(x,t)$.   It seems that the numerical errors of the real and imaginary parts of $A(x,t)$ might counteract each other for $|A(x,t)|$ in the case under consideration.

\begin{figure}[tbp]
\centering
	\subfigure{
		\label{level.sub.3}
		\includegraphics[width=0.5\linewidth]{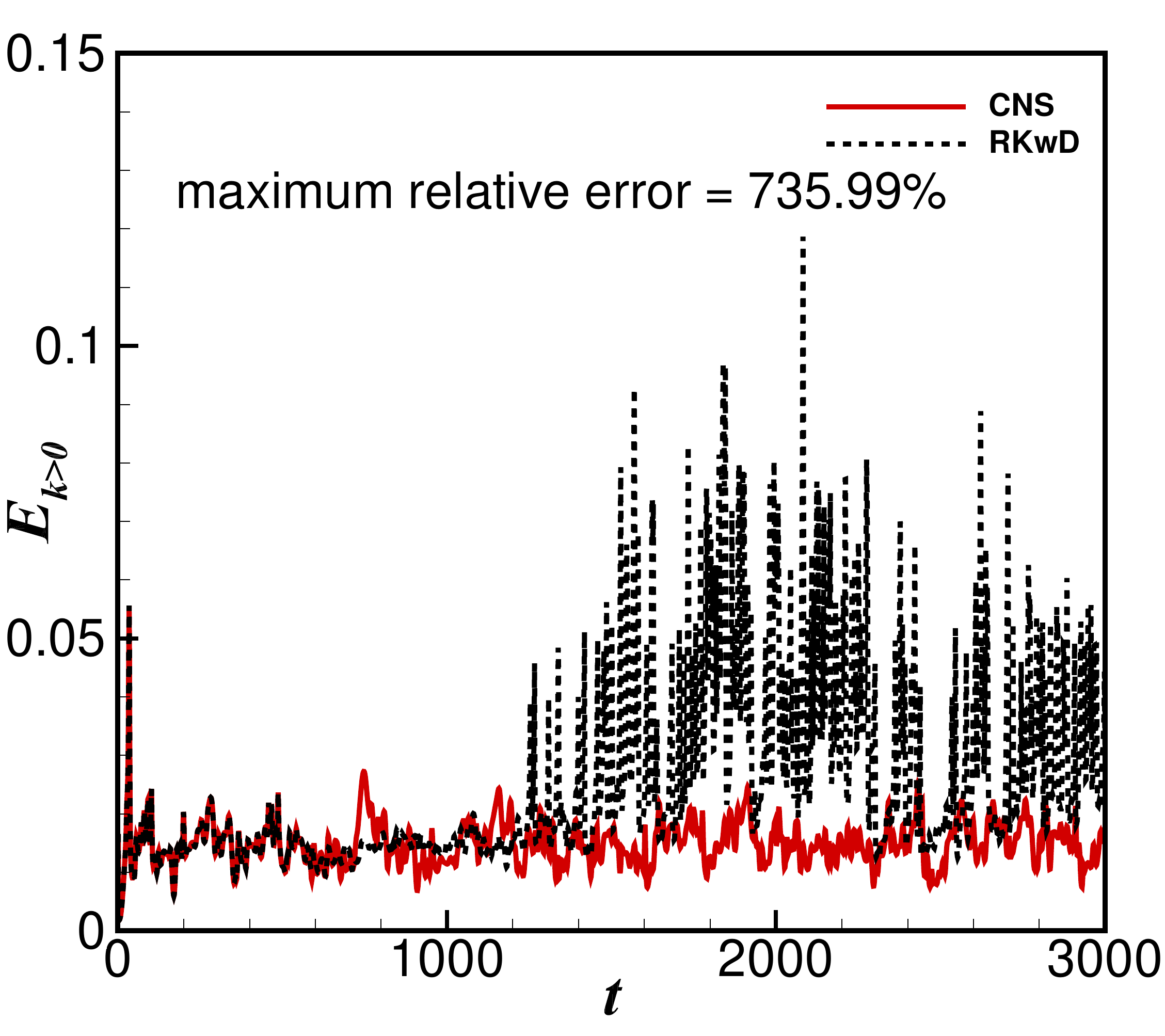}
\label{level.sub.4}
		\includegraphics[width=0.5\linewidth]{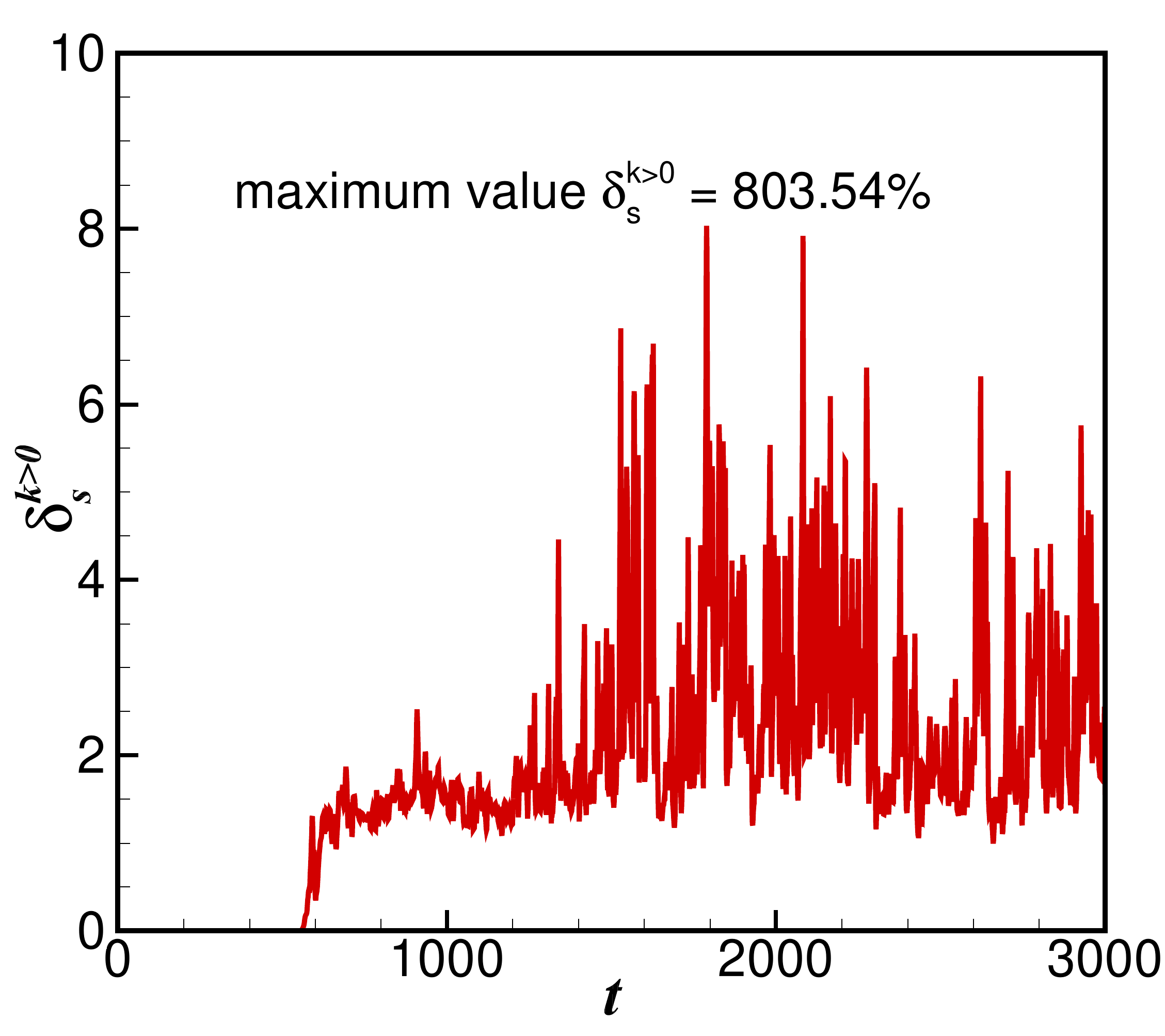}
}	
  \caption{ Left: Time histories of spatial spectrum-energy $E_{k>0}$ of  $|A(x,t)|$  except  the zero wave number component, obtained by CNS and RKwD in the chaotic case of $c_{1}=2$ and  $c_{3}=1$ with the initial condition (\ref{initial2gg}), with a maximum relative error of 735.99\%.  CNS results (red line); and RKwD results (black dashed line). Right: Time histories  of corresponding spectrum-deviation $\delta_s^{k>0}$  with a maximum value of $803.54\%$ }
  \label{spectrum-energy-|A|:k>0}
\end{figure}

Note that the constant term in the spatial Fourier spectrum has the wave number zero, i.e. $k = 0$, corresponding to the overall spatial average of the simulation, and thus does not contain any information about the {\em structure} of the solution.  In fact, it is the spatial Fourier spectrum {\em without} $k=0$ that contains information on the solution structure.  The spatial energy-spectrum without the wave number $k=0 $ of $|A(x,t)|$ obtained by the RKwD has a maximum relative error  that is 735.99\% of the total spatial spectrum-energy and a maximum relative error that is 803.54\% of the spatial spectrum-deviation, as shown in Fig.~\ref{spectrum-energy-|A|:k>0}.   This reveals that numerical noise has a {\em huge} influence on the solution structure of $|A(x,t)|$ (as may be not evident in Fig.~\ref{spectrum-energy-|A|}).

\begin{figure}[htbp]
\centering
  \subfigure[(a)]{
   \includegraphics[width=14cm,height=5.2cm]{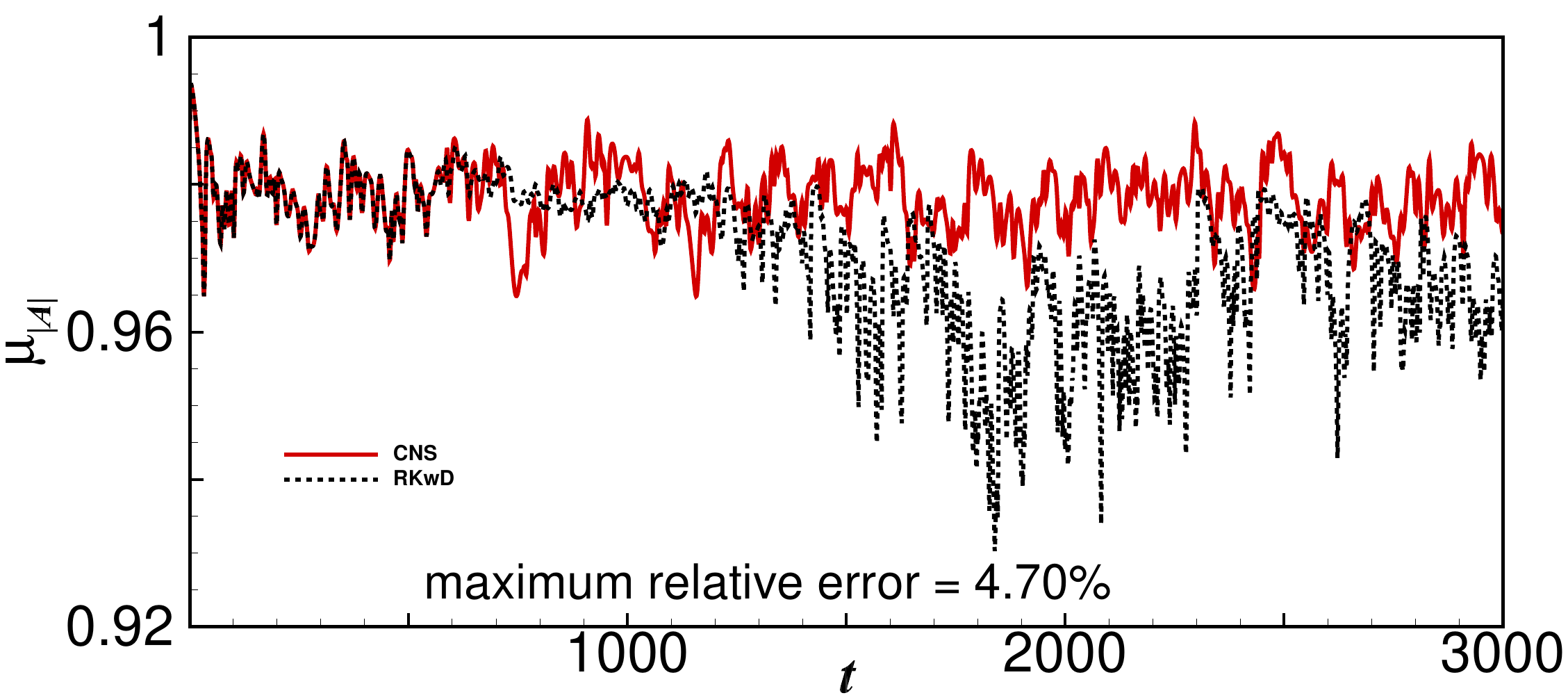}\
   }
    \subfigure[(b)]{
   \includegraphics[width=14cm,height=5.2cm]{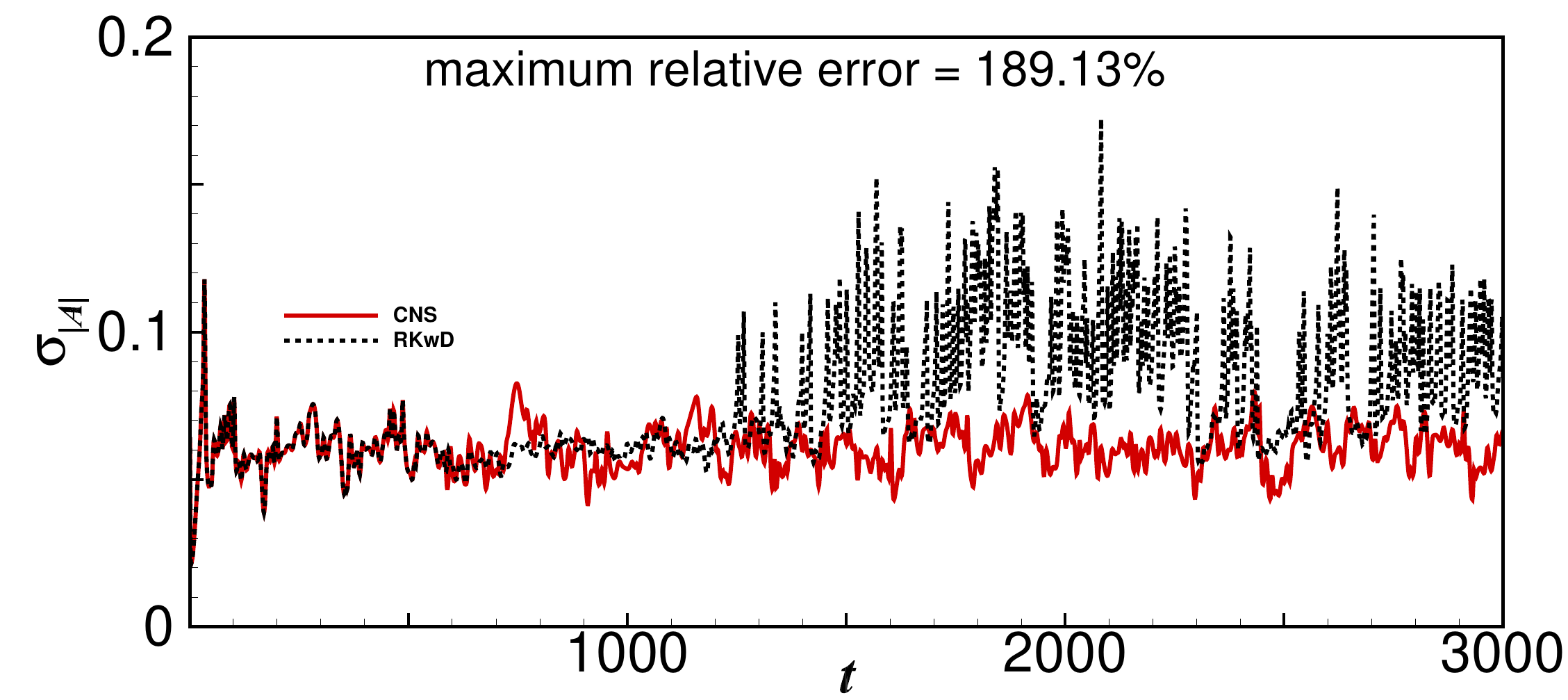}\
   }
\renewcommand{\figurename}{Fig.}
  \caption{Time histories of (a) spatial mean value $\mu_{|A|}(t)$  and (b) spatial  standard deviation $\sigma_{|A|}(t)$ of $\left|A(x,t)\right|$ obtained by CNS and RKwD in the chaotic case of $c_{1}=2$ and $c_{3}=1$ with initial condition (\ref{initial2gg}), with maximum  relative errors of $4.70\%$ for $\mu_{|A|}(t)$ and $189.13\%$ for $\sigma_{|A|}(t)$, respectively: CNS results (red line); and RKwD results (black dashed line).}
  \label{spatial-mean-deviation-|A|}
\end{figure}

\begin{figure}[htbp]
 \centering
  \subfigure[(a)]{
    \includegraphics[width=8.15cm,height=7.5cm]{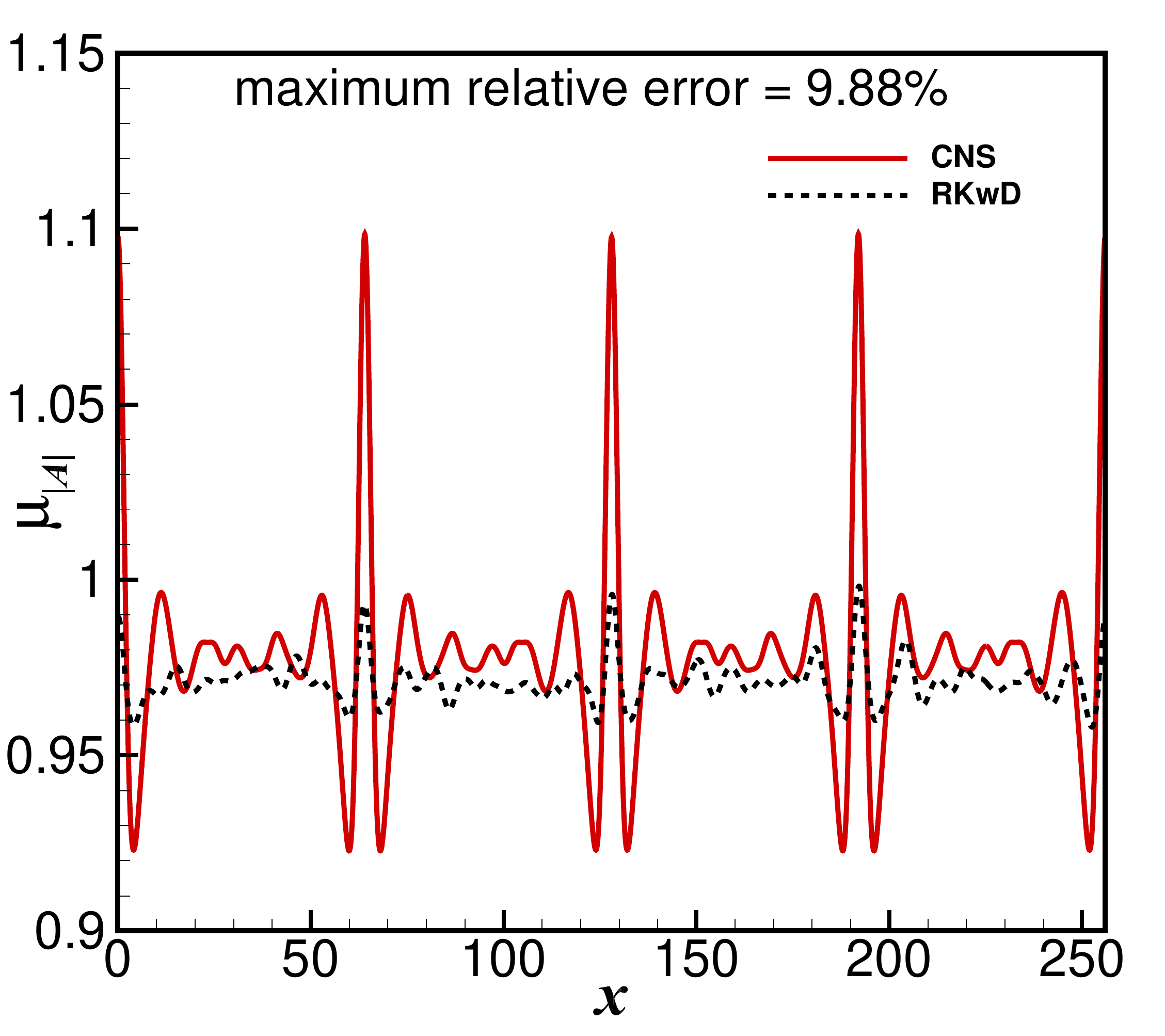}
  }
  \subfigure[(b)]{
    \includegraphics[width=8.15cm,height=7.5cm]{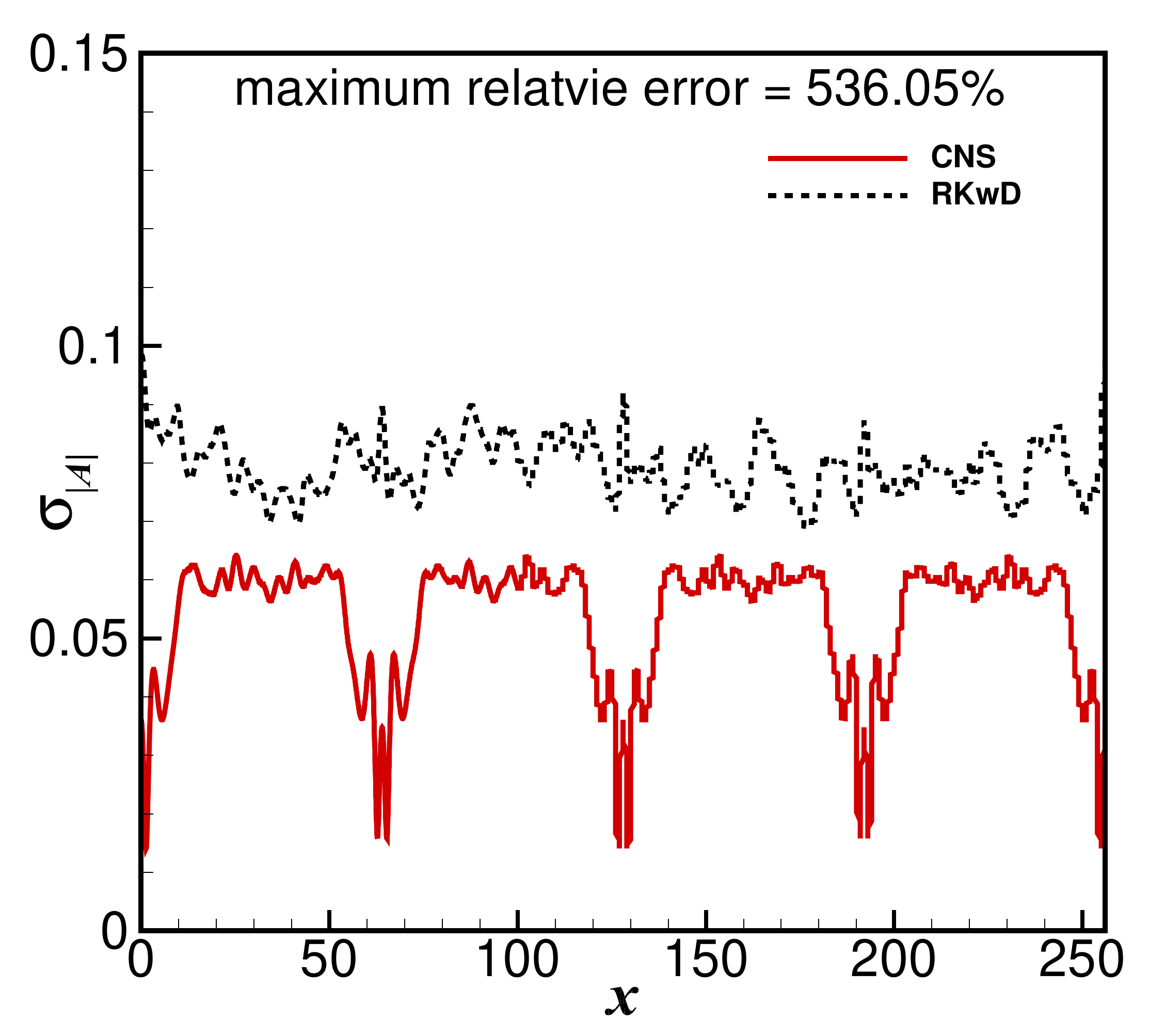}
  }
  \caption{Comparison of spatial profiles of (a) temporal  mean  $\mu_{|A|}(x)$  and (b) corresponding   spatial   standard deviation
  $\sigma_{|A|}(x)$ of $\left|A(x,t)\right|$, obtained by CNS and RKwD in the chaotic case of $c_{1}=2$ and $c_{3}=1$ with the initial condition (\ref{initial2gg}), with maximum relative errors of $9.88\%$ for $\mu_{|A|}(x)$ and $536.05\%$ for $\sigma_{|A|}(x)$, respectively. CNS results (red line); and RKwD results (black dashed line).}
 \label{temporal-mean-deviation-|A|}
\end{figure}

\begin{figure}[htbp]
\centering
  \subfigure[(a)]{
   \includegraphics[width=14cm,height=5.2cm]{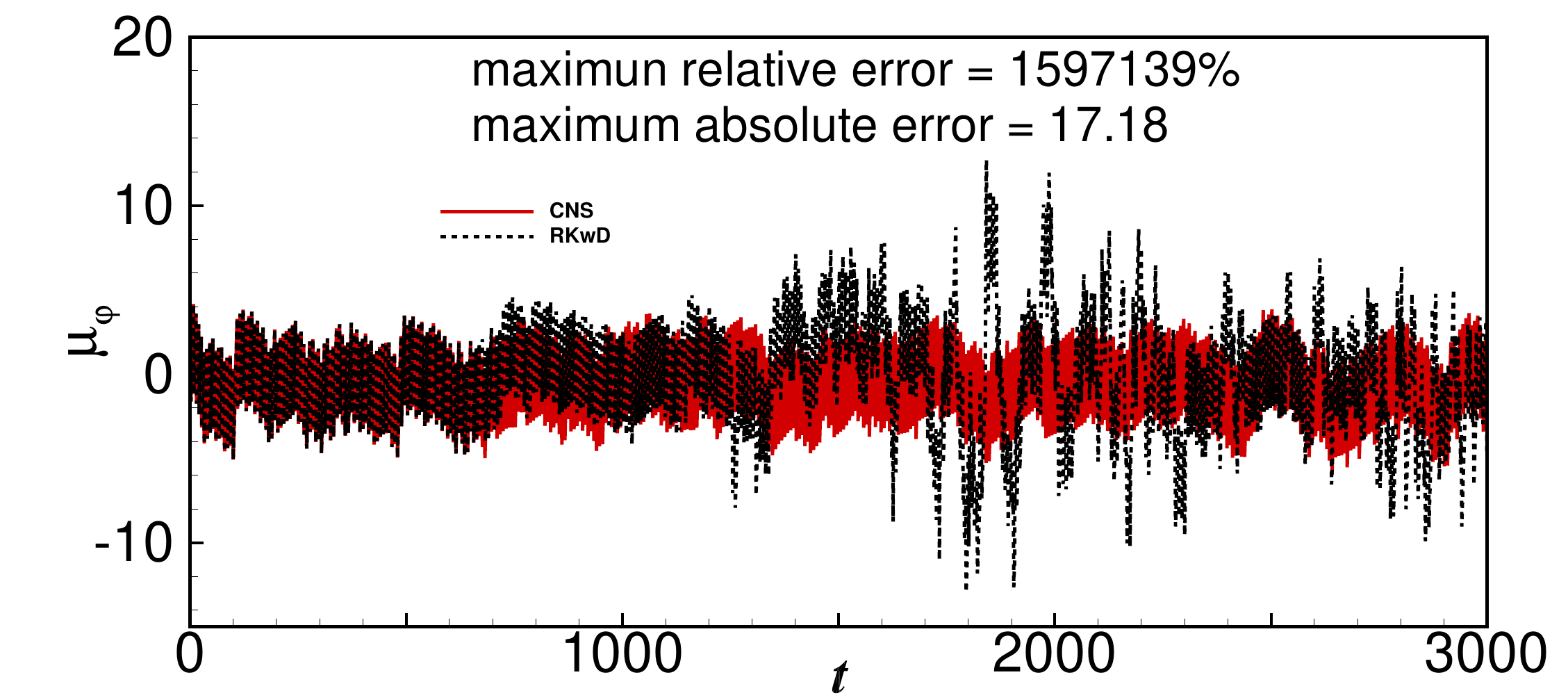}\
   }
    \subfigure[(b)]{
   \includegraphics[width=14cm, height=5.2cm]{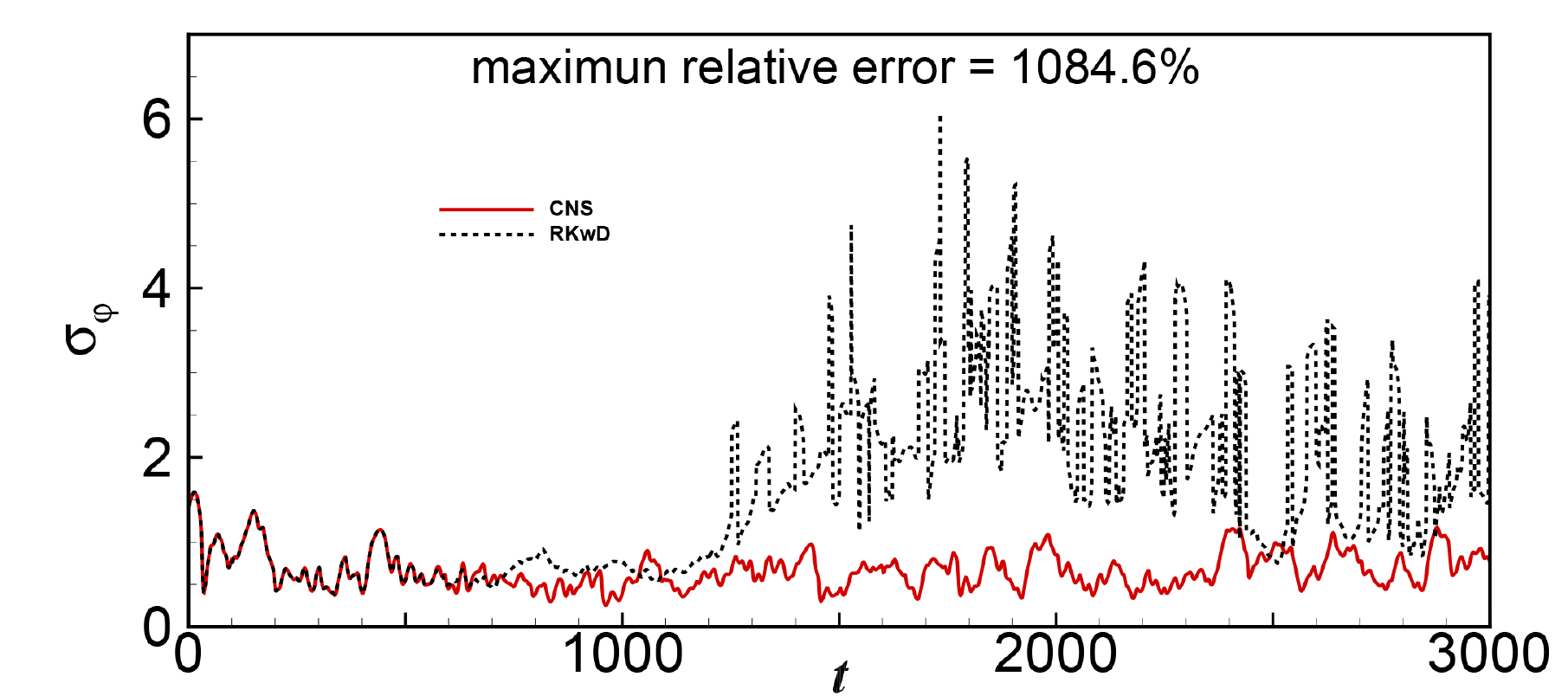}\
   }
\renewcommand{\figurename}{Fig.}
  \caption{Time histories of (a) spatial mean value $\mu_{\varphi}(t)$  and (b) spatial standard deviation $\sigma_{\varphi}(t)$ of phase of $A(x,t) $, obtained by CNS and RKwD in the chaotic case of $c_{1}=2$ and $c_{3}=1$ with the initial condition (\ref{initial2gg}): CNS results (red line); and RKwD results (black dashed line).}
  \label{spatial-mean-deviation-phi}
\end{figure}

\begin{figure}[tbhp]
\centering
  \subfigure[(a)]{
 \centering
    \includegraphics[width=8.15cm,height=7.5cm]{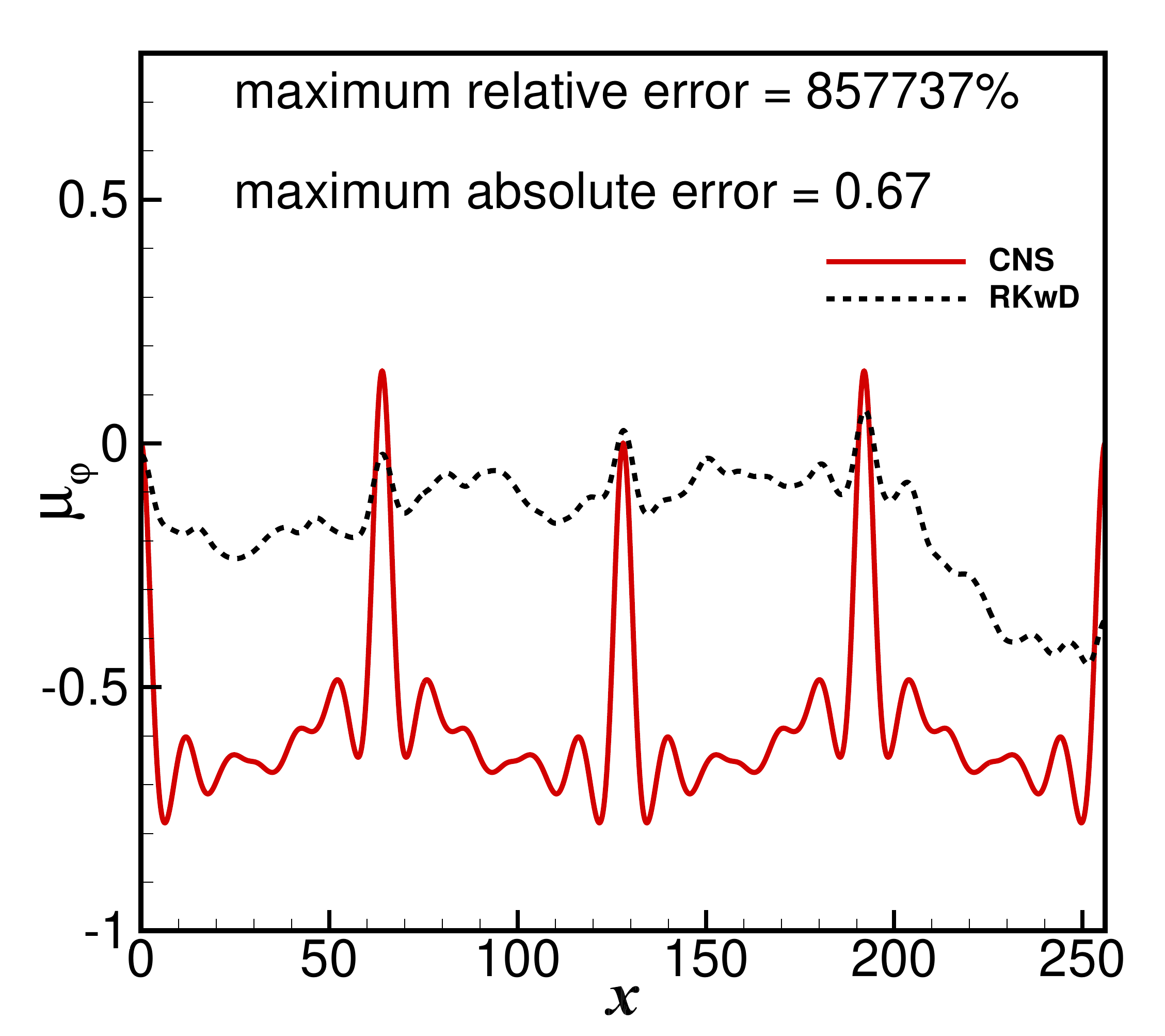}}
\centering
  \subfigure[(b)]{
 \centering
    \includegraphics[width=8.15cm,height=7.5cm]{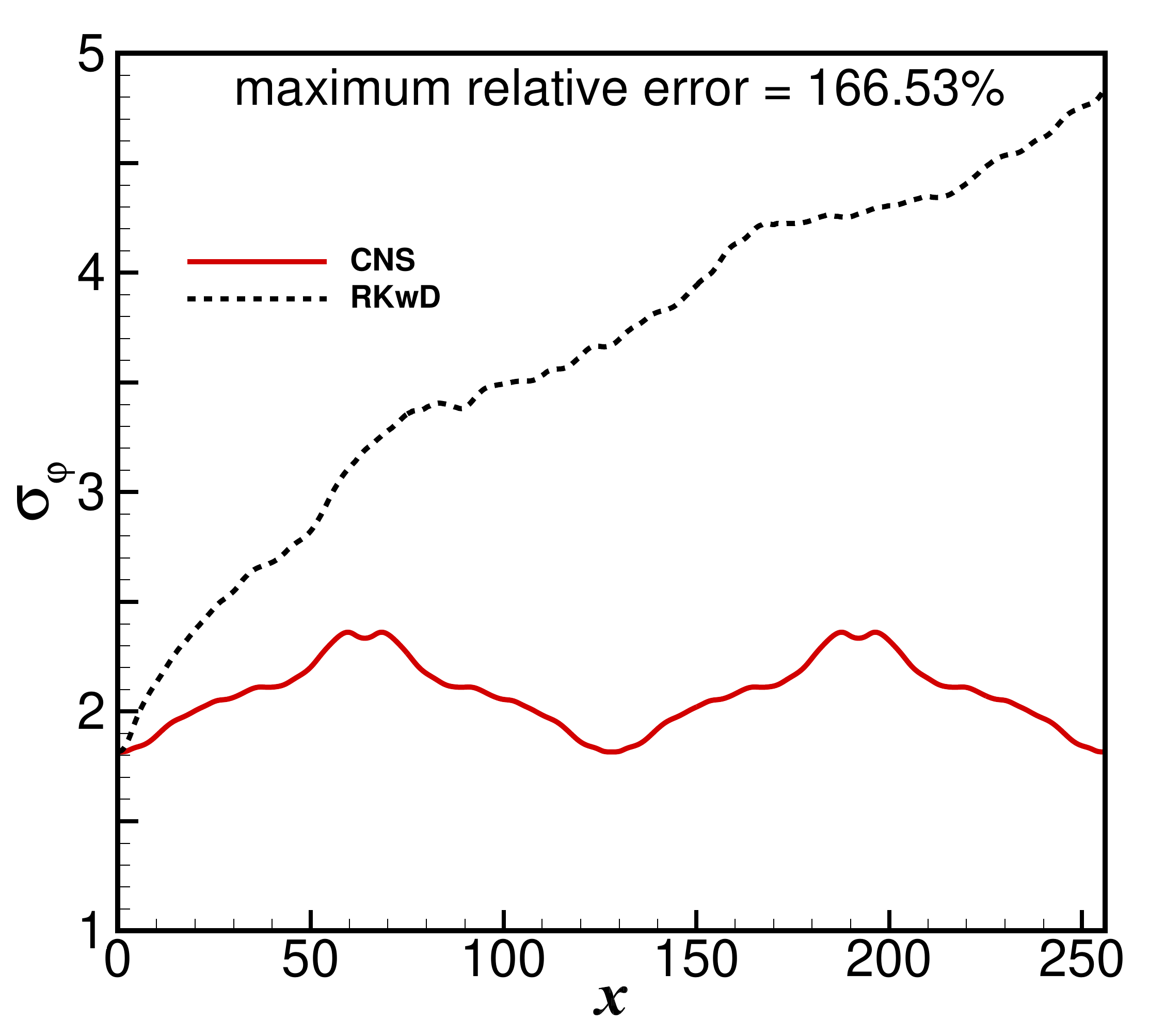}}
  \caption{Comparison of spatial profiles of (a) temporal  mean  $\mu_{\varphi}(x)$  and (b) corresponding  standard deviation $\sigma_{\varphi}(x)$ of the phase of $A(x,t)$, obtained by CNS and RKwD in the chaotic case of $c_{1}=2$ and $c_{3}=1$ with the initial condition (\ref{initial2gg}): CNS results (red line); and RKwD results (black dashed line).}
  \label{temporal-mean-deviation-phi}
\end{figure}

\begin{figure}[tbhp]
\centering
  \subfigure[]{
   \includegraphics[width=14cm,height=5.2cm]{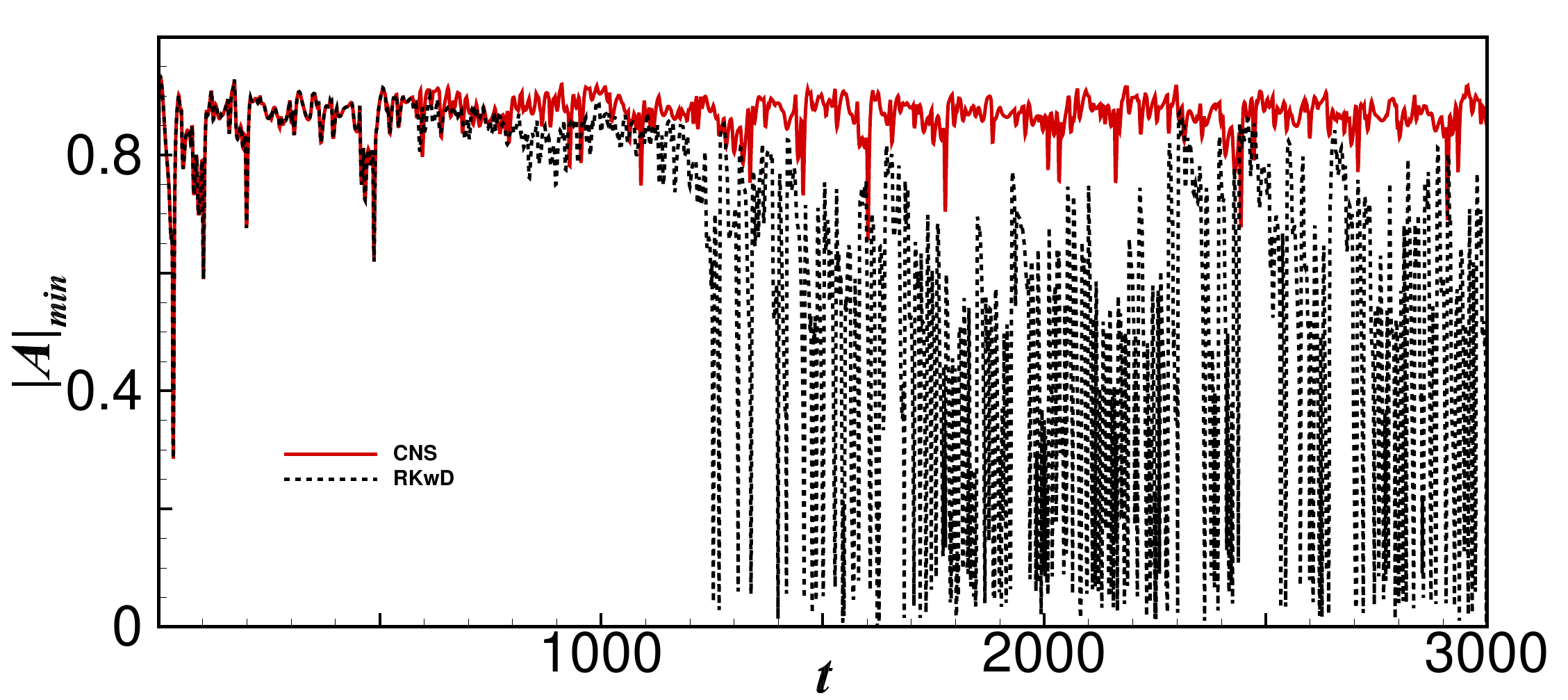}\
   }
\renewcommand{\figurename}{Fig.}
  \caption{Time histories of the spatial minimum of $|A(x,t)| $ obtained by CNS and RKwD over $t\in[0,3000]$ in the chaotic case of $c_{1}=2$ and $c_{3}=1$ with the initial condition (\ref{initial2gg}): CNS results (red line); and RKwD results (black dashed line).}
  \label{minimum-|A|}
\end{figure}

It is often assumed that numerical simulations of chaotic systems are sufficiently accurate in terms of their statistics even if their  spatio-temporal trajectories are completely different.  We now examine whether this is really true. First, let us  compare the spatial mean value
\[    \mu_{|A|}(t) = \frac{1}{L} \int_0^L |A(x,t)| dx   \]
of $\left|A(x,t)\right|$ and the corresponding  spatial standard deviation
 \[   \sigma_{|A|}(t) = \sqrt{\frac{1}{L} \int_0^L \Big[|A(x,t)|-\mu_{|A|}(t)\Big]^2 dx}  \; ,    \]
of the CNS and RKwD results (using $N=4096$)  in the chaotic case of $c_{1}=2$ and  $c_{3}=1$,  subject to the initial condition (\ref{initial2gg}).  Fig.~\ref{spatial-mean-deviation-|A|} shows that, for $t >580$, the  results given by RKwD are obviously different from those given by CNS, with a maximum relative error of 4.70\% in the spatial mean $\mu_{|A|}(t) $ and 189.13\% in the corresponding  spatial standard deviation $\sigma_{|A|}(t)$.  The differences grow considerably after $t > 1200$. This again confirms the deleterious effect of the contamination by numerical noise of computer-generated simulations of chaos {\em even} on the statistics, especially for long-duration simulations.

Secondly, let us examine the temporal mean
\[   \mu_{|A|}(x)  = \frac{1}{T} \int_0^{T} |A(x,t)| dt  \]
and the corresponding temporal standard deviation
\[   \sigma_{|A|}(x)  = \sqrt{ \frac{1}{T} \int_0^{T}\Big[  |A(x,t)| - \mu_{|A|}(x) \Big]^2 dt  }  \]
of  $|A(x,t)|$ given by the CNS and RKwD, where $T = 3000$.   The temporal statistic results given by the RKwD exhibit obvious differences from those by the CNS, with a maximum relative error  $9.88\%$ in $\mu_{|A|}(x)$ and $536.05\%$ in $\sigma_{|A|}(x)$, as can be seen in Fig.~\ref{temporal-mean-deviation-|A|}.  Notably, although the temporal mean  $\mu_{|A|}(x)$ and standard deviation $\sigma_{|A|}(x)$ obtained by the CNS have a kind of spatial symmetry, this fundamental statistical property of the true solution is lost in the RKwD simulation.

Turning thirdly to the spatial mean value
\[    \mu_{\varphi}(t) = \frac{1}{L} \int_0^L  \varphi(x,t) dx   \]
of the phase $\varphi(x,t)$ of $A(x,t)$ and its corresponding  spatial standard deviation
 \[   \sigma_{\varphi}(t) = \sqrt{\frac{1}{L} \int_0^L \Big[\varphi(x,t)-\mu_{\varphi}(t)\Big]^2 dx}  \; ,    \]
 obtained by the CNS and RKwD  results  in the chaotic case of $c_{1}=2$ and  $c_{3}=1$ with the initial condition (\ref{initial2gg}).
For $t > 580$, the results given by RKwD differ considerably from those given by CNS, with a maximum relative error of 1597139\%  and the maximum absolute error of 17.18 for the spatial mean $\mu_{\varphi}(t) $, maximum relative error of 1084.6\% for the corresponding  spatial standard deviation $\sigma_{\varphi}(t)$ (Fig.~\ref{spatial-mean-deviation-phi}).  The results  continue to diverge for $t > 1200$. This again confirms the great impact of numerical noise on computer-generated simulations of chaos {\em even} in terms of the statistics, particularly for long-duration simulations.

Next, we examine the temporal mean
\[   \mu_{\varphi}(x)  = \frac{1}{T} \int_0^{T} \varphi(x,t) dt  \]
of the phase $\varphi(x,t)$ of $A(x,t)$ and its corresponding temporal standard deviation
\[   \sigma_{\varphi}(x)  = \sqrt{ \frac{1}{T} \int_0^{T}\Big[  \varphi(x,t)| - \mu_{\varphi}(x) \Big]^2 dt  },   \]
where $T = 3000$. In Fig.~\ref{temporal-mean-deviation-phi} these temporal statistics of the RKwD numerical predictions display obvious differences from those by CNS, with maximum relative error of $857737\%$  and maximum absolute error of 0.67 for $\mu_{\varphi}(x)$ and maximum relative error $166.53\%$ for $\sigma_{\varphi}(x)$. The temporal mean  $\mu_{\varphi}(x)$ and standard deviation $\sigma_{\varphi}(x)$ obtained by CNS again demonstrate a kind of spatial symmetry correctly, unlike the corresponding results from the RKwD simulations.

\begin{figure}[tbhp]
	\centering
	\vspace{-0.35cm}
	\subfigtopskip=2pt
	\subfigbottomskip=2pt
	\subfigcapskip=-5pt

	\subfigure[(a)]{
		\label{level.sub.1}
		\includegraphics[width=0.5\linewidth]{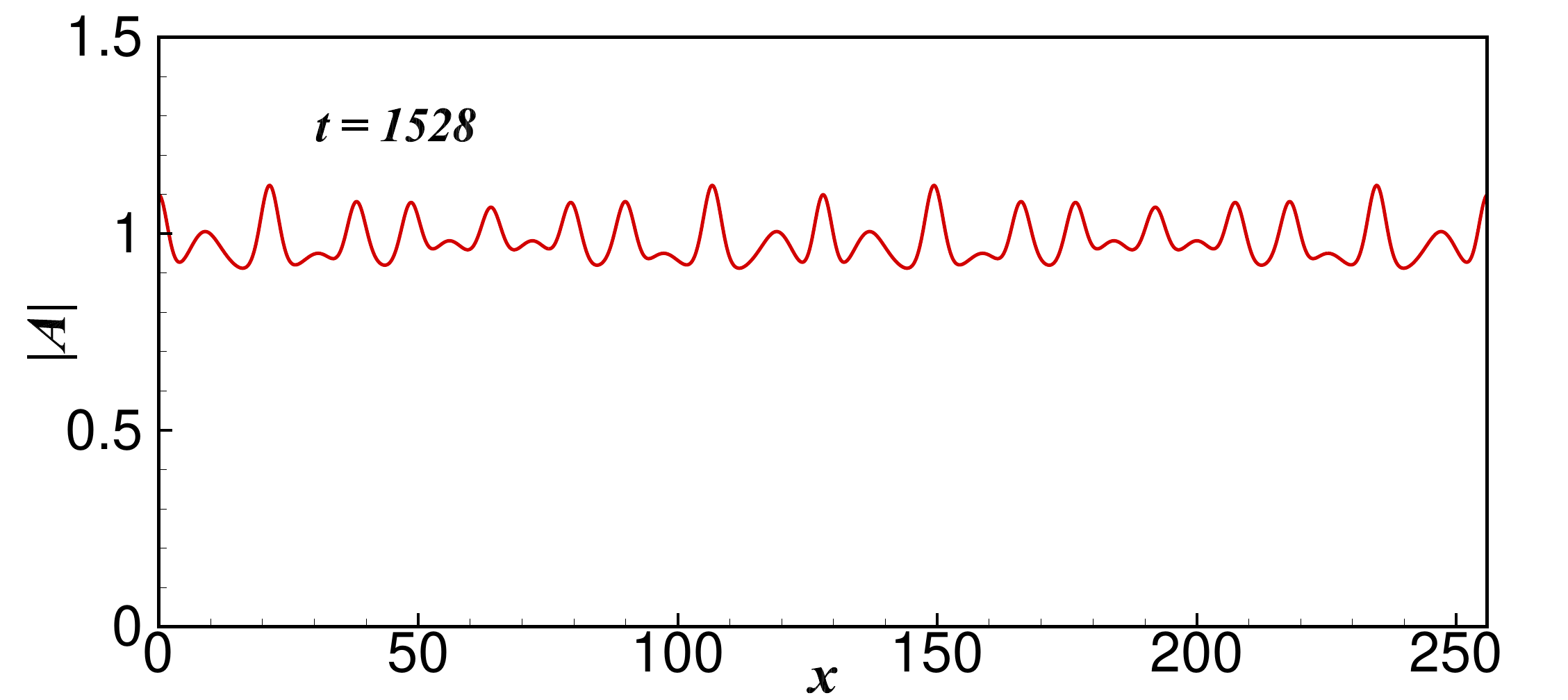}
\label{level.sub.2}
		\includegraphics[width=0.5\linewidth]{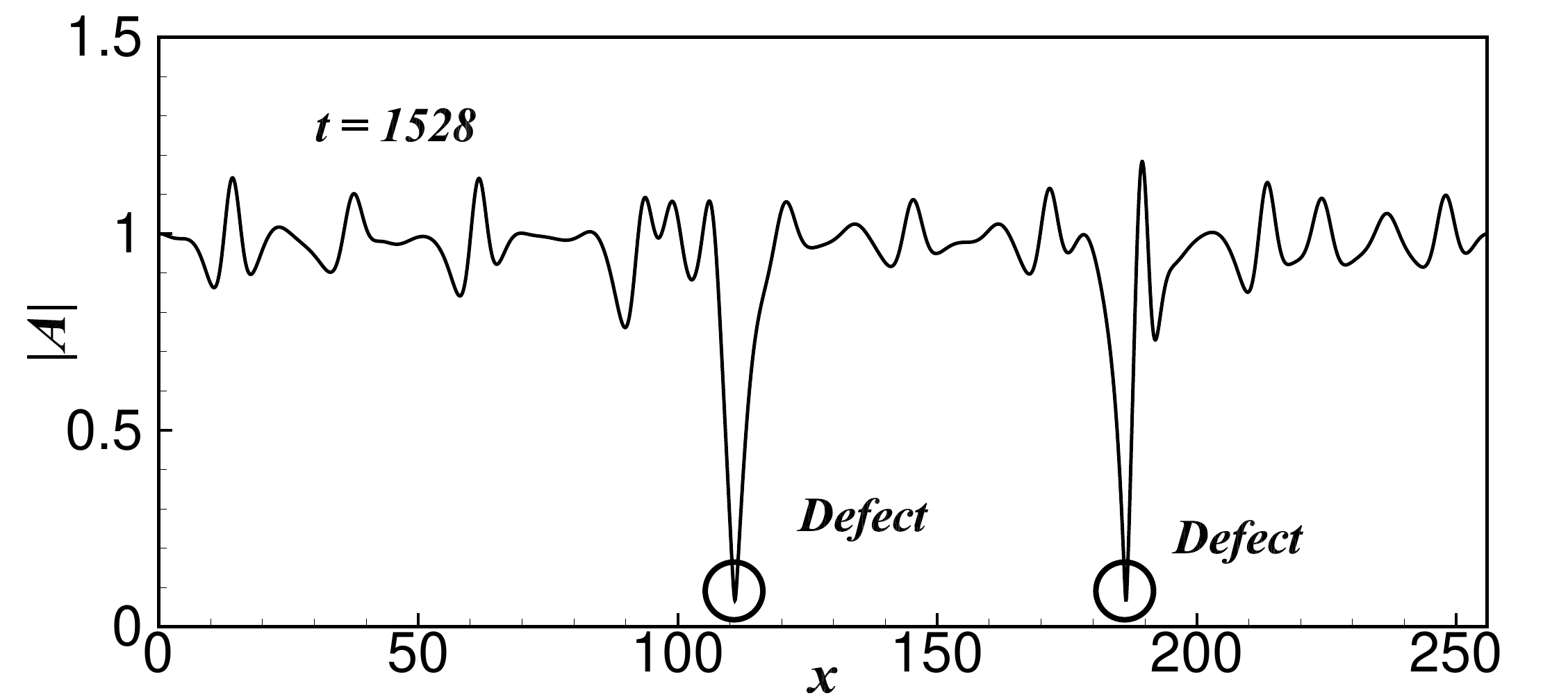}
}
	\subfigure[(b)]{
		\label{level.sub.3}
		\includegraphics[width=0.5\linewidth]{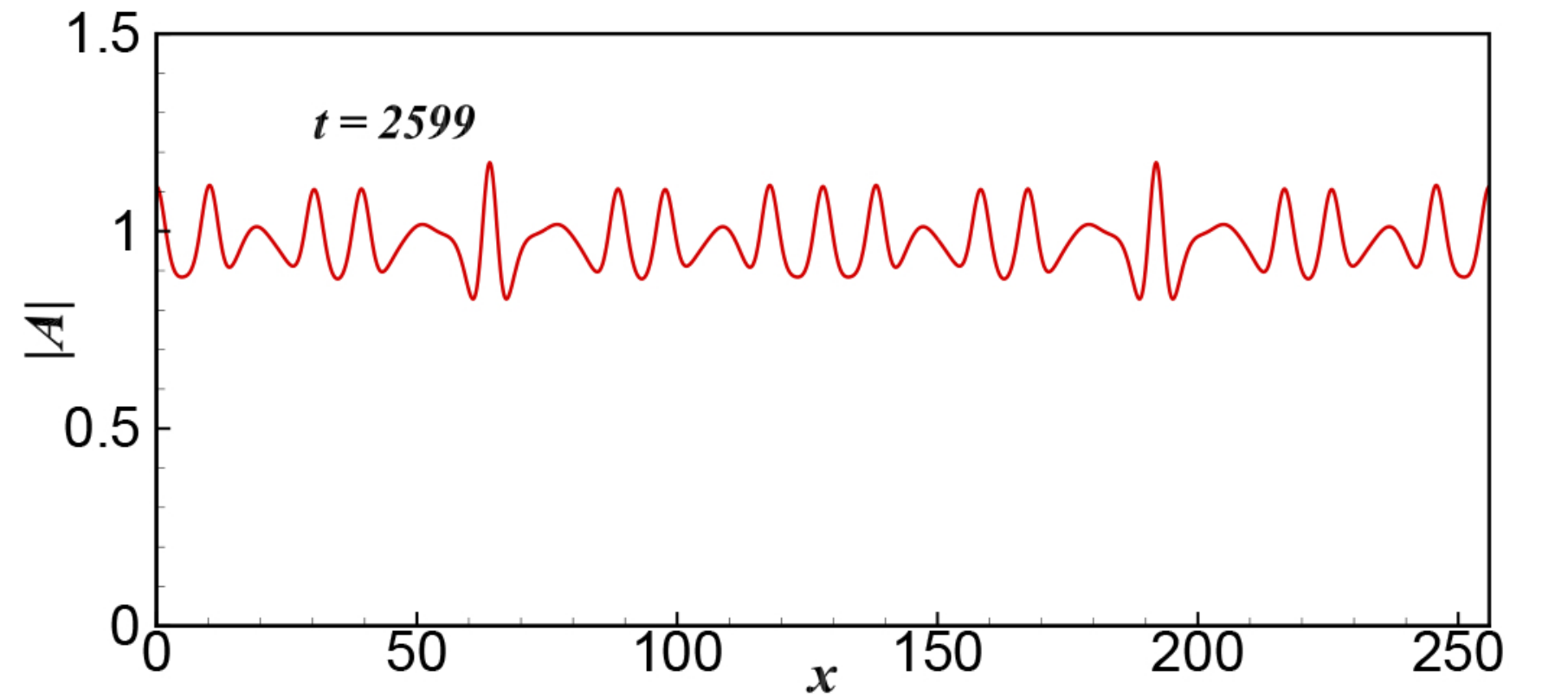}
\label{level.sub.4}
		\includegraphics[width=0.5\linewidth]{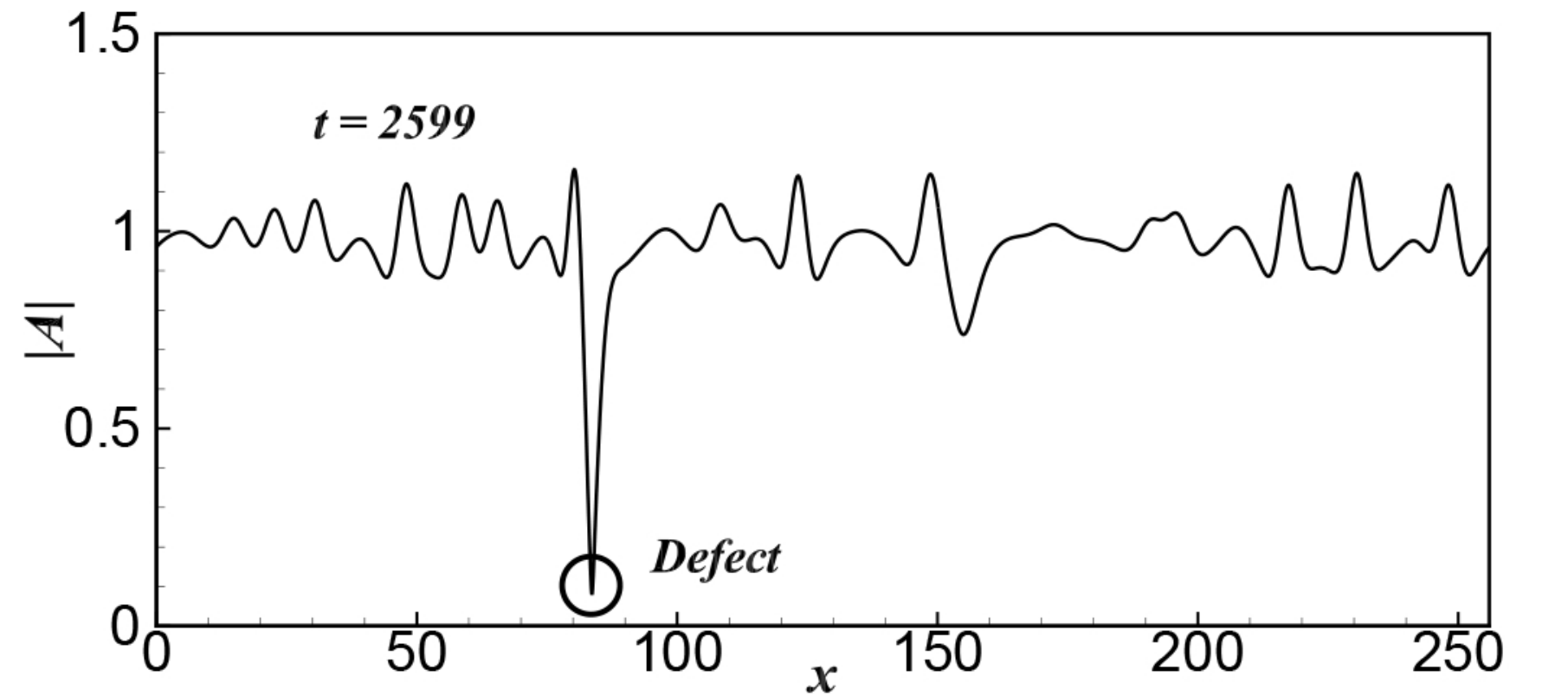}
}
	\subfigure[]{
 \qquad
 \quad

		CNS
 \qquad
  \qquad
  \qquad
  \qquad
  \qquad
  \qquad
  \qquad
  \qquad
  \qquad
RKwD
}
\caption{Spatial profiles of $|A(x,t)|$ obtained by CNS and RKwD  in the chaotic case of $c_{1}= 2$ and $c_{3}= 1$ with the initial condition (\ref{initial2gg}): (a) $t$ = 1528; and (b) $t$ = 2599. Left: CNS results (red line); Right: RKwD results (black line).}
	\label{|A|-time}
\end{figure}
\begin{figure}[tbhp]
	\centering
	\vspace{-0.35cm}
	\subfigtopskip=2pt
	\subfigbottomskip=2pt
	\subfigcapskip=-5pt
	\subfigure[(a)]{
		\label{level.sub.1}
		\includegraphics[width=0.5\linewidth]{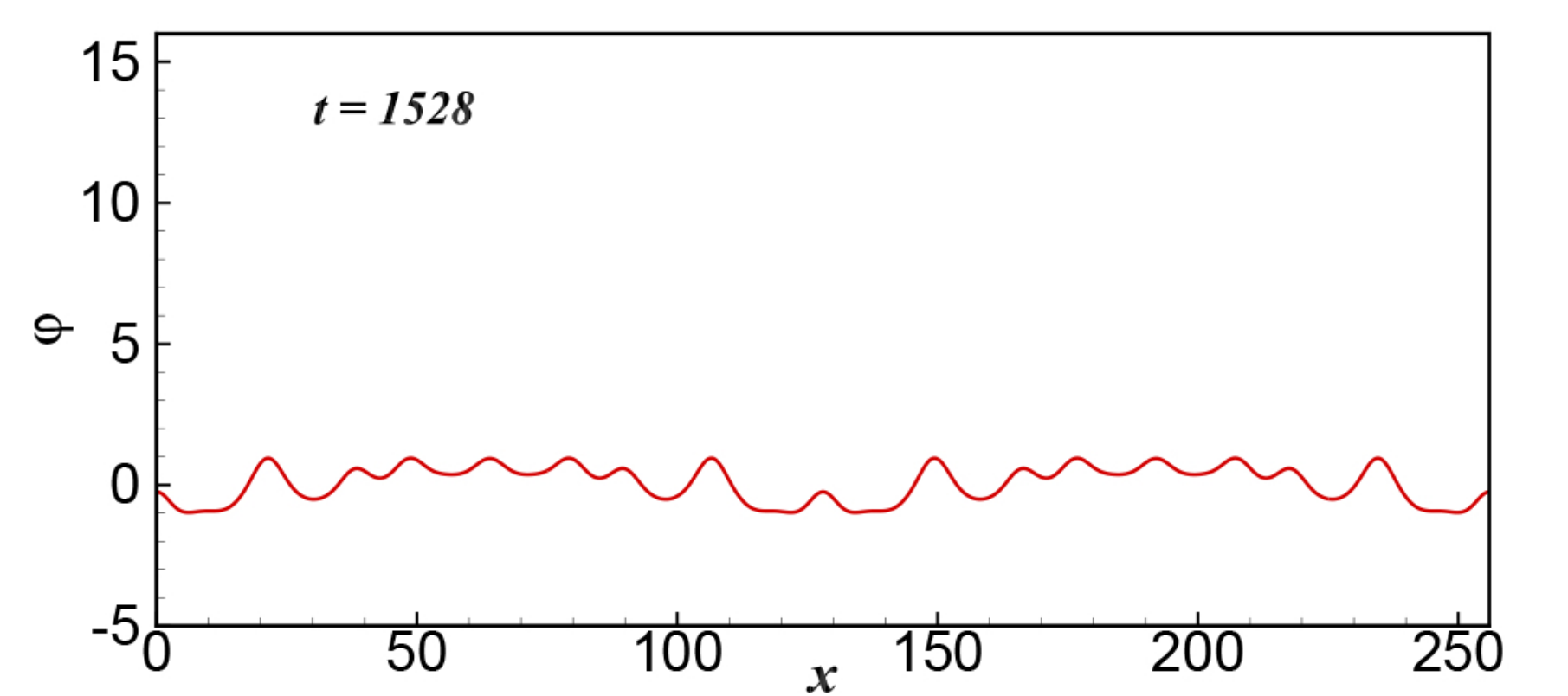}
\label{level.sub.2}
		\includegraphics[width=0.5\linewidth]{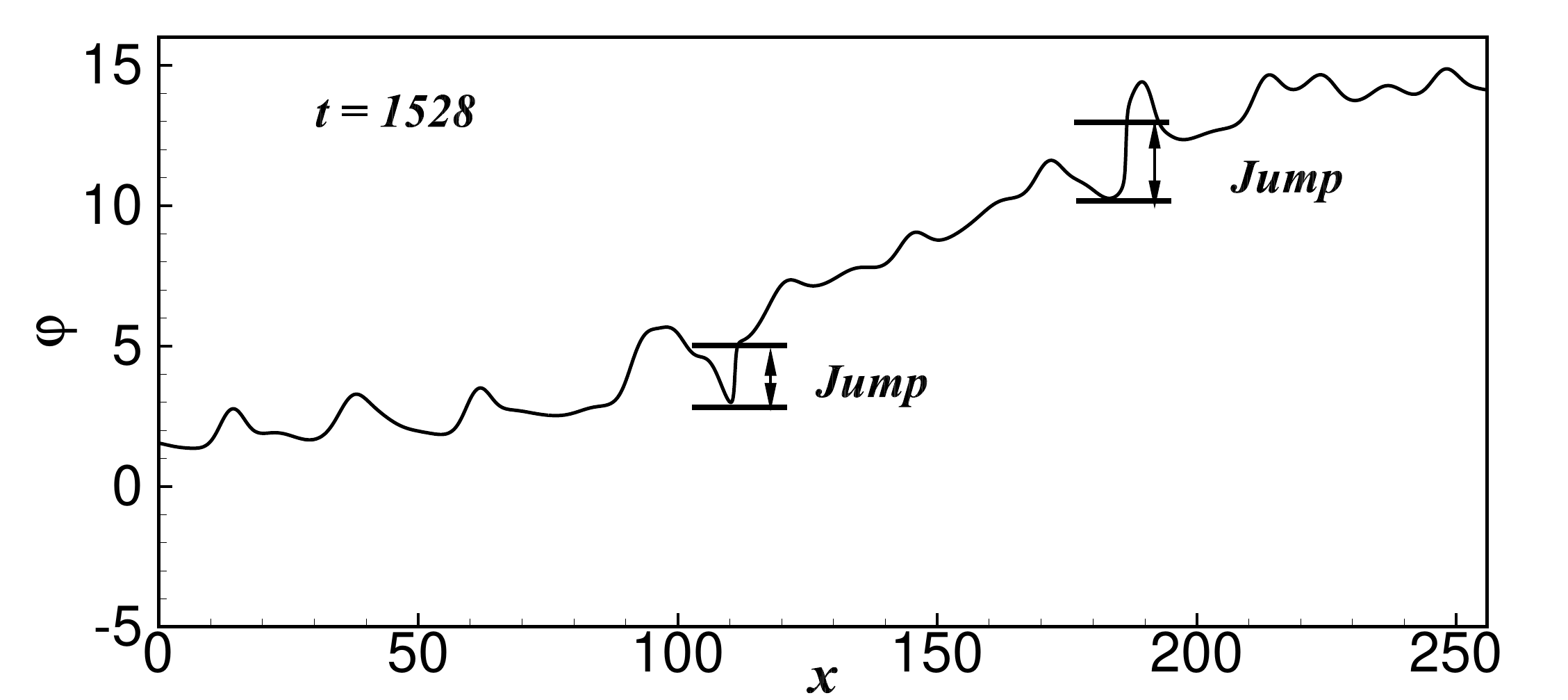}
}
	\subfigure[(b)]{
		\label{level.sub.3}
		\includegraphics[width=0.5\linewidth]{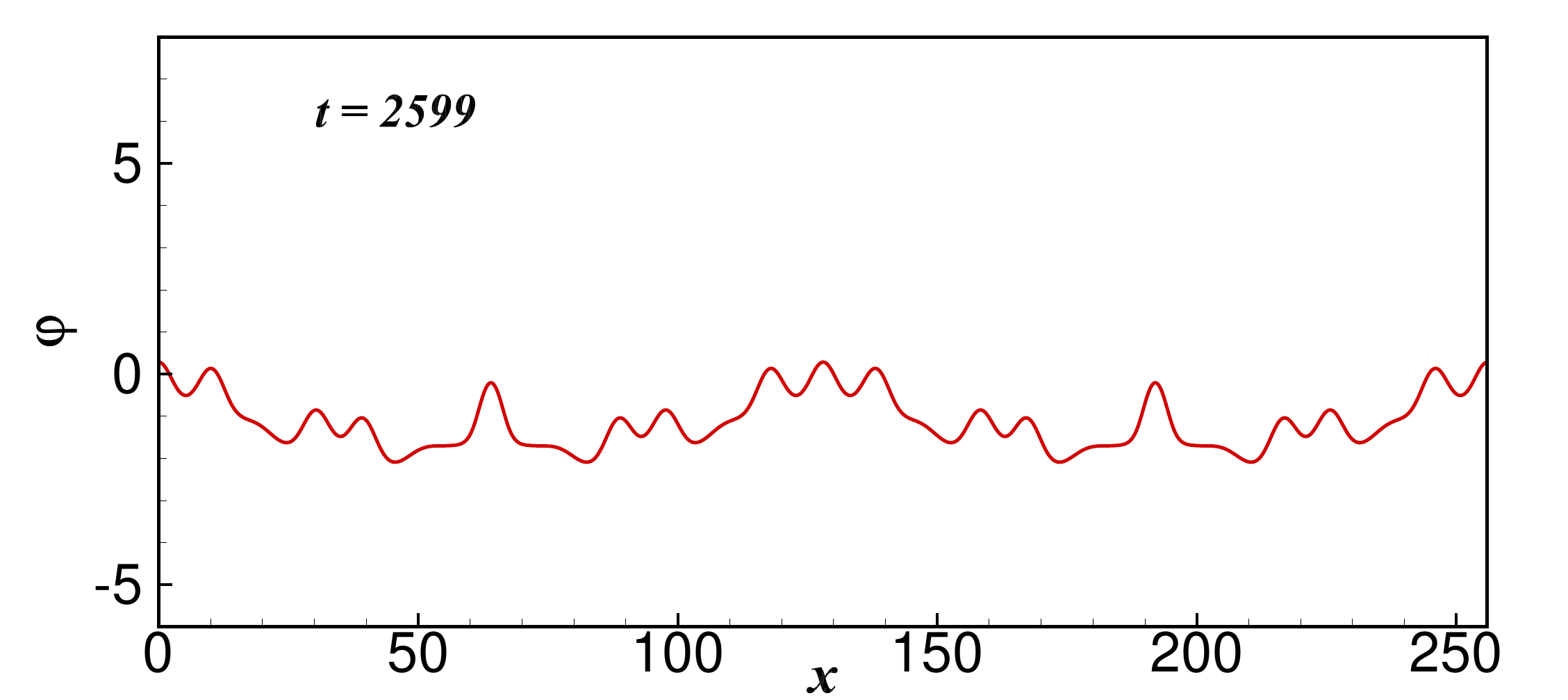}
\label{level.sub.4}
		\includegraphics[width=0.5\linewidth]{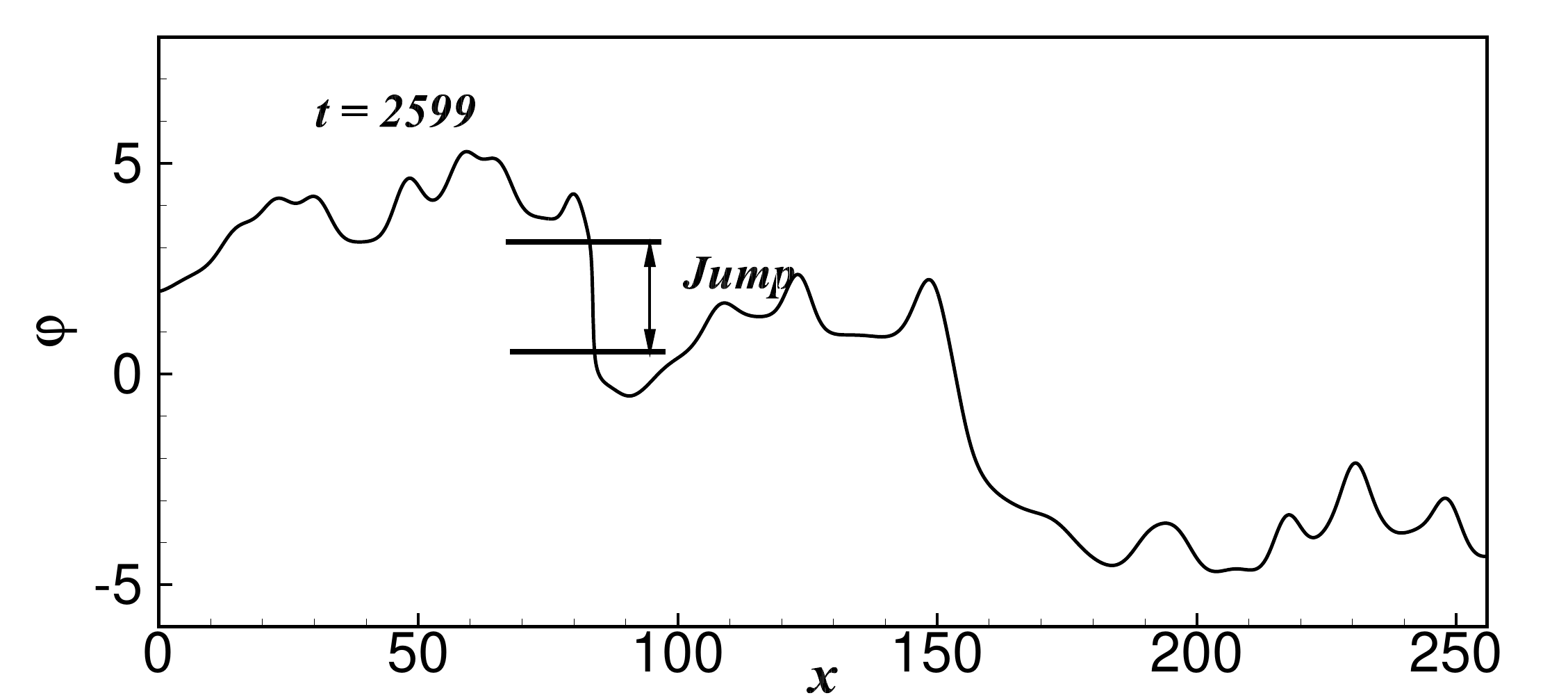}
}
	\subfigure[]{
 \qquad
 \quad

		CNS
 \qquad
  \qquad
  \qquad
  \qquad
  \qquad
  \qquad
  \qquad
  \qquad
  \qquad
RKwD
}
\caption{Spatial profiles of the phase of $A(x,t)$ obtained by CNS and RKwD  in the chaotic case of $c_{1}= 2$ and $c_{3}= 1$ with the initial condition (\ref{initial2gg}): (a) $t$ = 1528; and (b) $t$ = 2599. Left: CNS results (red line); Right: RKwD results (black line).}
	\label{phase-time}
\end{figure}

Fig.~\ref{spatial-mean-deviation-phi} shows that the spatial mean $\mu_\varphi(t)$ and the spatial standard deviation  $\sigma_\varphi(t)$ of the CNS result remain similar throughout the \emph {entire} interval $t\in[0,3000]$.  However, the corresponding RKwD results are highly non-stationary, with considerable changes between the values of spatial mean $\mu_\varphi(t)$ and spatial standard deviation  $\sigma_\varphi(t)$ before and after $t\approx1200$.  Similar behaviour is exhibited by the spatial mean $\mu_{|A|}(t)$ and the spatial standard deviation $\sigma_{|A|}(t)$ shown in Fig.~\ref{spatial-mean-deviation-|A|}, the spatial spectrum energy of $|A|$ except the constant term (corresponding to the wave number $k=0$) in Fig.~\ref{spectrum-energy-|A|:k>0}, the total spectrum energy of $|A|$ in Fig.~\ref{spectrum-energy-|A|},  and the total spectrum energy of the real and imaginary parts of $A(x,t)$ in Fig.~\ref{Energy-spectrum-Re-Im}.

\begin{figure}[tbhp]
\centering
  \subfigure[]{
    \includegraphics[width=8cm]{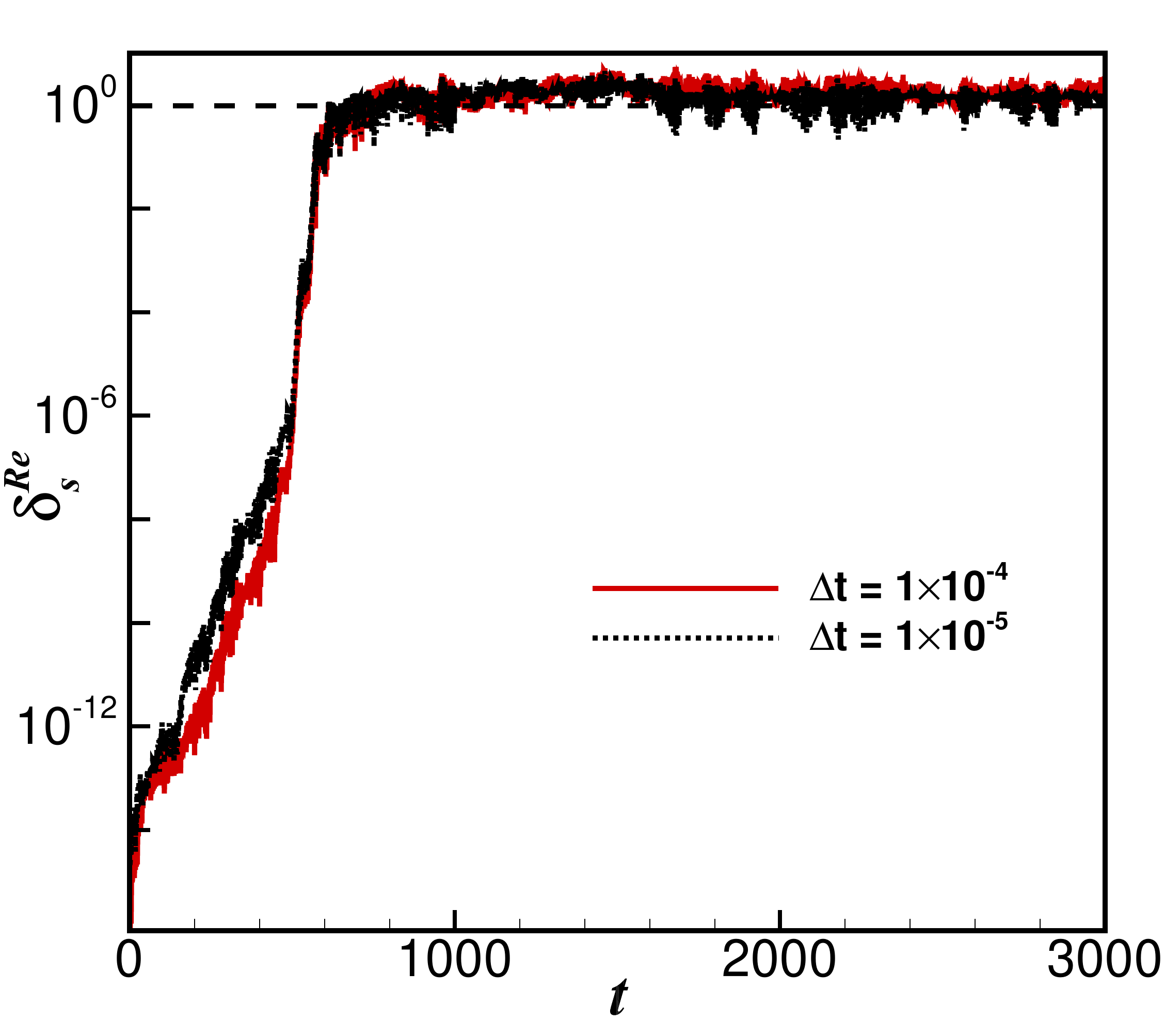}
  }
  \caption{Time histories of spectrum-deviation of the real part of the RKwD simulations in the chaotic case of $c_{1}=2$ and  $c_{3}=1$ with the initial condition (\ref{initial2gg}) using different values of time-step, with the CNS result as the benchmark:  $\Delta t=1\times10^{-4}$ (red solid line); and $\Delta t=1\times10^{-5}$ (black dashed line).}
  \label{spectrum-deviations-multiple-timesteps}
\end{figure}

The one-dimensional CGLE exhibits two distinct chaotic phases, namely  ``phase chaos''  when $A$ is bounded away from zero,  and ``defect chaos''  when the phase of $A$ exhibits singularities where $A \approx 0$, respectively \cite{sakaguchi1990breakdown, shraiman1992spatiotemporal, brusch2000modulated}.  Shraiman {\em et al} \cite{shraiman1992spatiotemporal}  pointed out  that   the crossover between phase and defect chaos is invertible when $c_{1} > 1.9$ (we consider $c_1 = 2$ in this section).   Let
\[  |A|_{min}(t) = \min \left\{\hspace{0.25cm} |A(x,t)|  \hspace{0.25cm} \Big|  \hspace{0.25cm}  x\in[0,L] \hspace{0.25cm}\right\}  \]
denote the spatial minimum of $|A(x,t)|$ at a given time $t$. Fig.~\ref{minimum-|A|} compares time histories of $|A|_{min}(t) $ given by CNS and RKwD. It can be seen that the CNS result of $|A|$ is invariably bounded away from zero,  implying that the solution is in the so-called {\em phase chaos} regime throughout the {\em entire} duration of the simulation, i.e.  $t\in[0,3000]$.   However, the  $|A|_{min}(t)$ given by RKwD often exhibits a sudden drop to a value very close to zero after $t  > 1200$ (Fig.~\ref{minimum-|A|}), corresponding to the so-called {\em defect chaos} displayed in Fig.~\ref{|A|-time} for $|A|$   and  Fig.~\ref{phase-time} for the phase $\varphi$ with occasional jumps in phase indicating the presence of singularities.  The RKwD simulation therefore \emph{incorrectly} predicts a crossover between phase and defect chaos in the interval $[0,3000]$, which does not occur in the \emph {reliable} CNS result. This illustrates the risk posed by numerical noise in computer-generated simulation of chaotic systems.

Note that both of CNS and RKwD use the same number of modes, $N=4096$, in the spatial Fourier expansion in space and thus have the same spatial truncation error. Given that CNS utilizes the temporal high-order Taylor series in multiple precision, the CNS algorithm should have a much smaller truncation error in time dimension  than the 4th-order Runge-Kutta method in double precision (RKwD).  It is found that using a smaller time-step (such as   $\Delta t=1\times10^{-5}$) can not improve the spectrum-deviations of the real part of $A(x,t)$ given by the RKwD, as evident in Fig.~\ref{spectrum-deviations-multiple-timesteps}.   This means that the temporal truncation error of RKwD is not the dominant source of its numerical noise.   Instead, the numerical noise of the RKwD result primarily arises from the round-off error  due to the use of double precision.  So, all of the above-mentioned comparisons illustrate the risk of using double precision in computer-generated simulation of spatio-temporal chaotic systems, particularly for a long-duration simulation.

\section{Concluding remarks and discussions \label{Conclusion_v2}}

Due to the famous butterfly-effect,  numerical noise caused by the truncation and round-off errors increases exponentially for chaotic system so that it is hard to obtain a reliable computer-generated simulation in a long-duration time.   The shadowing method \cite{ANOSOV1967, BOWEN1975, hammel1987numerical, hammel1988numerical, grebogi1990shadowing, sauer1991rigorous} works for a uniformly hyperbolic system,  but hardly any chaotic systems are uniformly hyperbolic \cite{dawson1994obstructions, do2004statistics}.   Particularly,  there is no practical method  to  efficiently gain reliable  computer-generated simulations in a long-duration time for spatio-temporal chaotic systems governed by nonlinear PDEs, to the best of our knowledge.

Unlike the shadowing method,  the strategy of the Clean Numerical Simulation (CNS)  is  to  greatly  reduce numerical noises (caused by both of temporal/spatial truncation error and round-off error) to such a tiny level that is required for a reliable simulation in a given interval of time $t\in[0,T_c]$.  Based on such a hypothesis that numerical noise of chaotic system increases exponentially,  the so-called ``critical predictable time'' $T_c$ is determined by comparing a simulation with an additional new one given by the same initial condition and physical parameters but smaller numerical noise.   In this way, we gain a ``clean'' numerical simulation in $[0,T_c]$, whose noise is below a given criterion of noise.   This kind of  ``clean'' numerical simulation should be close to the  true solution and thus can be used as a benchmark solution for many purposes, for example,  verifying numerical results given by a developing code,  studying the propagation of physical micro-level uncertainty of a chaotic dynamic system \cite{Liao2013On, liao2015inherent, Lin2017On},  finding new periodic orbits of the three-body problem \cite{Li2017More, li20171223, Li2019-NewAstronomy}, investigating the influence of numerical noise on chaotic simulations given by the traditional algorithms in double precision (as illustrated in this paper), and so on.

In this paper, a new CNS algorithm in physical space is proposed for spatio-temporal chaos, which is computationally much more efficient than its predecessor  in spectral space \cite{Lin2017On}.   To verify its computational performance, the new CNS algorithm  was used to solve the one-dimensional  complex Ginzburg-Landau equation (CGLE), a well known example of spatio-temporal chaos.   In the case of $c_1=2$, $c_3 = 1$ with the initial condition (\ref{initial2gg}),  the CNS result (using the spatial Fourier mode number $N = 4096$ and the number $N_s=105$ of significant digits in multiple precision) remained convergent over considerably long time interval $t\in[0,3000]$ throughout the whole spatial domain $x\in[0,L]$, where $L = 256$.   This CNS result was  considered reliable, and was therefore used as a benchmark.

By comparing this benchmark solution with that obtained using the 4th-order Runge-Kutta integration in double precision (RKwD), it was found that the two simulations only exhibited agreement over a small interval of time $t\in[0,580]$. The CNS solution retained  a kind of spatial symmetry and the phase chaos over the entire time interval $t\in[0,3000]$, unlike the RKwD simulation which lost the spatial symmetry when $t > 580$ and degenerated into a mixture of phase and defect chaos after $t>1200$.   Moreover, the odd wave number components in the spatial spectrum of the CNS benchmark solution invariably remained zero, i.e. did not attract energy, unlike the RKwD simulation where the odd wave number components progressively gained energy after $t>580$. This energy transfer in the RKwD simulation has arisen as a consequence of the double precision arithmetic where random information at the level of the last significant figure has grown exponentially to the macroscopic level, contaminating all wave numbers in the spatial spectrum.
Particularly,  the  RKwD simulation in a long-duration interval $t\in[0,3000]$ is obviously different from  the CNS result even in statistics!     The overall message of this paper, based on the foregoing findings, is that numerical noise from truncation and round-off errors in double precision arithmetic could lead to huge quantitative and {\em statistical} discrepancies in predictions on spatio-temporal chaos.

It again should be noted that the Lorenz equation essentially derives from a greatly simplified form of the Navier-Stokes equations that are commonly used to describe turbulent flows.  Given that chaos inherently has a close relationship to turbulence, it is important to check the risk  posed by the use of double precision for numerical simulations of turbulent flows, particularly in a rather long interval of time.

Currently,  the deep learning \cite{lecun2015} has been widely used to many complicated problems including the turbulent flows \cite{Raissi-2019-JFM, Raissi-2020-Science} and the three-body problems \cite{breen2019}.  What will happen when the deep learning is applied to solve spatio-temporal chaos whose statistic results are sensitive to numerical noises (as mentioned in this article) ?  This is an interesting but open question.        

Since rather small truncation error is required in time dimension,  a low order temporal algorithm is unacceptable in the frame of the CNS.  For example, it is found that the standard low order method such as the third-order exponential time differencing (ETD) method \cite{cox2002exponential, beylkin1998new}, which is a straightforward extension of the 4th-order Runge-Kutta integration and can provide a rather long time step-size for stiff systems,  requires a quite small time-step $\Delta t \approx 10^{-25} $ to obtain the reliable simulation in the time interval $t\in[0,3000]$ for the considered problem.   Its corresponding CPU time cost is too expansive.

Therefore, high order temporal algorithms must be considered in the frame of the CNS.  Although the exponential time differencing (ETD) method also have {high-order} scheme of multi-step type \cite{cox2002exponential, beylkin1998new},  the {high-order} Taylor series method  still has some irreplaceable advantages in practice.  First,  using the {high-order} Taylor series method is very convenient, because only initial condition is required for it.  But using the  multi-step type ETD is not easy, because a $k$th-order multi-step ETD  requires $k$ {\em previous} time step values.   However,  sufficiently  accurate  previous time step values are  difficult to obtain,  especially when the order $k$ is high.  In addition, this might bring additional numerical noise.    Secondly,  for the {high-order} Taylor series method,  the code is the {\em same} for arbitrary order $M$.   However,  using the multi-step type ETD,  one  had  to  modify  the  code  for  a  different order.   This is  rather  inconvenient, especially when the required order is very high, as illustrated in \cite{LIAO2014On}.   Furthermore,  according to our experience,  the  high-order Taylor series method seems more stable than the multi-step type ETD, especially when the order is rather high.

In this paper, a new CNS algorithm is proposed, which is based on the discretization of unknown variables in physical space, as illustrated in (\ref{discretization}).  The  Fourier collocation method with the FFT is used to calculate the spatial partial derivatives only.  As shown in \S 2.5,  the new CNS algorithm in physical space is computationally much more efficient than its predecessor  in spectral space described in \S 2.1.   Besides,  it gives as accurate simulation as its predecessor, as shown in \S 2.6.   Since the new CNS algorithm directly solves problems in physical space, it is unnecessary for us to use the Fourier pseudo-spectral method \cite{trefethen1991pseudospectra, trefethen2005spectra}, which is  equivalent to  the Fourier collocation method, as pointed by Peyret \cite{Peyret2002Spectral, Peyret2002SpectralBook}.   This also explains why the two CNS algorithms can give the simulations at the  {\em same} level of accuracy, as shown in \S 2.6.

 The CGLE has stiff solutions such as Bekki-Nozaki holes \cite{bekki1985formations},  and there exist isolated non-analyticities when defect chaos occur.    Note that the so-called Bekki-Nozaki holes  \cite{bekki1985formations} has a closed-form solution, which is certainly {\em not} chaotic.   However,  in the case of $c_1=2$, $c_3 = 1$ with the initial condition (\ref{initial2gg}),  the CNS benchmark solution in $t\in[0,3000]$ retains  the ``{\em phase chaos}'',  because  $|A|$ is {\em always} bounded away from zero, as shown in Figs. \ref{minimum-|A|} and \ref{|A|-time},  and besides its phase has {\em no} jumps at all, as shown in Fig. \ref{phase-time}.   This is quite different from the RKwD simulation, whose $|A|$  often exhibits a sudden drop to a value very close to zero after $t  > 1200$ (see Fig.~\ref{minimum-|A|}), corresponding to the {\em defect chaos} displayed in Fig.~\ref{|A|-time} for $|A|$   and  Fig.~\ref{phase-time} for the phase $\varphi$ with occasional jumps indicating the presence of {\em singularities}.   Obviously, the singularity (related to the defect chaos) of the RKwD simulation is a kind of {\em artefact}, caused by the numerical noises and the butterfly-effect of chaos.   Generally speaking, any singularities are difficult to solve numerically.   So, it is worth  investigating  whether  the  CNS  can  handle such kind of singularities in spatio-temporal chaos, when they indeed exist.

  According to our experience of the CNS on the Lorenz equation \cite{Liao2009On, LIAO2014On}, the three-body problem \cite{liao2014physical, liao2015inherent, Li2017More, li20171223, Li2019-NewAstronomy}, the Rayleigh-B\'{e}nard  turbulent flows (governed by the Navier-Stokes equation) \cite{Lin2017On} and so on,  a computer-generated simulation of a chaotic system often has a {\em finite} value of the critical predictable time $T_c$, which seems to be dependent upon values of physical parameters and level of numerical noises.   From the viewpoint of the CNS,  this  is  easy  to  understand, since numerical noise at a {\em specfied} level needs, more or less,  some time to propagate to a {\em critical} level, below which the simulation is {\em clean} and thus {\em reliable}. Obviously,  the lower the {\em specified} level of the numerical noise,  the larger the value of $T_c$, and besides, the lower the {\em critical} level of the noise,  the more accurate and the more reliable is the CNS result.  Here,  $T_c$ is determined by comparing two simulations, using the butterfly-effect of chaos (although it is traditionally often regarded to be negative).   The CNS is generally valid for most of chaotic systems, even if they are {\em not} hyperbolic.   So,  compared to the shadowing lemma \cite{ANOSOV1967, BOWEN1975},  the strategy of the CNS is more {\em practical},  though it loses somewhat mathematical rigour.    Note that,  using a simple model,  Yuan and Yorke \cite{yuan2000collapsing}  showed that a numerical artifact may persist  even for an arbitrary high numerical precision.   Indeed,  there is a long way to go to  reach  a  reliable computer-generated simulation  for  any spatio-temporal chaotic systems in a reasonably long interval of time within a complicated spatial domain.

\section*{Acknowledgment}
Thanks to the anonymous reviewers for their valuable suggestions and constructive comments. This work is financially supported by the National Natural Science Foundation of China (Approval No. 91752104)

\section*{References}

\bibliographystyle{elsarticle-num}

\bibliography{JCP2020}

\end{document}